\documentclass[twocolumn]{aastex62}
\pdfoutput=1 
\usepackage{amsmath,amstext,amssymb,mathtools,graphicx,color}
\usepackage[T1]{fontenc}
\usepackage{apjfonts}
\usepackage[figure,figure*]{hypcap}
\usepackage{chngpage}

\let\splitbox\relax
\usepackage{adjustbox}

\usepackage{graphicx}	
\usepackage{amsmath}	
\usepackage{amssymb}	
\usepackage{xcolor}
\usepackage{soul}

\newcommand{\SuperNu}{{\tt SuperNu}}
\newcommand{\WinNET}{{\tt WinNET}}
\newcommand{\gcc}{{\textrm{g}\ \textrm{cm}^{-3}}}
\newcommand{\erggs}{\textrm{erg}\ \textrm{g}^{-1}\textrm{s}^{-1}}

\received{March 31, 2020}
\submitjournal{ApJ}

\shorttitle{Axisymmetric Models of Kilonovae}
\shortauthors{O. Korobkin et al.}
\begin{document}


\title{Axisymmetric Radiative Transfer Models of Kilonovae}


\correspondingauthor{O. Korobkin}
\email{korobkin@lanl.gov}


\author[0000-0003-4156-5342]{Oleg Korobkin}
\affiliation{Center for Theoretical Astrophysics, Los Alamos National Laboratory, Los Alamos, NM, 87545, USA}
\affiliation{Joint Institute for Nuclear Astrophysics - Center for the Evolution of the Elements, USA}
\affiliation{Computer, Computational, and Statistical Sciences Division, Los Alamos National Laboratory, Los Alamos, NM, 87545, USA}

\author[0000-0003-3265-4079]{Ryan~T. Wollaeger}
\affiliation{Center for Theoretical Astrophysics, Los Alamos National Laboratory, Los Alamos, NM, 87545, USA}
\affiliation{Computer, Computational, and Statistical Sciences Division, Los Alamos National Laboratory, Los Alamos, NM, 87545, USA}

\author[0000-0003-2624-0056]{Christopher~L. Fryer}
\affiliation{Center for Theoretical Astrophysics, Los Alamos National Laboratory, Los Alamos, NM, 87545, USA}
\affiliation{Joint Institute for Nuclear Astrophysics - Center for the Evolution of the Elements, USA}
\affiliation{Computer, Computational, and Statistical Sciences Division, Los Alamos National Laboratory, Los Alamos, NM, 87545, USA}
\affiliation{The University of Arizona, Tucson, AZ 85721, USA}
\affiliation{Department of Physics and Astronomy, The University of New Mexico, Albuquerque, NM 87131, USA}
\affiliation{The George Washington University, Washington, DC 20052, USA}

\author[0000-0001-6893-0608]{Aimee~L. Hungerford}
\affiliation{Center for Theoretical Astrophysics, Los Alamos National Laboratory, Los Alamos, NM, 87545, USA}
\affiliation{Joint Institute for Nuclear Astrophysics - Center for the Evolution of the Elements, USA}
\affiliation{Computational Physics Division, Los Alamos National Laboratory, Los Alamos, NM, 87545, USA}

\author[0000-0002-3833-8520]{Stephan Rosswog}
\affiliation{The Oskar Klein Centre for Cosmoparticle Physics, Stockholm University, AlbaNova, Stockholm SE-106 91, Sweden}
\affiliation{Department of Astronomy, Stockholm University, AlbaNova, Stockholm SE-106 91, Sweden}

\author[0000-0003-1087-2964]{Christopher~J. Fontes}
\affiliation{Center for Theoretical Astrophysics, Los Alamos National Laboratory, Los Alamos, NM, 87545, USA}
\affiliation{Computational Physics Division, Los Alamos National Laboratory, Los Alamos, NM, 87545, USA}

\author[0000-0002-9950-9688]{Matthew~R. Mumpower}
\affiliation{Center for Theoretical Astrophysics, Los Alamos National Laboratory, Los Alamos, NM, 87545, USA}
\affiliation{Joint Institute for Nuclear Astrophysics - Center for the Evolution of the Elements, USA}
\affiliation{Theoretical Division, Los Alamos National Laboratory, Los Alamos, NM, 87545, USA}

\author[0000-0003-1005-0792]{Eve~A. Chase}
\affiliation{Center for Theoretical Astrophysics, Los Alamos National Laboratory, Los Alamos, NM, 87545, USA}
\affiliation{Computational Physics Division, Los Alamos National Laboratory, Los Alamos, NM, 87545, USA}
\affiliation{Center for Interdisciplinary Exploration and Research in Astrophysics (CIERA), Northwestern University, Evanston, IL, 60201, USA}
\affiliation{Department of Physics and Astronomy, Northwestern University, Evanston, IL, 60208, USA}

\author[0000-0002-5412-3618]{Wesley~P. Even}
\affiliation{Center for Theoretical Astrophysics, Los Alamos National Laboratory, Los Alamos, NM, 87545, USA}
\affiliation{Joint Institute for Nuclear Astrophysics - Center for the Evolution of the Elements, USA}
\affiliation{Computer, Computational, and Statistical Sciences Division, Los Alamos National Laboratory, Los Alamos, NM, 87545, USA}
\affiliation{Department of Physical Science, Southern Utah University, Cedar City, UT, 84720, USA}

\author[0000-0001-6432-7860]{Jonah Miller}
\affiliation{Center for Theoretical Astrophysics, Los Alamos National Laboratory, Los Alamos, NM, 87545, USA}
\affiliation{Computer, Computational, and Statistical Sciences Division, Los Alamos National Laboratory, Los Alamos, NM, 87545, USA}

\author[0000-0002-0637-0753]{G.~Wendell Misch}
\affiliation{Center for Theoretical Astrophysics, Los Alamos National Laboratory, Los Alamos, NM, 87545, USA}
\affiliation{Joint Institute for Nuclear Astrophysics - Center for the Evolution of the Elements, USA}
\affiliation{Theoretical Division, Los Alamos National Laboratory, Los Alamos, NM, 87545, USA}

\author[0000-0002-5936-3485]{Jonas Lippuner}
\affiliation{Center for Theoretical Astrophysics, Los Alamos National Laboratory, Los Alamos, NM, 87545, USA}
\affiliation{Joint Institute for Nuclear Astrophysics - Center for the Evolution of the Elements, USA}
\affiliation{Computer, Computational, and Statistical Sciences Division, Los Alamos National Laboratory, Los Alamos, NM, 87545, USA}

\begin{abstract}

The detailed observations of GW170817 proved for the first time
directly that neutron star mergers are a major production site
of heavy elements. The observations could be fit by a number of
simulations that qualitatively agree, but can quantitatively differ
(e.g. in total r-process mass) by an order of magnitude.
We categorize kilonova ejecta into several typical morphologies
motivated by numerical simulations, and apply a radiative transfer
Monte Carlo code to study how the geometric distribution of the ejecta
shapes the emitted radiation.
We find major impacts on both spectra and light curves.
The peak bolometric luminosity can vary by two orders of magnitude
and the timing of its peak by a factor of five.
These findings provide the crucial implication that the ejecta masses
inferred from observations around the peak brightness are uncertain
by at least an order of magnitude.
Mixed two-component models with lanthanide-rich ejecta are particularly
sensitive to geometric distribution.
A subset of mixed models shows very strong viewing angle dependence due
to lanthanide ``curtaining,'' which persists even if the relative mass
of lanthanide-rich component is small.
The angular dependence is weak in the rest of our models, but different
geometric combinations of the two components lead to a highly diverse set
of light curves.
We identify geometry-dependent {P Cygni} features in late spectra that
directly map out strong lines in the simulated opacity of neodymium,
which can help to constrain the ejecta geometry and to directly probe
the r-process abundances.
\end{abstract}


\section{Introduction}
\label{sec:intro}

The origin of the rapid neutron capture (r-process) elements is one of the longest-standing unsolved problems in nuclear astrophysics with
the most popular candidate sites being stellar collapse and compact
object mergers \citep{cowan19}.
Neutron star mergers have long been argued to be sources of r-process elements~\citep{lattimer74,eichler89,rosswog98,rosswog99,freiburghaus99},
mostly because their neutron richness effortlessly leads to the
production of platinum-peak elements; this has been a major challenge for
other production sites. Such rare events (compared to supernovae) that
eject large amounts of r-process per occurrence are also supported
by geological evidence \citep{hotokezaka15,wallner15}.
However, without {\em direct} observational
evidence, our understanding of the enrichment of the universe
by r-process elements was primarily based on theoretical models~\citep[for a review, see][]{cote17}.

This situation changed dramatically with the gravitational and electromagnetic wave detection
of a nearby neutron star merger, GW170817~\citep{abbott17a}.
The optical and infrared emission from this merger \citep{chornock17,cowperthwaite17, kasliwal17b, kilpatrick17, mccully17, rosswog18, smartt17, tanvir17, troja17, villar17} matches theoretical
expectations that much of the ejecta is r-process material
\citep{lattimer74, eichler89, rosswog98, rosswog99, freiburghaus99, goriely11, roberts11, korobkin12, wanajo14}.

The high r-process yields estimated from GW170817 combined with
the large merger rate predicted by its detection~\citep{abbott17h},
indicate that neutron star mergers could be the dominant
site for galactic r-process elements~\citep{cote18,rosswog18} and, in fact,
could actually produce more r-process than is observed.
Current chemical evolution models~\citep{cote19,wehmeyer19}, however,
argue that an additional component is needed, at least at early times, to find
satisfactory agreement with the observed r-process abundances.

In a supernova context, it has been known for a long time
that inferring masses from the optical emission is difficult and
that a given light curve can be matched by a wide range of ejecta
masses~\citep[see][for example]{delarosa17}. Differences in the modeling
methods (e.g. opacity implementations, transport schemes) and the applied
microphysics (e.g. opacities, shock physics, equilibrium assumptions)
both contribute to these errors.
The situation is similar for GW170817, where a range of masses
has been inferred by different groups \citep{cote18,ji19}.
To better observationally constrain neutron star merger yields, kilonova emission models need to be refined, with a particular emphasis on exploring what role the ejecta geometry plays in shaping the electromagnetic emission.
This is the subject that we address here.

\begin{figure}
	\includegraphics[width=\columnwidth]{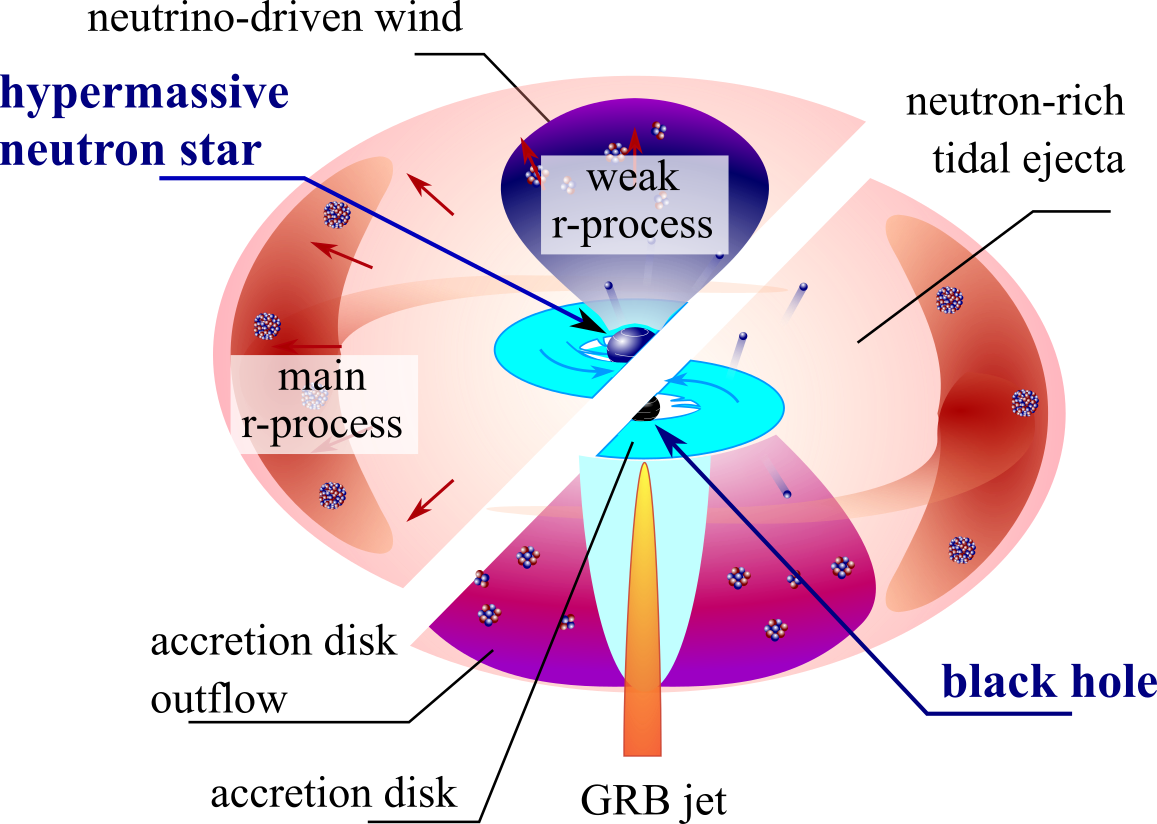}
    \caption{Sketch of the standard kilonova scenario, involving either a
    hypermassive neutron star (upper left) or a black hole as
    (lower right) a central object. Various types of ejected matter
    are depicted, contributing to the main and weak r-process production
    \citep[see, e.g.][for a review on mass ejection in NSMs]{shibata19}.}
    \label{fig:kilonova-sketch}
\end{figure}

Recent studies \citep{bauswein13, fernandez13, rosswog14, just15, fernandez15, radice16, sekiguchi16, baiotti17, rosswog17, shibata17, fahlman18, metzger18a, papenfort18, siegel18, radice18, miller19b, krueger20} have revealed different ejecta channels and morphologies.
These sources include tidal tails with low electron fractions ($Y_e<0.2$), polar shocked material (sometimes referred to as squeeze ejecta), magnetar or neutrino-driven wind ejecta (all with $Y_e \sim 0.25-0.45$) and disk wind material with electron fractions ranging from $0.1$ to $0.4$~\citep{metzger19}.
A number of models have invoked two \citep{evans17, tanaka17, tanvir17, troja17, kawaguchi18} or even three \citep{cowperthwaite17, perego17, villar17} ejecta components to explain the electromagnetic observation of GW170817.
A typical kilonova scenario with multicomponent ejecta mechanisms is sketched in Figure~\ref{fig:kilonova-sketch}.
The components may include toroidal tidally expelled highly neutron-rich ejecta, which produces the red (near-IR; nIR-IR) emission in the kilonova, or a medium neutron-rich wind emerging from the surface of a transient hypermassive neutron star and/or an accretion disk \citep[see][for a review on mass ejection in kilonovae]{shibata19}.

It appears that the models ignoring the multidimensional character of kilonova ejecta tend to infer higher ejected masses than multidimensional models.
On the other hand, the models with a true multidimensional treatment of radiative transfer \citep{tanvir17,troja17,kawaguchi18} seem to favor lighter and faster neutron-rich ejecta accompanied by a heavy ``wind'' with moderately neutron-rich composition \citep[shown in Fig.~4 of][]{ji19}.
This scenario is also supported by numerical simulations \citep[e.g.][]{perego14, radice16, dietrich17, shibata19}.

The neutron-rich dynamical ejecta, or the neutron-rich outflows from the post-merger accretion disk \citep{janiuk14, miller19b} are expected to contain a large fraction of high-opacity lanthanides \citep{barnes13, tanaka13, fontes15a, gaigalas19, fontes19}.
The properties of this lanthanide-rich outflow dictate how much light in optical and red bands will be blocked; in some cases, lanthanide-rich geometric configurations can produce an effective lanthanide ``curtain'' \citep{kasen15,wollaeger18,nativi20}.
In such cases, the kilonova can be highly sensitive to the viewing angle.

Kilonova light curves and spectra are influenced by both ejecta morphology and microphysics, for example
 complicated, wavelength-dependent opacities \citep{kasen17, even20}, nuclear heating \citep{lippuner15}, and thermalization \citep{barnes16, rosswog17, hotokezaka19}.
Opacities, nuclear heating, and thermalization are all sensitive to the density, which is specified by geometric distribution of mass.
As we shall see, deviations from the spherical shape are responsible for modified light curves not only via modified surface area or ``curtaining,'' but mainly through an indirect effect on the microphysics because of their different density.

Previously, \cite{kasen17} studied different ejecta geometries: one with a light r-process composition focused around the axis to represent wind, and one with a heavy r-process composition and oblate ellipsoidal morphology to represent dynamical ejecta.
In these models, the dynamical ejecta with an ellipsoid axis aspect ratio of four produces the angular variation on the order of a factor of two.
In the wind ejecta, when focused into $\sim$45$^\circ$ conical regions about the axis,~\cite{kasen17} find an angular variation of $\sim$20\% in peak luminosity.
The relatively minor angular variation in peak luminosity of the wind is a result of the less extreme variation in the projected surface area of the wind, relative to the dynamical ejecta, so that the high lanthanide opacity of the low-electron-fraction ejecta played a diminished role.

In several recent works \citep{barbieri19,darbha20,zhujp20}, sensitivity of the light curves to inclination angle was investigated using uniform gray opacity with a variety of nonspherical 2D and 3D analytically prescribed shapes.
The resulting light curves exhibit only mild variability, within a factor of $\sim$2-3, because the projected area of the photosphere, which serves as a proxy for angular dependence, does not vary substantially.
However, the location of the photosphere itself, and more importantly, its temperature, strongly depend on the properties of material-specific opacities.
Here we explore the differences between detailed and gray-opacity prescription and conclude that more realistic detailed opacity enhances morphological variability in the light curves beyond the projected-area effects.

In \cite{bulla19}, it was demonstrated that a multidimensional two-component model deviates strongly from a simple sum of single-component ones. However, bound--bound opacities were approximated using a time-dependent analytic fit. This excludes temperature feedback on the opacity, which is also substantial, as we show here.
\cite{kawaguchi20} performed advanced special-relativistic radiative transfer simulations of axisymmetric two-component models of kilonova from dynamical ejecta with morphology fits to numerical relativity \citep{kiuchi17, hotokezaka18} and spherical post-merger outflows. They used a full suite of line r-process opacities \citep{tanaka20} and found several strong effects of multidimensional, multicomponent interaction, such as preferential diffusion toward polar regions in the presence of toroidal optically thick dynamical ejecta and significant dependence on inclination angle due to the ``lanthanide curtaining.'' We describe similar effects observed in our models, but also take a broader set of possible morphologies and compare the light curves to gray-opacity models in order to understand their limitations.

In this paper, we study the importance of morphology in combination with its effects on microphysics, focusing on the geometrical distributions that depart from spherical symmetry.
Ejecta morphologies are generated from the family of Cassini ovals in 2D axisymmetric geometry.
For these morphologies, we demonstrate the enhancement to the radiative flux when departing from spherical symmetry.
We then proceed to construct two-component models, representing different possible configurations of low-electron-fraction ($Y_e$) dynamical ejecta/accretion disk outflows, and high-$Y_e$ wind.
For each of the components, we employ a respective composition and nuclear heating representative of low-$Y_e$ or high-$Y_e$ post-nucleosynthetic conditions.
Here we demonstrate the following effects of radiative interaction between two superimposed components: lanthanide curtaining, photon reprocessing, and photon redirection.
Our other recent work provides preliminary studies focusing specifically on the effects of composition~\citep{even20} and additional energy sources \citep{wollaeger19}.

This paper is organized as follows.
In Section~\ref{sec:methods}, we discuss the numerical methods and approximations made in our simulations.
In Section~\ref{sec:models}, we describe the functional forms and compositions of our model ejecta.
In Section~\ref{sec:results}, we present light curves and spectra for several of our one- and two-component models, and discuss the effect of morphology and superimposing of the components on these observables.
We conclude in Section~\ref{sec:conclusions} with a summary of the results and discussion of the possible impact of morphology on the interpretation of observations.
In Appendix~\ref{sec:model_tables}, we provide absolute AB magnitudes for each model, in the $r$ and $J$ bands at days 1, 4, and 8.
In Appendix~\ref{sec:rad_structure}, we give a detailed analysis of the radiative structure of single-component axisymmetric morphologies and different factors affecting their electromagnetic emission.

\section{Methods}
\label{sec:methods}

In this section, we review our approach for modeling kilonovae, which involves computing multiwavelength opacity tables and using them to simulate radiative transfer through homologously expanding ejecta.
Our code uses the same method and opacity implementation as described in~\cite{wollaeger18, wollaeger19}, \cite{even20}, and \cite{fontes19}.
Below we briefly review this methodology and then focus primarily on the improvements to the opacity and radiative transfer introduced for this paper.

\subsection{Radiative transfer modeling}

We employ the multidimensional Monte Carlo radiative transfer software \SuperNu\footnote{\url{https://bitbucket.org/drrossum/supernu/overview}} \citep{wollaeger14}
to synthesize the spectra from our models.  \SuperNu\ uses a Monte Carlo method for thermal radiative transfer that is semi-implicit in time, which introduces an effective scattering term~\citep{fleck71}.  A discrete diffusion Monte Carlo (DDMC) optimization is also implemented to replace multiple small scattering steps with larger jumps between spatial cells
\citep{abdikamalov12, densmore12, cleveland14}. DDMC has recently been generalized to include leakage probabilities out of lines for Lyman-$\alpha$ transport~\citep{smith18}.  The geometries available are spherical, axisymmetric, and Cartesian, each in 1D, 2D, or 3D.  In the calculations presented here, we have incorporated an improved treatment of Doppler shift in DDMC (R.~T.~Wollaeger et al 2020, in preparation), that is consistent with the opacity-regrouping algorithm described by~\cite{wollaeger14}.
This Doppler shift algorithm has permitted more accurate application of DDMC on the binned opacities
described by~\cite{fontes19}.

\begin{figure*}
\begin{tabular}{ccccc}
    \includegraphics[height=0.35\columnwidth]{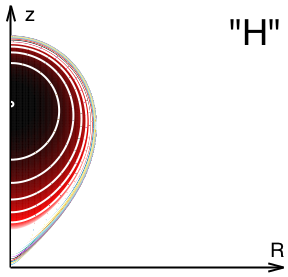} &
    \includegraphics[height=0.35\columnwidth]{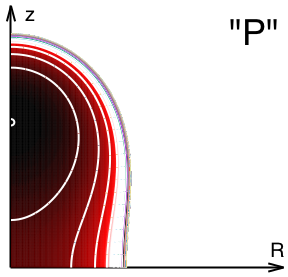} &
    \includegraphics[height=0.35\columnwidth]{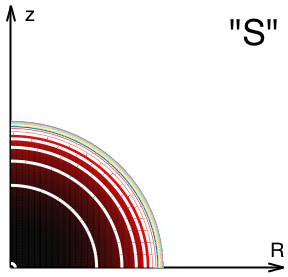} &
    \includegraphics[height=0.35\columnwidth]{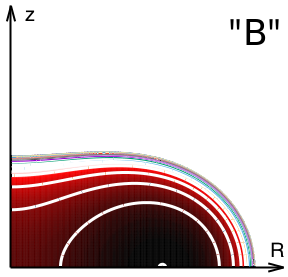} &
    \includegraphics[height=0.35\columnwidth]{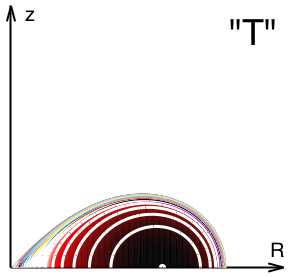}
\end{tabular}
\caption{Maps of density of the basic building-block morphologies used in this work, shown in the upper polar half-plane (the $R-z$ plane, with $R$ being the cylindrical radius): "hourglass" (H), "peanut" (P), "sphere" (S), "biconcave" (B), and "torus" (T).
The contours are equi-spaced by 0.5 dex in log space, down from the density maximum.}
\label{fig:basic_morphologies}
\end{figure*}
Originally intended for supernova transients~\citep{vanrossum16,kozyreva17,wollaeger17},
\SuperNu\ has been applied to modeling kilonovae in 1D and 2D axisymmetric geometry~\citep{evans17, kasliwal17a, tanvir17, troja17, wollaeger18, wollaeger19, even20}.
These studies varied the mass and velocity of wind and dynamical ejecta components but always assumed a spherical wind superimposed with ejecta from SPH simulations~\citep{rosswog14}.
The morphologies we explore in this paper more systematically test the multidimensional functionality of \SuperNu.

\subsection{Choice of composition and opacities}

We use tabulated opacities generated with the LANL suite of atomic physics codes~\citep{fontes15a}.
As with~\cite{fontes17} and \cite{wollaeger18}, the tables are calculated under the assumption of local thermodynamic equilibrium (LTE), where ionization and excitation populations can be obtained from Saha--Boltzmann statistics.
For this work, we use an updated approach to the tabulation of opacity, which employs the binning of lines instead of smearing them over multiple wavelength points \citep{fontes19}.
In conjunction with the higher group resolution in \SuperNu\, this approach permits more resolved spectra for lower-velocity outflows.
For the radiative transfer, the frequency values for each density and temperature point of these tables are mapped to a logarithmic 1024-group wavelength grid from $10^3$ to $1.28\times10^5$~\AA.

All models in this study use one of the two compositions, representing a low- and high-electron-fraction ($Y_e$) abundance pattern.
The low-$Y_e$ abundance pattern follows the solar r-process residuals and is the same as that in \cite{even20}, their Fig.~1(b).
It includes all lanthanides, uranium representing the actinides, and lighter r-process elements represented by Se, Br, Te, Pd, Zr, and Fe.
The high-$Y_e$ abundance pattern is the ``wind 2'' composition studied in \cite[][see their Fig.~3]{wollaeger18}.
Throughout this paper, we will refer to these two cases as \emph{``low-$Y_e$''} and \emph{``high-$Y_e$''} compositions.
Sometimes, for simplicity, we refer to these components as \emph{``dynamical ejecta''} and \emph{``wind,''} respectively, but in general, low-$Y_e$ material does not necessarily come from the dynamical ejection \citep{fernandez13}, while high-$Y_e$ material does not have to be generated in the wind \citep{wanajo14}.

Nuclear composition influences the amount and type of radioactive nuclear heating produced in the ejecta over time \citep{lippuner15,zhuyl18,zhuyl20}.
We use detailed time-dependent nuclear heating output from a nucleosynthesis network which calculates separate contributions from different types of radiation ($\alpha$-, $\beta$-, $\gamma$-radiation, and fission fragments) and their thermalization.
The latter is estimated using phenomenological formulae from \cite{barnes16}
\citep[as in][]{rosswog17,wollaeger18}.

\subsection{Ejecta Geometries}
\label{sec:models}

We adopt a suite of ejecta geometries motivated by numerical simulations of dynamical ejecta from binary neutron stars and winds from accretion disks and post-merger hypermassive neutron star. We focus on single- and two-component models with uniform composition across individual components.

\subsubsection{Single-component models}
\label{sec:singlecomp}

\begin{figure}
\begin{tabular}{c}
  \includegraphics[width=\columnwidth]{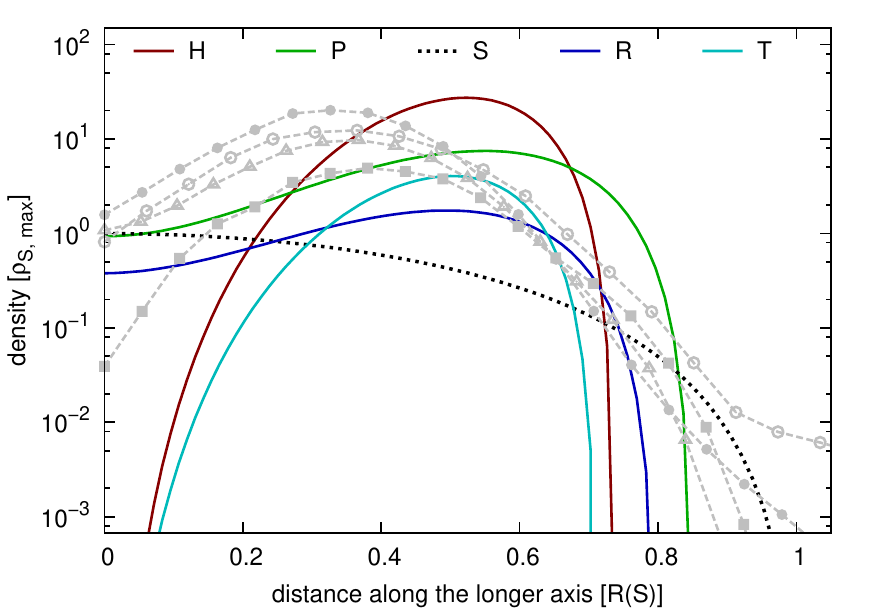}
\end{tabular}
\caption{Density profiles of the basic building-block morphologies along the
axis through the $\rho_{\rm max}$. Morphologies are
constrained to have the same mass and average (RMS) expansion velocity.
In gray: density profiles along the radial lines through the maximum density
of the dynamical ejecta from neutron star mergers~\citep{rosswog14}.
} 
\label{fig:density_SHPRT}
\end{figure}

\begin{figure}
  \centering
  \includegraphics[width=0.80\columnwidth]{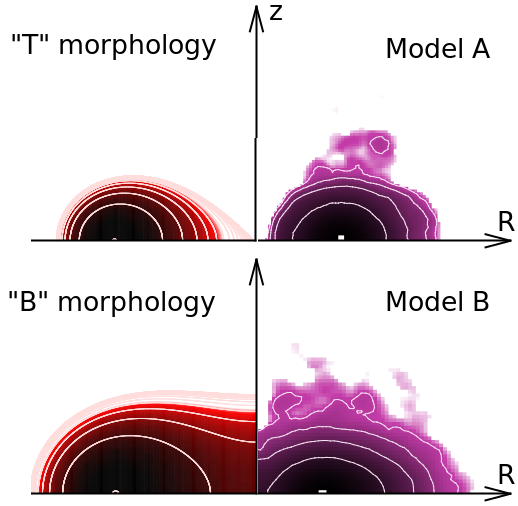}
  \caption{Morphologies ``T'' (toroidal) and
    ``B'' (biconcave; left panels) in comparison with toroidal axisymmetrized
    models from \cite{rosswog14}
    \citep[right panel; also used in][]{wollaeger18,heinzel21}.
  } 
  \label{fig:abinitio}
\end{figure}

\begin{table}
\centering
  \caption{Nondimensional shape factors $C_v$ and $C_\rho$ (see Eq.~(\ref{eq:CvCrho})).}
  \begin{tabular}{lll}
  \hline\hline
  Morphology & $C_v$ & $C_\rho$ \\
  \hline
  ``H'' -- hourglass  &  1.028850  &  5.367093 \\
  ``P'' -- peanut     &  1.079134  &  0.183475 \\
  ``B'' -- biconcave  &  0.986012  &  0.0222324 \\
  ``T'' -- torus      &  1         &  0.810569 \\
  ``S'' -- sphere     &  1.914854  &  0.223138 \\
  \hline\hline
  \end{tabular}
  \label{tab:CvCrho}
\end{table}

Although the variety of possible morphologies is limitless, we have picked a
few morphologies that are motivated by the astrophysical setting and easy to set up.
Figure~\ref{fig:basic_morphologies} depicts the five baseline axisymmetric
morphologies: two prolate (H "hourglass," P "peanut"), one spherical (S),
and two oblate (B "biconcave," T "torus").
The density for each model is given in velocity space axisymmetric coordinates ${\{v_r,\theta\}}$ using the level function for a family of Cassini ovals:
\begin{align}
    \rho(v_r,\theta)  = \rho_0 \left(\frac{t}{t_0}\right)^{-3} \times
      \begin{cases}
      (q - \bar{v}_r^4 - 2\bar{v}_r^2\cos{2\theta})^3\;\;\;{\rm prolate\,shapes}, \\
      (q - \bar{v}_r^4 + 2\bar{v}_r^2\cos{2\theta})^3\;\;\;{\rm oblate\,shapes},
      \end{cases}
    \label{eq:cassini}
\end{align}
The value $q=0$ gives a "figure-eight" shape, which if rotated around the $z$-axis
produces our "H" and "T" morphology for the prolate and oblate cases, respectively.
We use ${q = 1}$ for the "P" and ${q = 1.5}$ for the "B" morphologies (see Fig.~\ref{fig:basic_morphologies}). The spherical morphology "S" is specified by the ``cubed inverted parabola'' density profile, introduced in \cite{wollaeger18}:
\begin{align}
    \rho(v_r) = \rho_0 \left(\frac{t}{t_0}\right)^{-3}
                \left(1 - \bar{v}_r^2\right)^3\;\;\;\text{-- spherical shapes.}
    \label{eq:cubedparabola}
\end{align}
In equations (\ref{eq:cassini}-\ref{eq:cubedparabola}), $\rho_0$ sets the density scale, $\bar{v}_r=v_r/v_0$, and $v_0$ sets the velocity scale.
These dimensional constants are connected with the ejecta mass $m_{\rm ej}$ and rms expansion velocity $v_{\rm ej}$ as follows:
\begin{align}
    v_0 &= C_v v_{\rm ej}, &
    \rho_0 &= C_\rho \frac{m_{\rm ej}}{v_{\rm ej}^3 t_0^3},
    \label{eq:CvCrho}
\end{align}
where the nondimensional shape factors $C_v$ and $C_\rho$ are listed in Table~\ref{tab:CvCrho} for each of the morphologies.

Models with the same mass and average expansion velocity can possess very different density profiles.
The latter are shown in Figure~\ref{fig:density_SHPRT} for our five basic morphologies along their axes that pass through $\rho_{\rm max}$.
The maximum densities vary by nearly two orders of magnitude between S and H morphologies, with the S(H) morphology having the lowest(highest) maximum density.
For comparison with numerical simulations, Figure~\ref{fig:density_SHPRT} also presents spherically averaged density profiles (in gray) of the dynamical ejecta from neutron star merger simulations of \cite{rosswog14} \citep[also used in][see their Fig.~2]{wollaeger18}.
Differences in the density from simulations vary by an order of magnitude.
Although variability in our models may be somewhat exaggerated, it is not unreasonable to expect it for other ejecta generation scenarios.
Our choice of morphologies attempts to capture this diversity.

The spherical and prolate morphologies are motivated by simulations of post-merger outflows~\citep[e.g.][]{perego14,martin15}, which in general tend to concentrate toward the polar regions.
Another case of prolate morphology is the ``squeezed'' collisional ejecta in the polar regions~\citep{oechslin07,sekiguchi16,kasen17}.
The oblate morphologies are motivated by simulations of tidally expelled dynamical ejecta, which are naturally concentrated toward the merger plane~\citep{rosswog14,sekiguchi16}.
Figure~\ref{fig:abinitio} compares the adopted toroidal morphologies ``T'' and ``B'' to the simulated models A and B from \cite{rosswog14} \citep[also used in][]{wollaeger18,heinzel21}.

Our one-component models assume uniform composition and are constructed by rotating Cassini ovals around vertical axis as specified above.
Although we expect multiple components in the true outflows, by simulating these one-component models, we can better understand the impact of geometry, such as the effects of the surface area or high-density regions.
This allows for a better assessment of the impact of superimposing morphologies (e.g. lanthanide curtain).
In addition, these nonspherical one-component models can be directly compared to other models in the literature~\citep[such as those studied in][]{perego17,darbha20,zhujp20}.

\subsubsection{Two-component models}

\begin{figure}
\begin{tabular}{cc}
    \includegraphics[width=0.45\columnwidth]{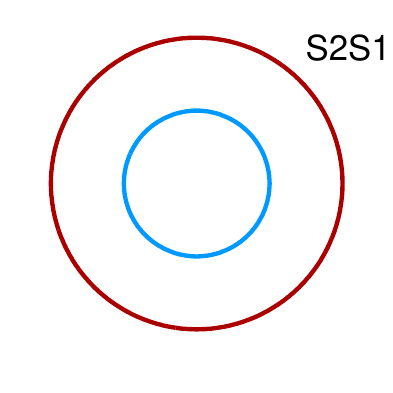} &
    \includegraphics[width=0.45\columnwidth]{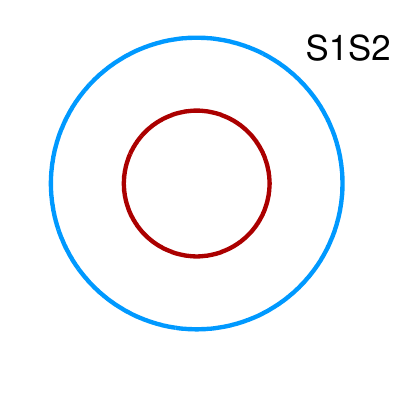}
    \\
    \includegraphics[width=0.45\columnwidth]{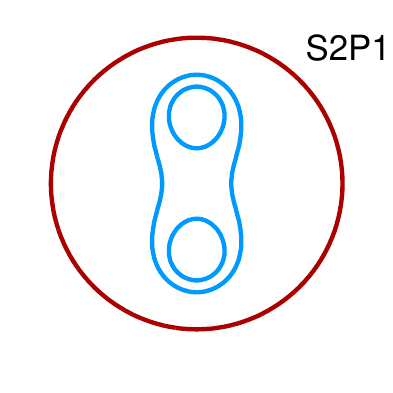} &
    \includegraphics[width=0.45\columnwidth]{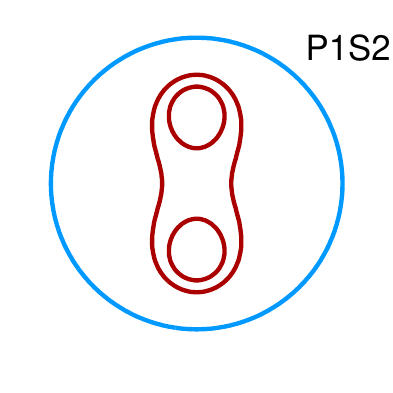}
    \\
    \includegraphics[width=0.45\columnwidth]{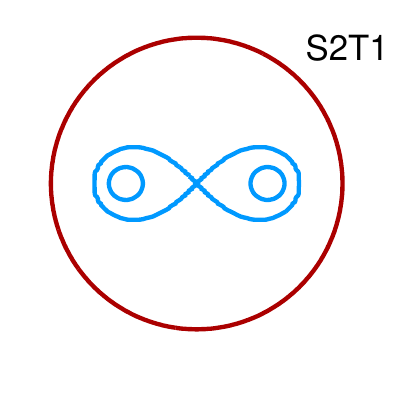} &
    \includegraphics[width=0.45\columnwidth]{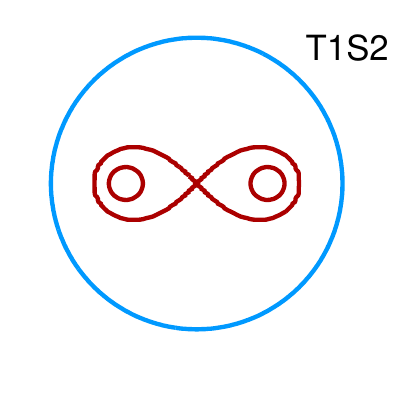}
    \\
    \includegraphics[width=0.45\columnwidth]{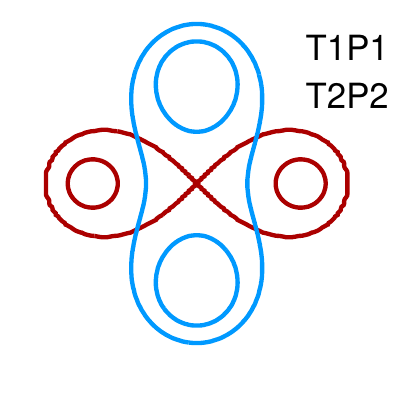} &
    \includegraphics[width=0.45\columnwidth]{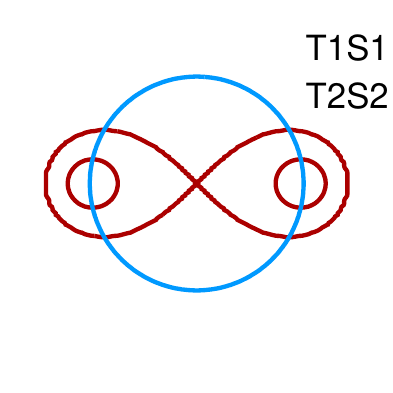}
    \\
    \includegraphics[width=0.45\columnwidth]{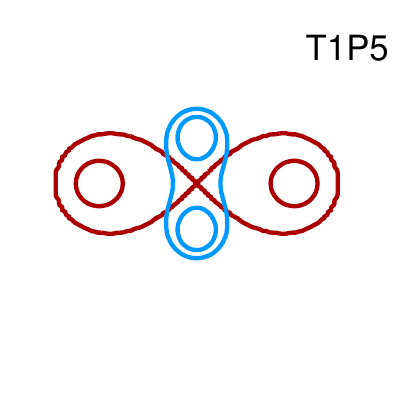} &
    \includegraphics[width=0.45\columnwidth]{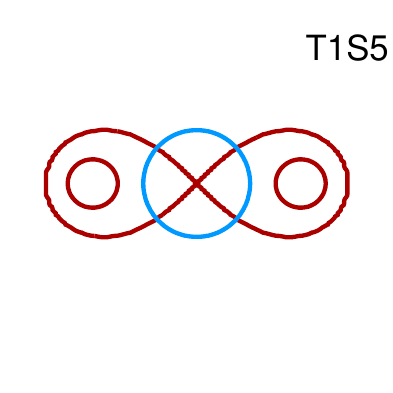}
\end{tabular}
\caption{Schematics of the adopted two-component combinations of morphologies
in the meridional plane.
The first letter represents abbreviation of the neutron-rich (low-$Y_e$) outflow
morphology, while the second letter stands for the higher-$Y_e$ outflow shape.
The low-$Y_e$ (high-$Y_e$) component is represented by the red (blue) contour.
The first (second) digit encodes mean expansion velocity of the low-$Y_e$
(high-$Y_e$) component: ${1 = 0.1\,c}$, ${2 = 0.2\,c}$, and ``5'' designates
${0.05\,c}$. This is reflected in the relative sizes of the components. }
\label{fig:twoc_morphologies}
\end{figure}

We next combine our one-component morphologies to have a suite of two-component (mixed) models, with overlapping high-$Y_e$ and low-$Y_e$ compositions.
In the mixed models, we only pick three basic morphologies: S, P, and T (combinations with H instead of P and B instead of T produce very similar results).
The selected combinations are outlined in Fig.~\ref{fig:twoc_morphologies}, with the low-$Y_e$ and high-$Y_e$ compositions shown in red and blue, respectively.
In the model names, the first and second letters stand for morphology of the low-$Y_e$ and high-$Y_e$ components, respectively, and the number designates the first significant digit in the mean velocity of expansion for the corresponding component.
More exactly, "1" designates a mean expansion velocity of ${0.1\,c}$, "2" designates ${0.2\,c}$, and "5" designates ${0.05\,c}$.
For example, {\tt T1S2} stands for a model with a toroidal low-$Y_e$ component with a mean expansion velocity ${0.1\,c}$ and a spherical high-$Y_e$ component with a mean expansion velocity ${0.2\,c}$.
We consider a variety of masses for the blue and red components, with the baseline, default mass for either of the components being ${0.01\ M_\odot}$.

Each superposition of the components represents a specific scenario in outflow configurations.
For instance, the {\tt TS} ({\tt TP}) morphologies can model a case when tidal dynamical ejecta has a toroidal shape, while the secondary wind outflow is spherical (axially focused).
Indeed, recent ab initio simulations of remnant accretion disks produce axially elongated outflows with spatially variable neutron richness, such that high-$Y_e$ ejecta expands along the axis, while low-$Y_e$ ejecta tends toward the equatorial plane \citep{fernandez13, miller19b}.
The bi-spherical models {\tt S1S2} ({\tt S2S1}) represent the simpler scenario of isotropic two-component outflow, in which high-$Y_e$ (low-$Y_e$) overtakes the other component, forming an outer envelope and obscuring it~\citep{fernandez16o}.
Some studies \citep{wanajo14} have suggested that the neutron fraction in a fast-moving interaction component, also known as ``squeezed'' dynamical ejecta \citep{goriely11,korobkin12}, can be reset to higher values by the intense heating and positron captures, which become available in abundance at high temperatures.
Morphologies {\tt T1S2}, {\tt S1S2} and {\tt P1S2} represent this scenario in our models.
The models {\tt P1S2} and {\tt S2P1} represent polar outflow of one type, enveloped by a spherical outflow of the other type.
The {\tt S2P1} scenario can be realized if initial low-$Y_e$ ejecta is fast and isotropic \citep{sekiguchi16}, and secondary high-$Y_e$ ejecta is slow and prolate.
The slow component can be realized by neutrino-driven wind with prolate morphology, either from a transient hypermassive neutron star~\citep{perego14}, or an accretion disk~\citep{miller19b}.
Overall, these configurations represent a reasonably exhaustive set of outflow scenarios that can happen in neutron star mergers.
However, they are still limited to axisymmetry.
This may be too constraining for kilonovae from neutron star--black hole mergers, where even more diversity is expected due to the strong non-axisymmetric nature of the ejecta in some cases~\citep{kyutoku13,perego17,zhujp20}.

\begin{figure}
\begin{tabular}{c}
  \includegraphics[width=\columnwidth]{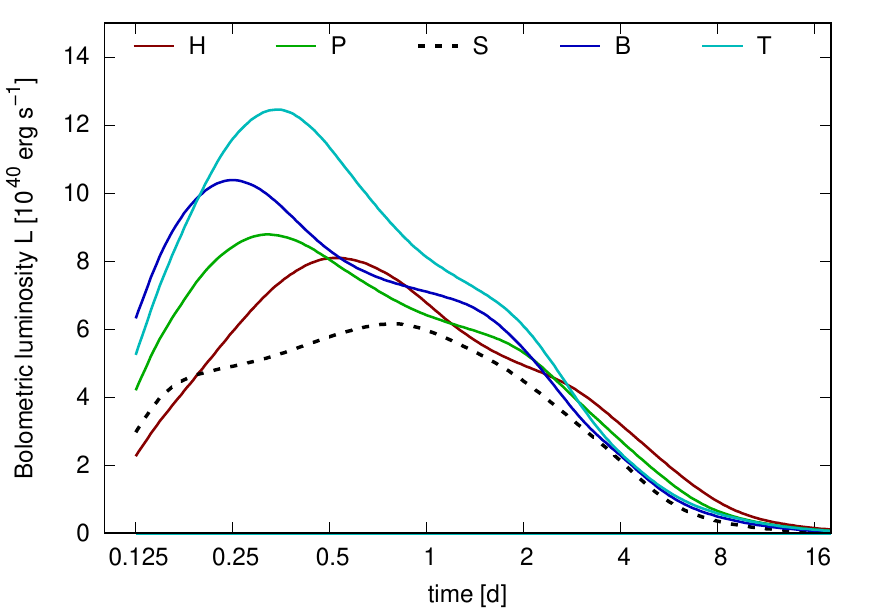}
\end{tabular}
\caption{Integral bolometric luminosity (summed over
  all angular bins, as a function of time for the five single-component
  morphologies and the low-$Y_e$ (neutron-rich, solar r-process residuals)
  composition. Full network heating, density-dependent thermalization, and
  detailed suite of opacities is used.
} 
\label{fig:onec_luminosities}
\end{figure}

\begin{figure}
\begin{tabular}{c}
  \includegraphics[width=\columnwidth]{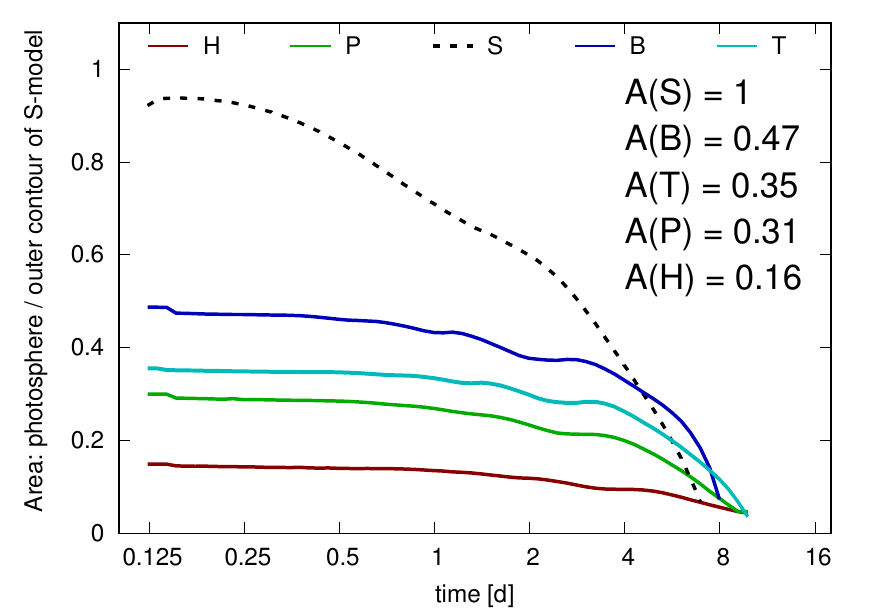} \\
\end{tabular}
\caption{Area of the receding diffusion surface relative to the area of the
outflow boundary of equivalent spherical model with the same mass and average
expansion velocity, as a function of time, for the range of single-component
models with low-$Y_e$ composition.
The labels show the area of the outer surface of the morphologies relative to the
spherical model.
} 
\label{fig:onec_areas}
\end{figure}

\begin{figure*}
\begin{tabular}{cc}
  \includegraphics[width=0.48\textwidth]{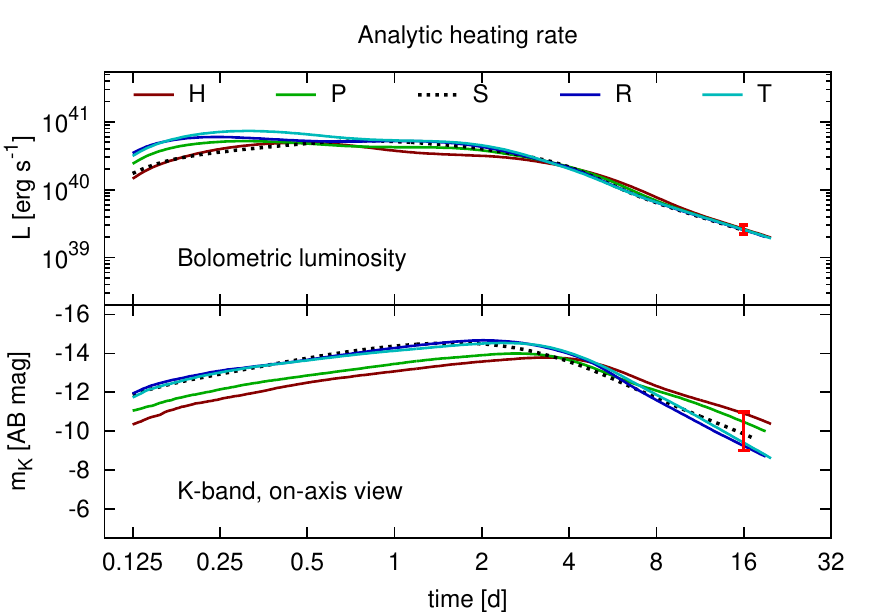} &
  \includegraphics[width=0.48\textwidth]{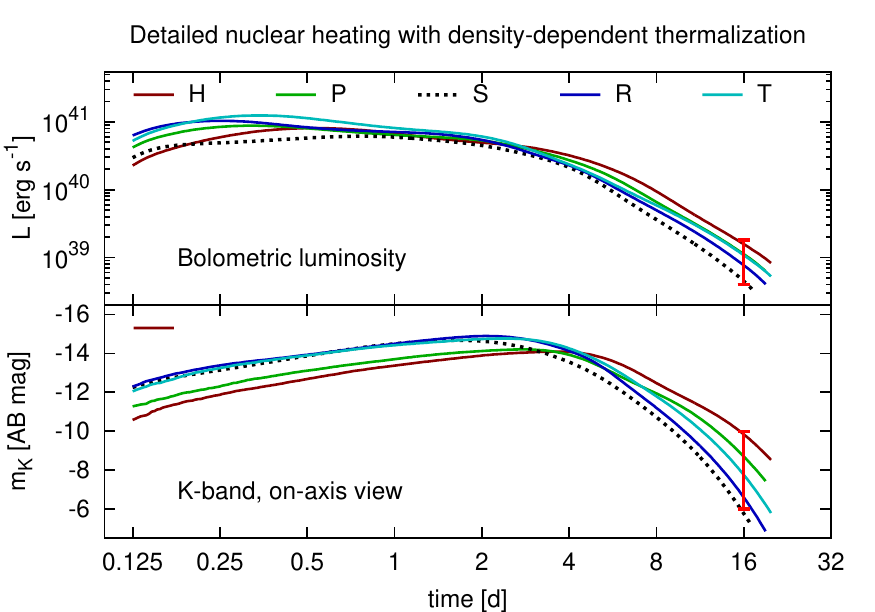}
\end{tabular}
\caption{Angle-integrated bolometric luminosity (top panels) and $K$-band magnitudes for single-component morphologies of the same mass and velocity, computed with analytic heating prescription and uniform thermalization (left) vs detailed nuclear heating input with density-dependent thermalization (right).
At late times, such as 16 days (highlighted in red), bolometric luminosity is not dependent on geometry when an analytic heating rate is used.
For detailed nuclear heating with density-dependent thermalization (right panels), there is significant morphology-dependent variability in bolometric luminosity.
Unlike bolometric light curves, the broadband ones, such as the $K$ band displayed here, show strong dependence on morphology in both cases (see the main text for discussion).
} 
\label{fig:late_times}
\end{figure*}

\section{Results}
\label{sec:results}

Here we explore the range of possible kilonova signals based on the selected suite of ejecta morphologies and two representative compositions, separately expected to produce the ``blue'' and ``red'' kilonova.
For the majority of simulations, we take the mass of either one or both components to be $0.01\,M_\odot$, in order to put more focus on the effects of ejecta geometry.
Below, we first study single-component models and then proceed to two-component models.

\subsection{Single-component Models}
\label{sec:single-component}

Figure~\ref{fig:onec_luminosities} shows the evolution of angle-integrated bolometric luminosity for the one-component axisymmetric morphologies.
The departure from spherical symmetry can have a large impact on the peak time and luminosity: {\tt B} and {\tt T} morphologies turn out to be brighter than the spherical model by a factor of two, and only need half of the time to reach their peak luminosity.
The trend in brightness and peak epoch is set by the temperature of the outer radiative layer and, to a lesser degree, by the ejecta surface area.
This can be inferred from Figure~\ref{fig:onec_areas} which shows the ratio of the diffusion surface area to the area of the outer boundary of equivalent spherical model, as a function of time.
It is clear that the surface area \emph{is not} the dominant factor in deciding the peak brightness and epoch.
The diffusion surface here is defined as enclosing the opaque ``core'' of the ejecta where photons are escaping slower than local expansion and are therefore trapped~\citep{grossman14}.
As the expansion progresses, the diffusion surface (along with the photosphere) recedes inwards until the entire bulk becomes exposed and transparent.
To better understand the uncertainties in what sets the brightness of single-component models, we present a more extended discussion of these aspects in Appendix~\ref{sec:rad_structure}.

\begin{table*}
  \centering
  \scriptsize
  \caption{Ejecta masses inferred using equation~\ref{eq:mass_from_tLpeak}, for single-component and mixed morphologies.
  The second column shows the mass (masses) of the single component (low-$Y_e$ component + high-$Y_e$ component) in percents of the Solar mass (${0.01\,M_\odot}$).}

\begin{tabular}{rc|ccc|ccc|ccc}
\hline\hline
    \multicolumn{2}{c}{Model}
    & \multicolumn{3}{|c}{top orientation}
    & \multicolumn{3}{|c}{side orientation}
    & \multicolumn{3}{|c}{average}
\\
morphology & $m_l + m_h$
      & $L_{\rm peak}$ & $t_{\rm peak}$ & mass
      & $L_{\rm peak}$ & $t_{\rm peak}$ & mass
      & $L_{\rm peak}$ & $t_{\rm peak}$ & mass
\\
$\,$ & $[0.01 M_\odot]$
      & [$10^{40}$ erg/s]  & [d]  & $[0.01 M_\odot]$
      & [$10^{40}$ erg/s]  & [d]  & $[0.01 M_\odot]$
      & [$10^{40}$ erg/s]  & [d]  & $[0.01 M_\odot]$
\\
\hline
 low-$Y_e$ composition:
   H1 & 1 &   5.4 & 1.12 &  2.6 &    6.5 & 1.09 & 3.2 &     6.4 & 1.11 & 3.2 \\
   P1 & 1 &   5.7 & 0.72 &  1.4 &    8.1 & 0.70 & 2.1 &     7.4 & 0.70 & 1.9 \\
   B1 & 1 &  12.1 & 0.59 &  2.5 &    7.0 & 0.54 & 1.2 &     9.0 & 0.56 & 1.7 \\
   T1 & 1 &  14.7 & 0.75 &  4.7 &    7.0 & 0.69 & 1.7 &    10.4 & 0.73 & 3.0 \\
   H2 & 1 &   9.1 & 0.50 &  1.4 &    7.1 & 0.52 & 1.1 &     8.1 & 0.52 & 1.3 \\
   P2 & 1 &   9.2 & 0.31 &  0.7 &    8.3 & 0.32 & 0.6 &     8.7 & 0.32 & 0.7 \\
   B2 & 1 &  11.5 & 0.26 &  0.7 &    9.5 & 0.24 & 0.5 &    10.3 & 0.24 & 0.6 \\
   T2 & 1 &  14.8 & 0.37 &  1.6 &   10.2 & 0.31 & 0.8 &    12.4 & 0.34 & 1.1 \\
   S2 & 1 &   6.1 & 0.78 &  1.7 &    6.1 & 0.78 & 1.7 &     6.1 & 0.78 & 1.7 \\
high-$Y_e$ composition:
   H1 & 1 &  24.3 & 0.79 &  9.3 &   29.7 & 0.83 &12.5 &    29.2 & 0.82 &12.0 \\
   P1 & 1 &  27.1 & 0.56 &  6.3 &   41.8 & 0.57 &10.6 &    38.0 & 0.57 & 9.6 \\
   B1 & 1 &  68.9 & 0.47 & 14.5 &   34.1 & 0.43 & 5.4 &    47.7 & 0.46 & 8.9 \\
   T1 & 1 &  68.9 & 0.54 & 17.9 &   30.7 & 0.50 & 6.0 &    48.3 & 0.53 &11.3 \\
   H2 & 1 &  47.7 & 0.40 &  7.3 &   40.2 & 0.47 & 7.6 &    44.2 & 0.44 & 7.7 \\
   P2 & 1 &  52.1 & 0.28 &  4.6 &   53.6 & 0.32 & 5.9 &    54.5 & 0.31 & 5.7 \\
   B2 & 1 &  80.4 & 0.26 &  7.0 &   54.8 & 0.21 & 3.3 &    65.3 & 0.23 & 4.7 \\
   T2 & 1 &  83.4 & 0.30 &  9.0 &   51.6 & 0.25 & 4.0 &    67.2 & 0.28 & 6.2 \\
   S2 & 1 &  41.9 & 0.37 &  5.5 &   41.9 & 0.37 & 5.5 &    41.9 & 0.37 & 5.5 \\
\hline
low-$Y_e$ + high-$Y_e$:
  P1S2 & $1+1$ &   106.3 & 0.24 &  8.9 &   105.7 & 0.24 &   8.8 &   105.9 & 0.24 &  8.8 \\
  S2P1 & $1+1$ &     5.1 & 4.06 & 17.7 &     4.2 & 5.07 &  19.8 &     4.5 & 4.69 & 18.9 \\
  S1S2 & $1+1$ &   103.4 & 0.23 &  7.6 &   103.4 & 0.23 &   7.6 &   103.4 & 0.23 &  7.6 \\
  S2S1 & $1+1$ &     4.6 & 4.12 & 16.3 &     4.6 & 4.12 &  16.3 &     4.6 & 4.12 & 16.3 \\
  S2T1 & $1+1$ &     4.3 & 4.91 & 19.8 &     4.8 & 4.37 &  18.4 &     4.6 & 4.53 & 18.8 \\
  T1P1 & $1+1$ &    49.2 & 0.65 & 15.8 &    43.7 & 0.67 &  14.3 &    44.9 & 0.66 & 14.4 \\
  T2P1 & $1+1$ &    67.1 & 0.64 & 22.0 &    14.7 & 0.60 &   3.3 &    41.1 & 0.62 & 12.0 \\
  T2P2 & $1+1$ &   103.1 & 0.31 & 12.6 &    65.8 & 0.37 &   9.4 &    78.3 & 0.35 & 10.5 \\
  T2P1 & $1+1$ &    90.1 & 0.81 & 45.1 &    19.4 & 0.80 &   7.1 &    54.8 & 0.80 & 24.6 \\
  T1S1 & $1+1$ &    66.5 & 0.39 & 10.3 &    55.6 & 0.29 &   5.4 &    58.8 & 0.32 &  6.8 \\
  T1S2 & $1+1$ &   107.2 & 0.25 &  9.0 &   105.3 & 0.24 &   8.7 &   106.1 & 0.24 &  8.8 \\
  T2S1 & $1+1$ &    76.7 & 0.47 & 16.1 &    12.6 & 0.30 &   0.9 &    37.6 & 0.43 &  6.1 \\
  T2S2 & $1+1$ &   110.4 & 0.25 &  9.8 &    82.3 & 0.21 &   5.2 &    92.8 & 0.23 &  6.8 \\
  T2P5 & $1+1$ &    64.2 & 1.07 & 46.7 &     3.8 & 0.33 &  0.27 &    32.9 & 1.11 & 22.1 \\
  T2S5 & $1+1$ &    45.8 & 1.26 & 39.8 &     3.8 & 0.32 &  0.25 &    19.9 & 1.34 & 16.3 \\

  T2P2 & $1+2$ &   135.6 & 0.38 & 23.4 &    93.4 & 0.44 &  18.4 &   108.5 & 0.41 & 20.2 \\
  T2S1 & $1+2$ &   104.8 & 0.57 & 32.1 &    17.1 & 0.35 &   1.7 &    51.1 & 0.51 & 11.6 \\
  T2S2 & $1+2$ &   164.7 & 0.29 & 19.1 &   131.3 & 0.23 &  10.7 &   143.6 & 0.26 & 13.7 \\

  T2P1 & $1+3$ &   106.9 & 0.93 & 68.5 &    23.1 & 0.94 &  11.2 &    65.0 & 0.93 & 37.7 \\
  T2P2 & $1+3$ &   161.3 & 0.44 & 35.6 &   111.2 & 0.50 &  28.3 &   129.3 & 0.46 & 29.9 \\
  T2S1 & $1+3$ &   125.2 & 0.65 & 48.1 &    20.4 & 0.36 &   2.2 &    61.1 & 0.58 & 17.4 \\
  T2S2 & $1+3$ &   204.9 & 0.31 & 28.3 &   168.8 & 0.25 &  16.4 &   182.1 & 0.28 & 20.3 \\

  T2P5 & $1+0.5$ &    48.2 & 0.85 & 23.0 &     3.8 & 0.32 &   0.2 &    24.6 & 0.87 & 10.8 \\
  T2P1 & $1+0.5$ &    50.8 & 0.51 & 11.4 &    11.6 & 0.47 &   1.7 &    31.5 & 0.50 &  6.2 \\
  T2S5 & $1+0.5$ &    35.0 & 0.99 & 19.9 &     3.8 & 0.33 &  0.26 &    15.4 & 1.05 &  8.3 \\
  T2S1 & $1+0.5$ &    45.9 & 0.63 & 13.7 &     3.8 & 0.31 &  0.25 &    19.7 & 0.61 &  4.8 \\
\hline\hline
\end{tabular}
  \label{tab:mass}
\end{table*}

Our single-component simulations assume the same mass, $0.01\,M_\odot$, and yet produce a wide range of peak times and luminosities, as shown in Table~\ref{tab:mass}.
Several groups have developed formulae that relate the peak luminosity and time of peak luminosity to the mass, expansion velocity, and opacity of the ejecta~\citep{grossman14,wollaeger18}.
To understand how the variation in the light curve leads to errors in the mass estimates, we have inverted the pair of formulae, Eqs.~(27)--(28) from \cite{wollaeger18}, deriving the mass inferred from the light curves produced in the models:
\begin{equation}
  \label{eq:mass_from_tLpeak}
  m_2 = \left(\frac{t_{\rm peak}}{1\,d}\right)^{1.5}
        \left(\frac{L_{\rm peak}}{2.8\times10^{40}\,{\rm erg}\,{\rm s}^{-1}}\right)^{1.2}
        \left(\frac{\kappa}{10\,{\rm cm}^2\,{\rm g}^{-1}}\right)^{0.2}.
\end{equation}
where $t_{\rm peak}$ and $L_{\rm peak}$ are the time and bolometric luminosity, respectively, at the peak, $\kappa$ is the average opacity, and the resulting mass is in units of $0.01\,M_\odot$.

We can now quantify the uncertainty of the ejecta mass that can be inferred from the peak times and luminosities using Eq.(\ref{eq:mass_from_tLpeak}).  Kilonova observations rising above the afterflow of gamma-ray bursts (GRBs) at late times will consist of only a few observations, and comparisons to integrated solutions such as equation (\ref{eq:mass_from_tLpeak}) are the only option for inferring ejecta mass.
If more observations become available, more accurate methods, such as spectral fitting or light-curve matching, are definitely more appropriate.
However, it is far more likely that the majority of kilonovae will be caught either when they are near their peak brightness, or at late times in follow-up observations.
Specifically, for off-axis events, it is more likely that the kilonova will be observed near peak brightness, while for GRB follow-ups, it is likely to be observed at late times after the GRB afterglow dims.
As we demonstrate below, the late-time light curves are at least as sensitive to morphologies as the peak-time light curves.

Inserting the peak parameters of our model light curves, we see that a mass that could be inferred from an observation using this formula can range over several orders of magnitude (see Table~\ref{tab:mass}).
The inferred mass only weakly depends on the effective opacity: for the calculations in Table~\ref{tab:mass}, we assumed $\kappa = 10{\rm \, cm^2 g^{-1}}$.

\begin{figure}
\begin{tabular}{c}
  \includegraphics[width=\columnwidth]{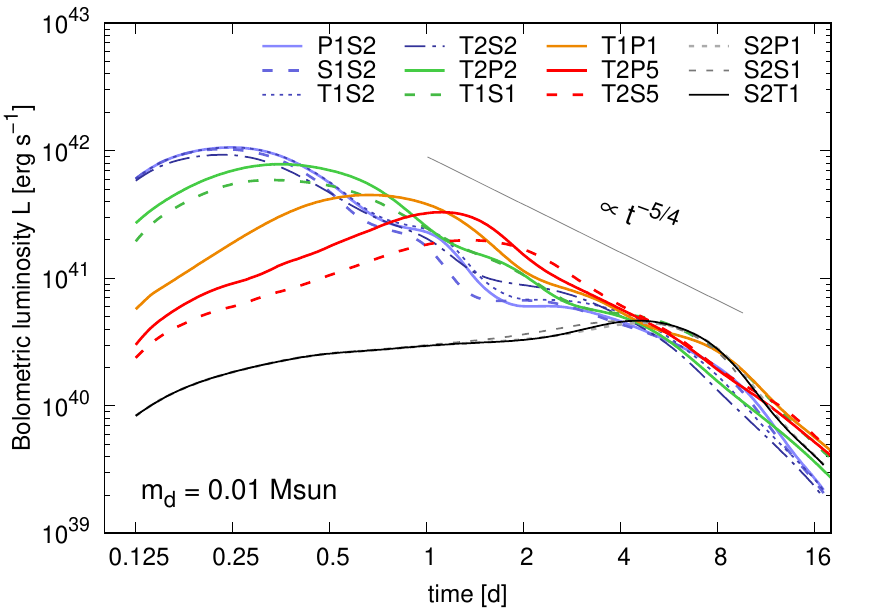} \\
  \includegraphics[width=\columnwidth]{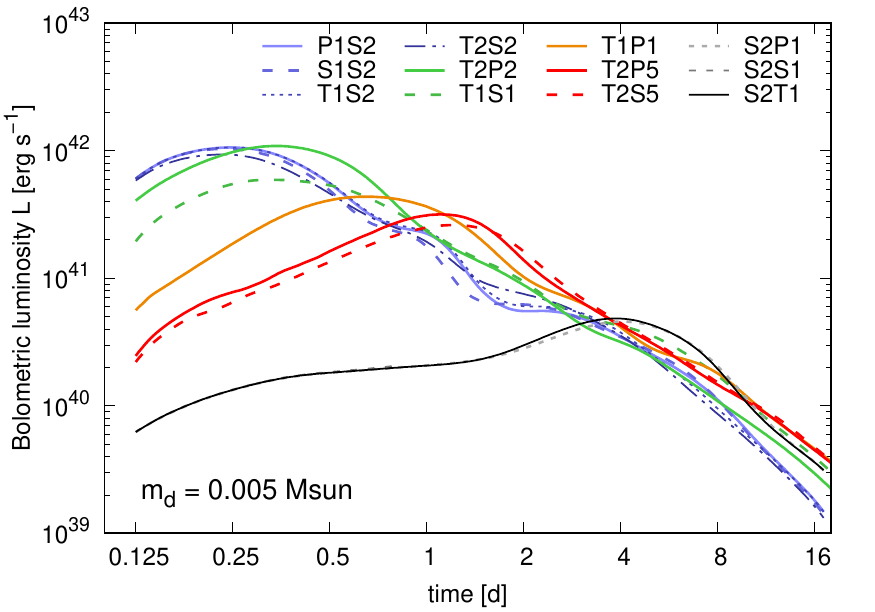} \\
  \includegraphics[width=\columnwidth]{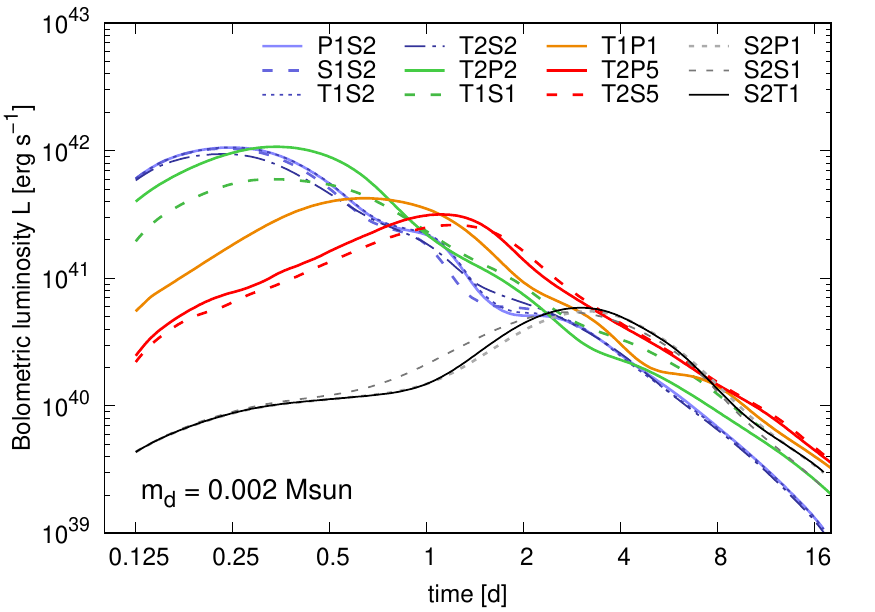}
\end{tabular}
\caption{Angle-integrated bolometric luminosity
as a function of time for twelve mixed morphologies,
and three different masses of the low-$Y_e$ (neutron-rich) component.
The top plot shows $\propto\,t^{-5/4}$ trend.
} 
\label{fig:twoc_luminosities}
\end{figure}

Since the bolometric light curves in Fig.~\ref{fig:onec_luminosities} are, in the absolute sense, converging at late times, it may seem that a better strategy to constrain the masses would be to use the late-time light curves.
However, it turns out that relative luminosities and broadband magnitudes at late times are similarly sensitive, if not more sensitive, to morphologies than the peak times.
Figure~\ref{fig:late_times} shows the bolometric light curves on a logarithmic scale, as well as the light curves in the infrared $K$-band, for two models of heating: analytic heating prescription (left panels) and full network heating with density-dependent thermalization (right panels).
Because morphologies have different densities, the density-dependent thermalization in the latter introduces variability in the late-time light curves by about half an order of magnitude for the different models considered here \citep[see also][on the importance of thermalization at late times]{kasen19,barnes20}.

Furthermore, even if the bolometric light curves converge for different morphologies, the broadband light curves do not have to.
Fig.~\ref{fig:late_times}, left panel, shows the bolometric and $K$-band light curves for the case when an analytic power-law heating is prescribed, which does not depend on density.
The broadband light curves still show a substantial degree of variability: by about 2~mag in $K$-band at 16~days.
This is because the same luminosity can be produced at different temperatures, and the latter depends on the cooling history of the ejecta.
Because optically thin ejecta are cooled inefficiently, their temperature maintains strong dependence on the previous state, and by extension, on the morphology.
This temperature variability for the same bolometric luminosity is an additional factor of uncertainty in the determination of the masses.
Notice that in this work, we make the assumption of LTE, both in the calculation of opacities and in the local emissivity.
This assumption is likely to be violated at late times, making the notion of a single temperature meaningless, and adding to the uncertainty in masses. If the late-time conditions that determine the atomic level populations are optically thin, then the charge state distribution will likely shift toward more neutral species compared to the LTE case, due to a reduction in the effective ionization resulting from radiative processes. This difference typically leads to absorption line features that occur at lower photon energies for non-LTE conditions, with a corresponding shift in the emission features. It is difficult to make more specific, quantitative predictions about non-LTE effects on the photometry without doing actual simulations, which we reserve for future work.
Another uncertainty comes from the problem of reconstructing bolometric light curves from photometry, which will be scarce at late times and highly dependent on transient broadband spectral features and large deviations from the blackbody.
If one adds uncertainties in the unknown nuclear heating, which may reach an order of magnitude~\citep{zhuyl20,barnes20}, in combination with the morphology-specific variability, mass inference from late-time bolometric luminosity becomes quite problematic.
Nevertheless, these complexities should not deter observers from obtaining the late-time data.  On the contrary, accurate deep photometry and spectra at late times are crucial to improve our understanding of the masses and morphologies of the ejecta.

We conclude that without strong observational constraints on the geometry of the ejecta, one can artificially infer larger masses from nonspherical explosions, which have higher densities and correspondingly higher temperatures, or higher apparent surface areas (see discussion in Appendix~\ref{sec:rad_structure}).
If we cannot constrain the composition, the uncertainty can be even larger.

\subsection{Two-component Models}

Combining two components with different morphologies enables a much wider class of kilonovae.
The range of mass estimates that can be inferred from limited information such as peak brightness and epoch now spans almost three orders of magnitude (see Table~\ref{tab:mass}).
The range of inferred masses can vary strongly even within the same model.
For example, in the model {\tt T2P5} with fast toroidal lanthanide-rich component and slow high-$Y_e$ outflow, the mass estimate ranges between 0.002 and 0.23\,$M_\odot$, depending on orientation.
In the side view, the toroidal component with lanthanide-rich composition has a very high opacity that completely obscures the bluer and brighter secondary outflow, which in turn leads to a smaller inferred mass.
This is the lanthanide ``curtaining'': high-opacity lanthanide-rich ejecta in front of more luminous outflow act as a ``curtain,'' hiding the blue kilonova for a range of viewing angles \citep{kasen15}.
At the same time, the top view leads to overestimated mass because of the brighter transient, since the secondary outflow is revealed and the toroidal component has a large projected area.

In Fig.~\ref{fig:twoc_luminosities}, we plot the angle-integrated bolometric luminosity versus time for all mixed models.
It shows that the impact of morphology can be a dominant indicator of the brightness, even when summed over viewing angle.
In every model where the low-$Y_e$ ejecta is more extended than the high-$Y_e$ ejecta ({\tt S2S1}, {\tt S2P1} and {\tt S2T1}), the high lanthanide opacities block all of the optical emission, causing the peak luminosity to be an order of magnitude dimmer than the rest of the models.
On the other hand, models in which the high-$Y_e$ wind component is spherical and more extended ({\tt S1S2}, {\tt P1S2} and {\tt T1S2}) are the brightest, but also peak too soon.

The rest of the models lie in between these two extremes.
The panels in Figure~\ref{fig:twoc_luminosities} correspond to three different masses of the lanthanide-rich component: 0.01, 0.005, and 0.002~$M_\odot$, while the mass of the high-$Y_e$ component is kept fixed at ${0.01\,M_\odot}$.
In every panel, the models that lie between the two extremes span a wide range of peak times (from 0.2 to 6~days) and two orders in luminosity.
For all models, there is a trend of an approximate inverse correlation between peak time and luminosity:
\begin{equation}
    L_{\rm peak} \propto t_{\rm peak}^{-5/4},
\end{equation}
which also follows from Eq.~(\ref{eq:mass_from_tLpeak}) for fixed mass and opacity.
This result shows that the effect of mixing morphologies is degenerate with the expansion velocity or effective opacity.

\begin{figure*}
\begin{tabular}{cc}
  \includegraphics[width=0.5\textwidth]{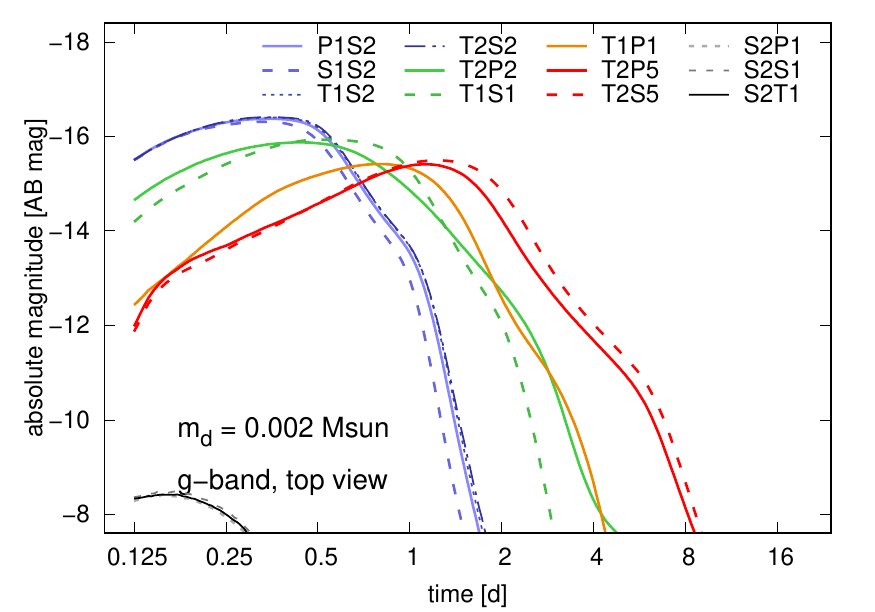} &
  \includegraphics[width=0.5\textwidth]{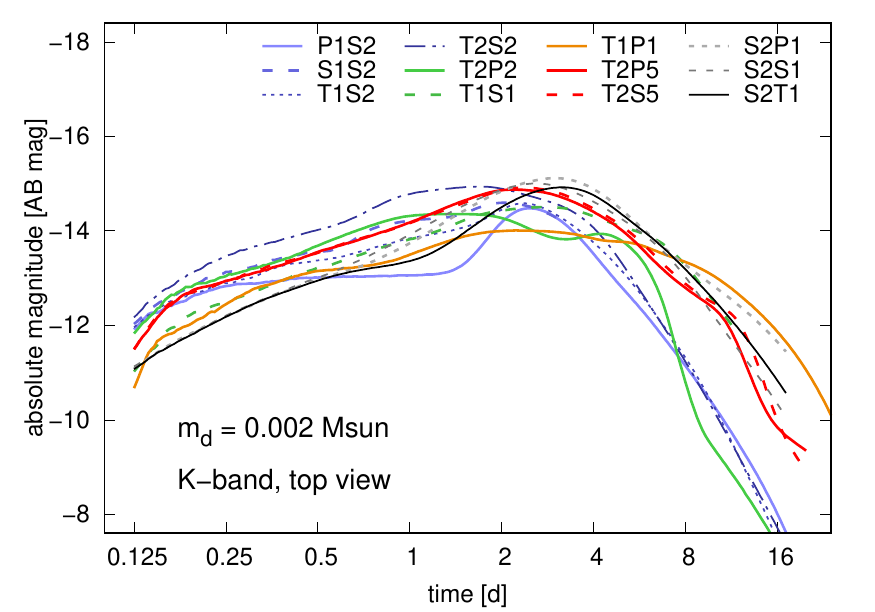} \\
  \includegraphics[width=0.5\textwidth]{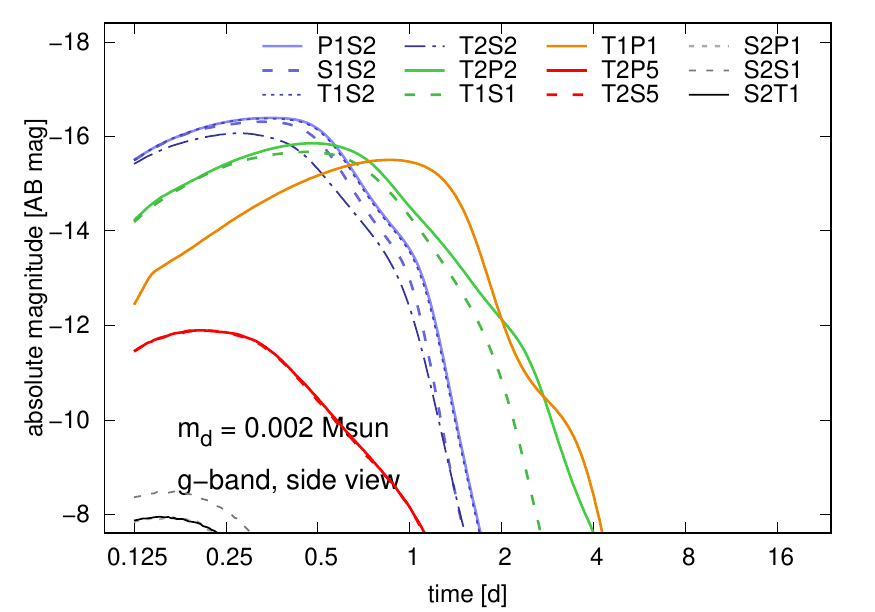} &
  \includegraphics[width=0.5\textwidth]{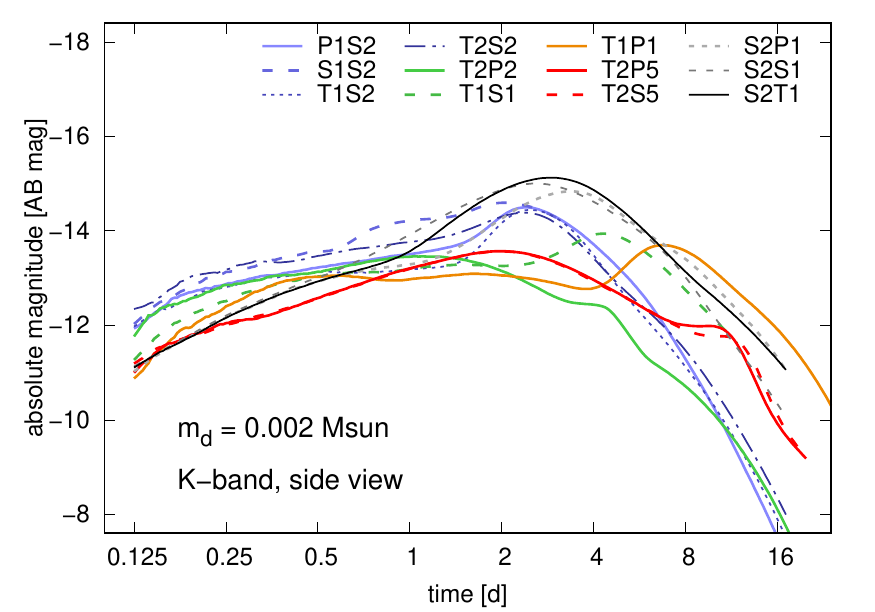}
\end{tabular}
\caption{Broadband light curves ($g$ and $K$ bands) for the 12 mixed
  morphologies with the lanthanide-rich (low-$Y_e$) component mass
  $0.002\,M_\odot$.
  The top row shows the on-axis, or "top" orientation, and the bottom panels show the "side" orientation toward the observer.
} 
\label{fig:twoc_broadband}
\end{figure*}

Notice that for the cases considered, the relative mass of the lanthanide-rich component has very little effect on the light curves.
The only exception is the position of the luminosity peak for models {\tt S2P1}, {\tt S2S1} and {\tt S2T1}, in which this component is spherical and more extended, hiding the high-$Y_e$ outflow.
This again illustrates that in many cases, the mass of the ejecta is subdominant in determining the light curves compared to the morphology.
The only cases where the effect of the lanthanide-rich component is most pronounced are when it completely blocks the blue component.
If the lanthanide-rich component is the interior one, or if it fails to block emission from the other component, its effect on light curves is minimal.
A lower mass of the lanthanide-rich component in these models leads to a brighter and earlier peak, due to the fact that with lower mass it becomes transparent sooner, exposing the brighter high-$Y_e$ outflow.
This effect is again degenerate with changing the expansion velocity or opacity, which further complicates ejecta mass estimates.

\begin{figure*}
\begin{tabular}{ccc}
    \hspace{-4mm}
    \includegraphics[width=0.34\textwidth]{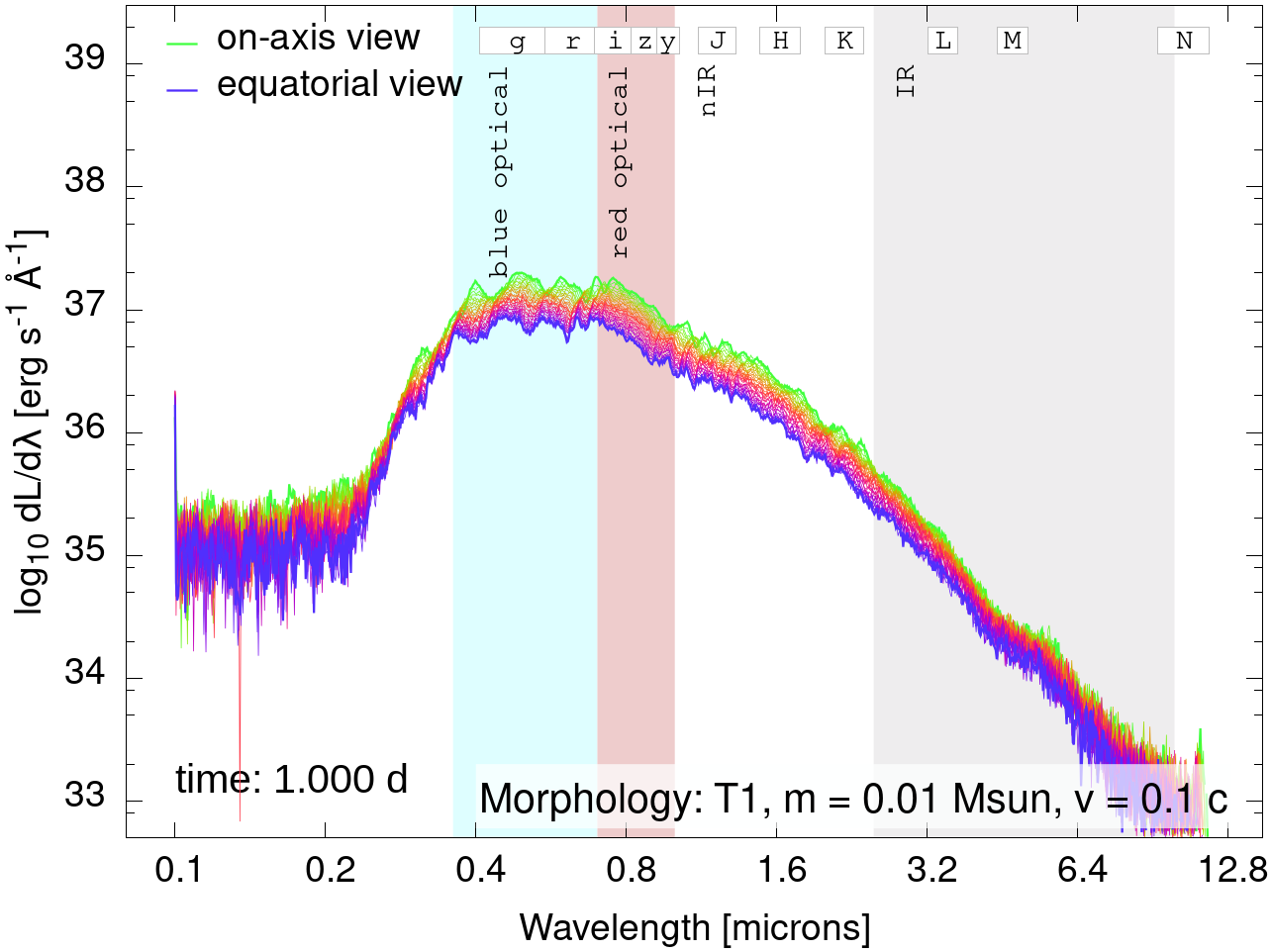} &
    \hspace{-4mm}
    \includegraphics[width=0.34\textwidth]{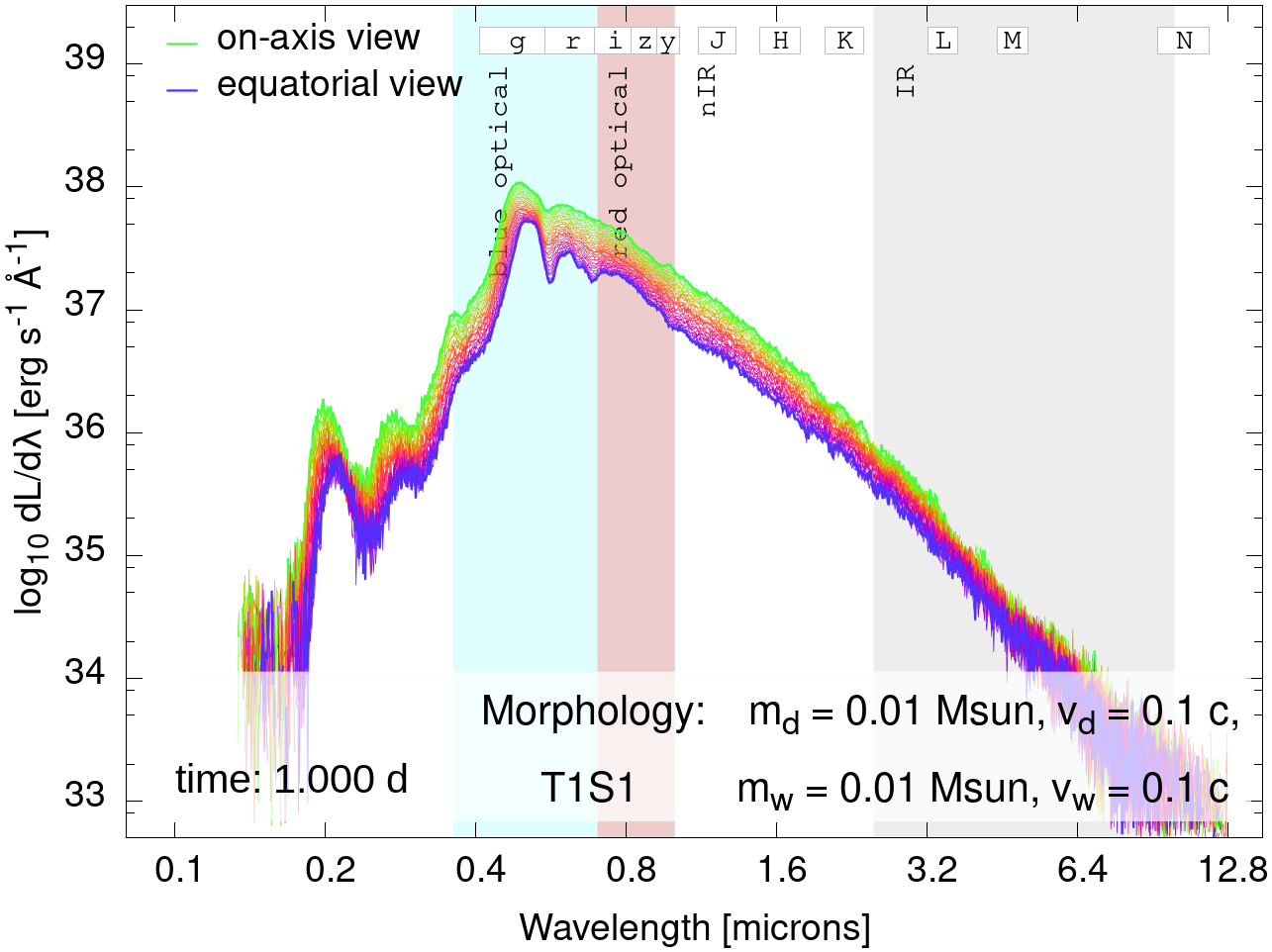} &
    \hspace{-4mm}
    \includegraphics[width=0.34\textwidth]{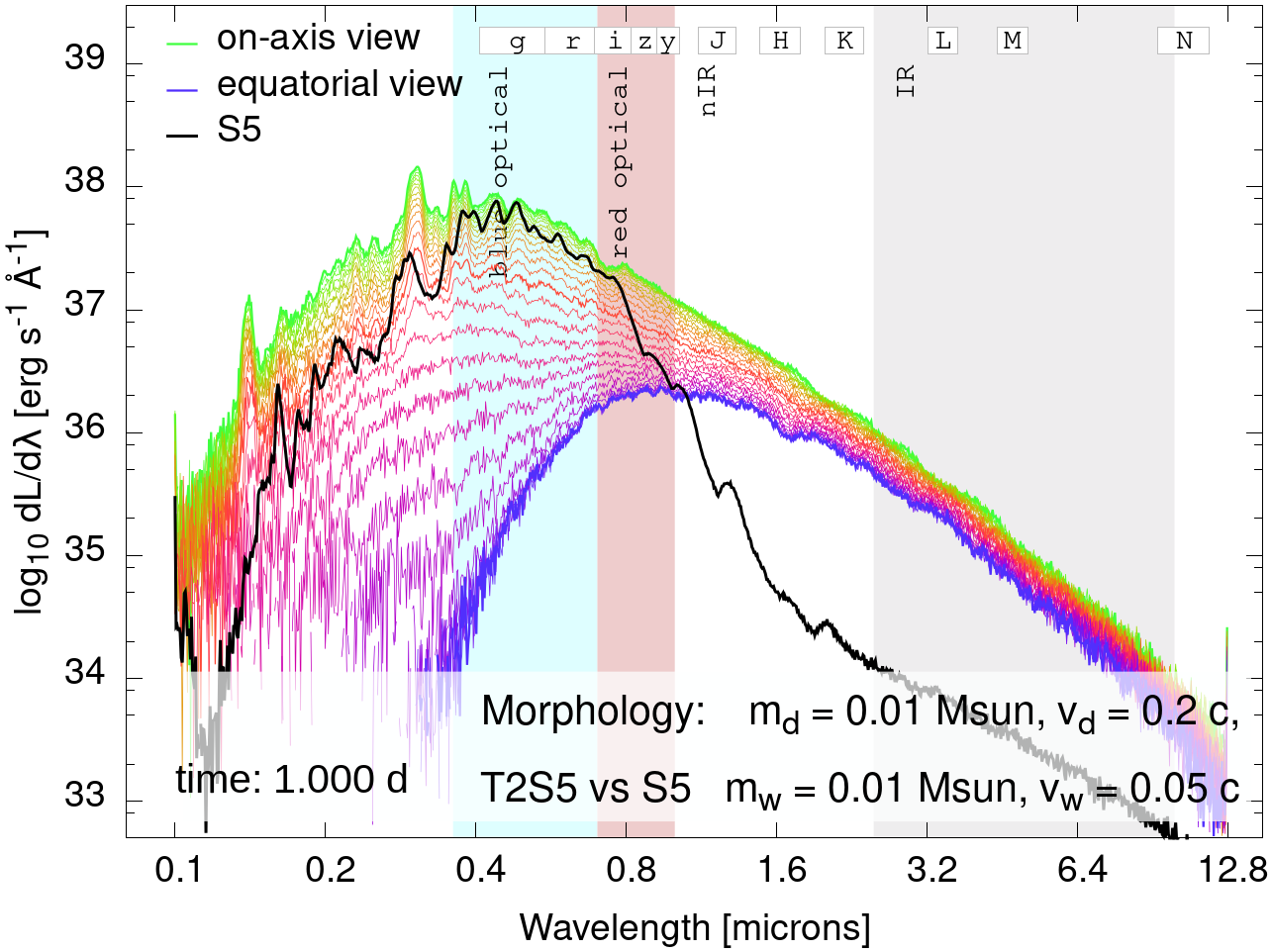}
    \\
    \hspace{-4mm}
    \includegraphics[width=0.34\textwidth]{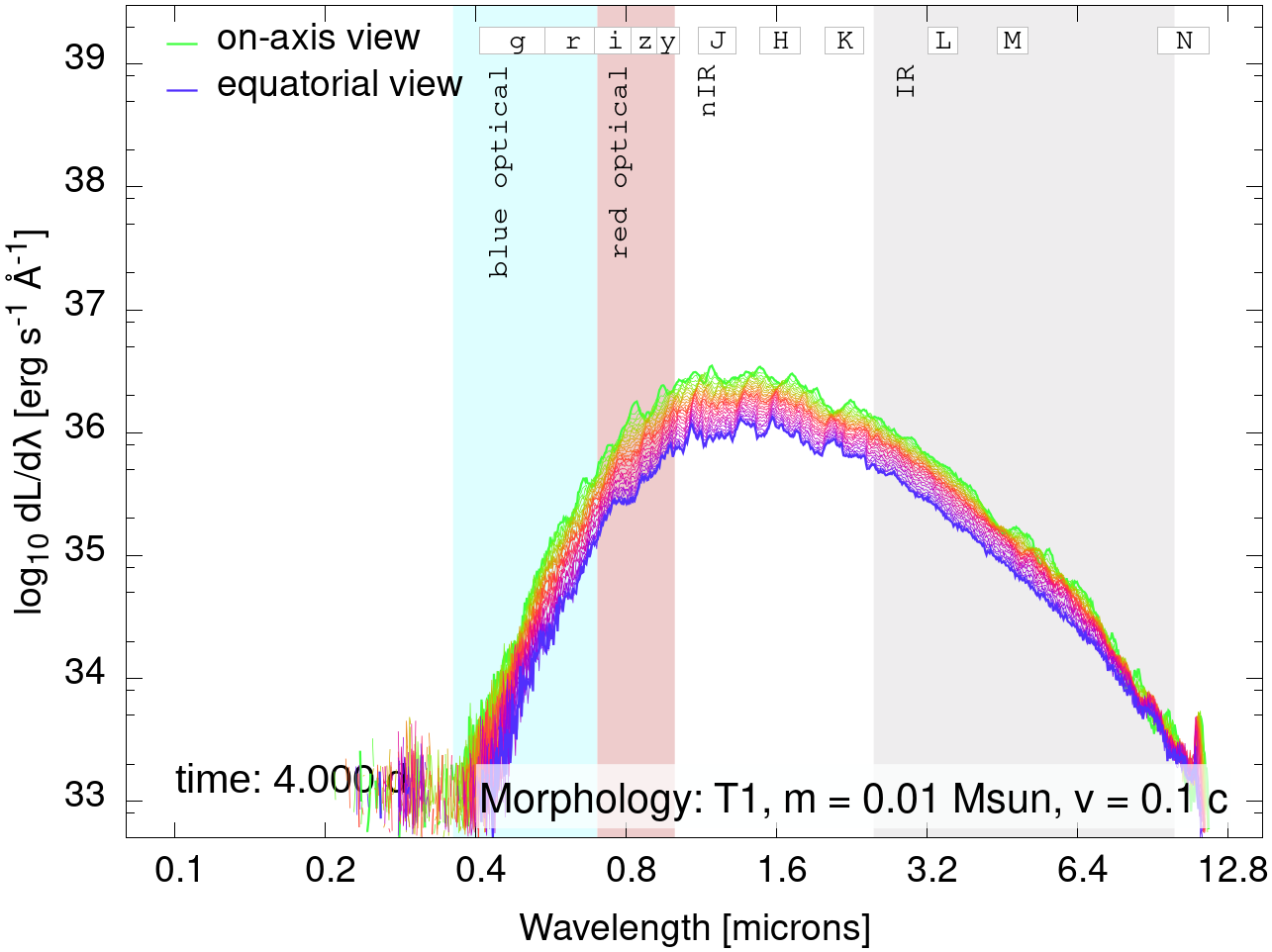} &
    \hspace{-4mm}
    \includegraphics[width=0.34\textwidth]{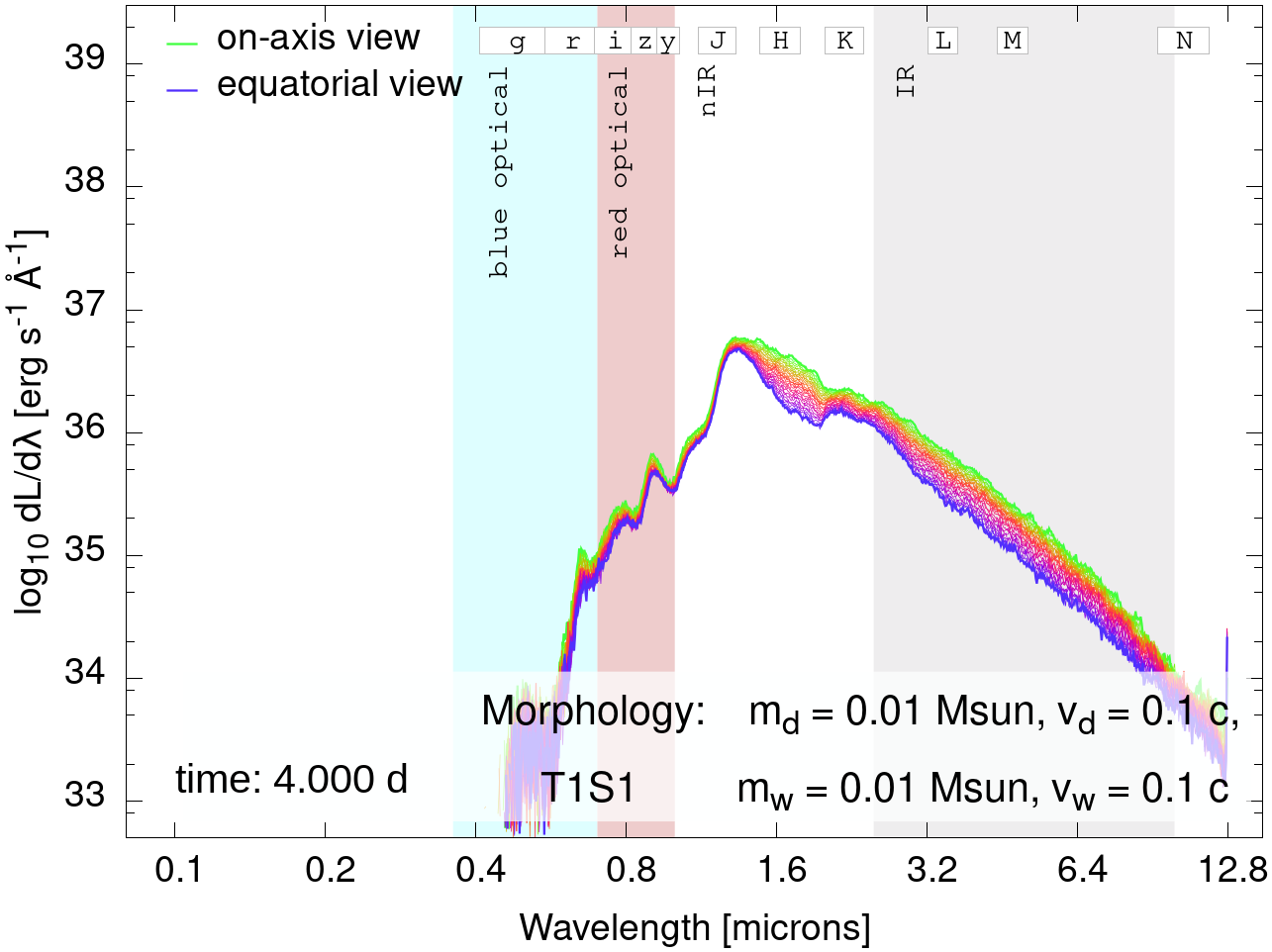} &
    \hspace{-4mm}
    \includegraphics[width=0.34\textwidth]{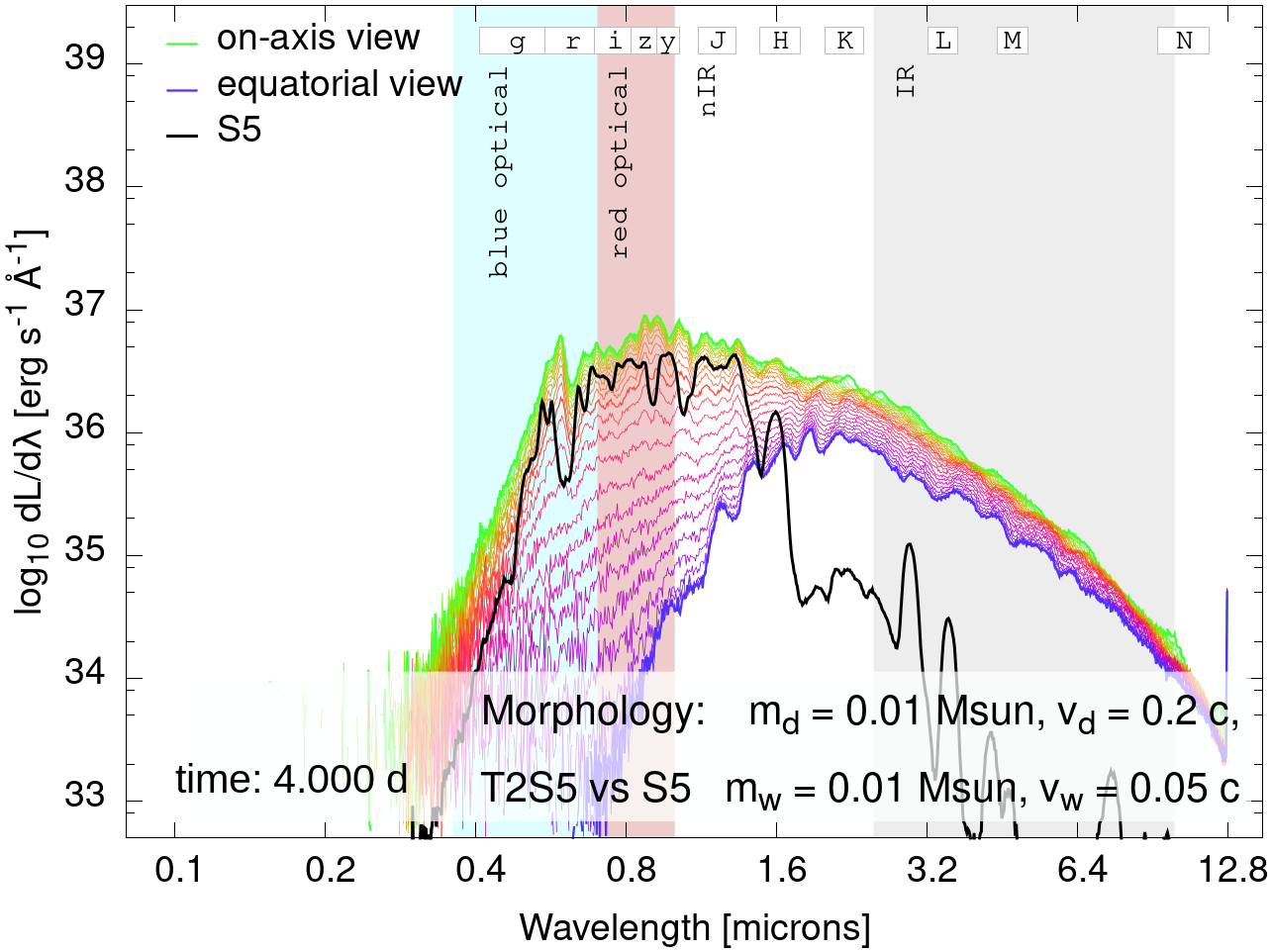}
    \\
    \hspace{-4mm}
    \includegraphics[width=0.34\textwidth]{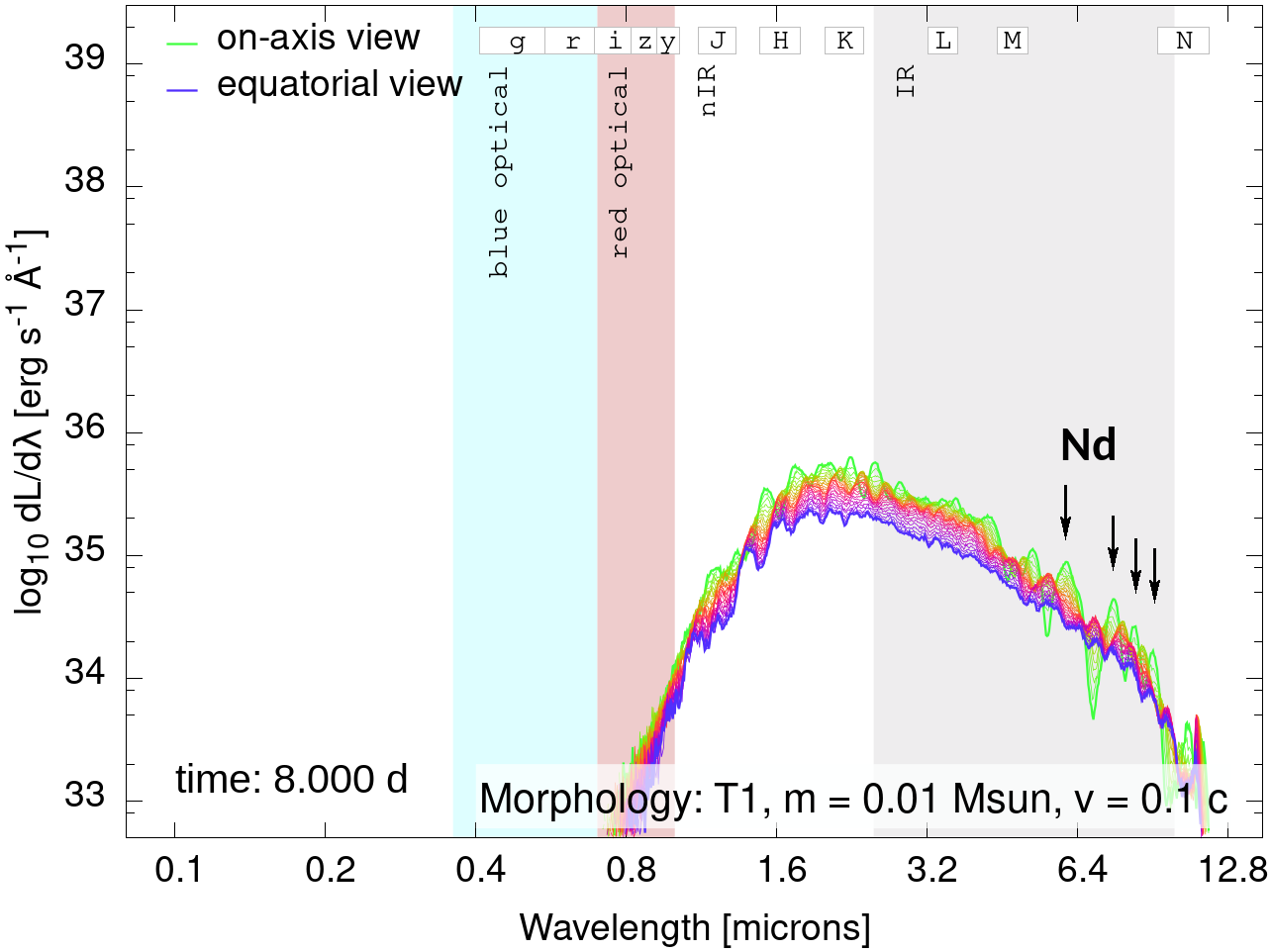} &
    \hspace{-4mm}
    \includegraphics[width=0.34\textwidth]{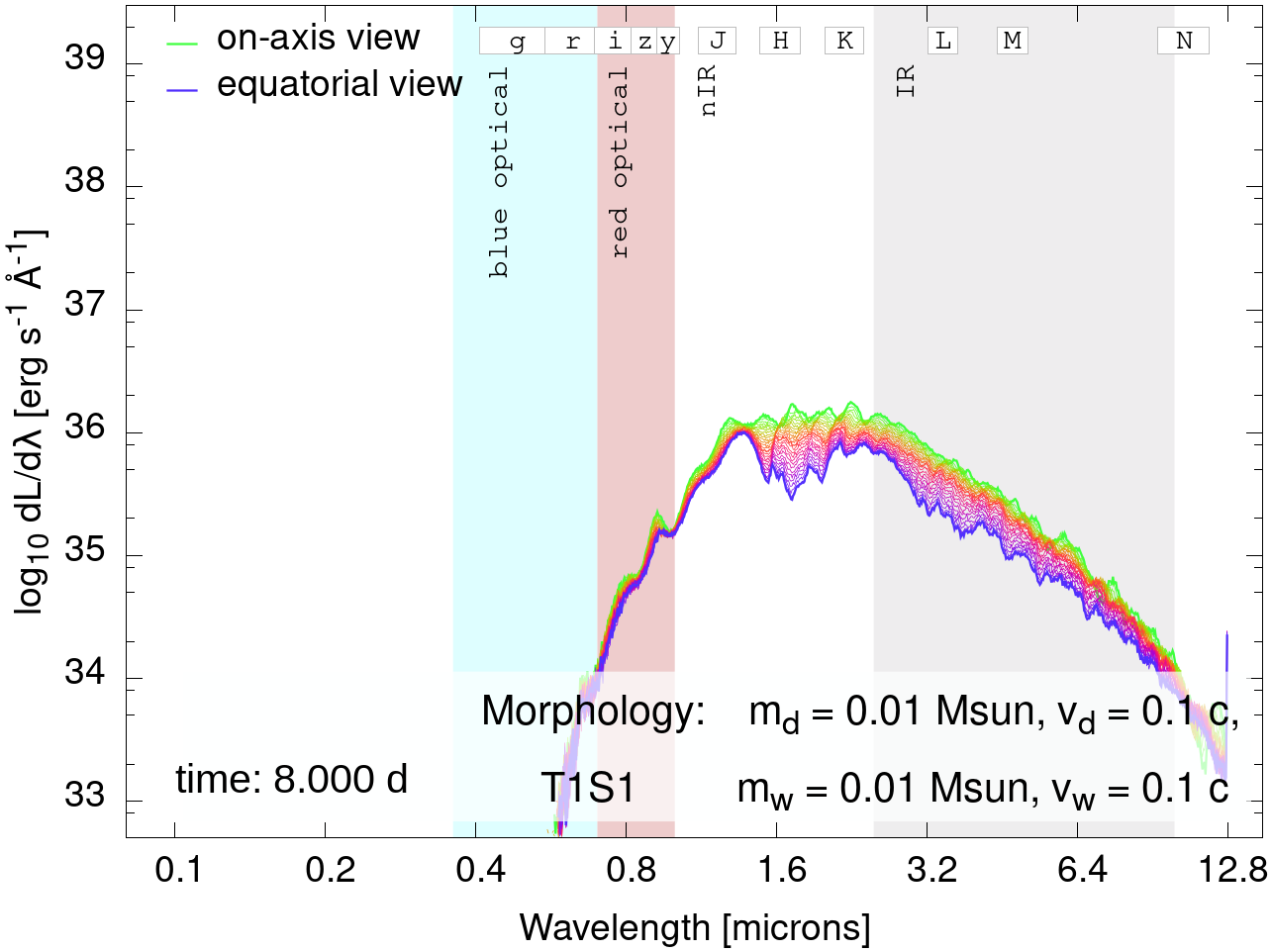} &
    \hspace{-4mm}
    \includegraphics[width=0.34\textwidth]{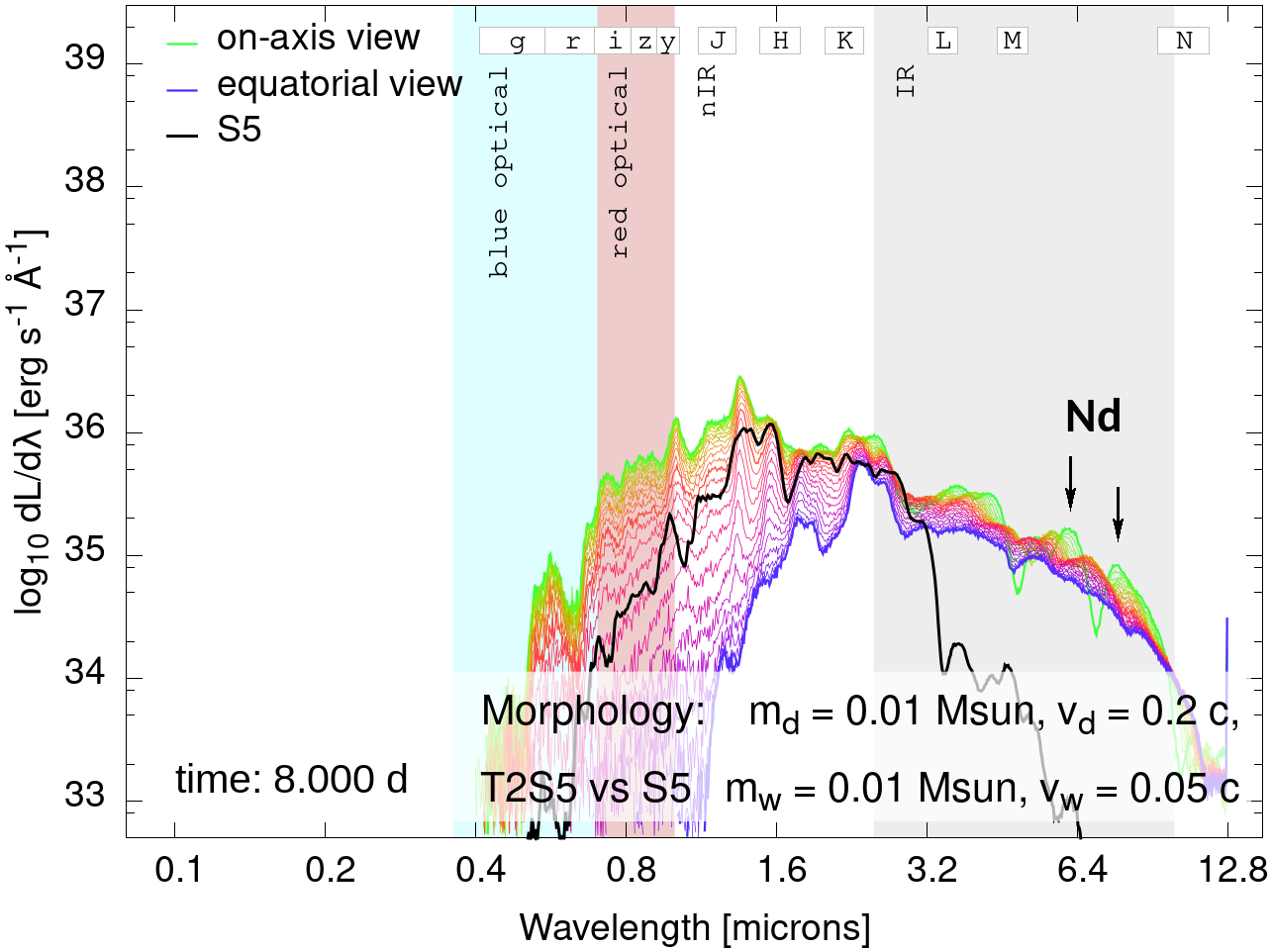}
\end{tabular}
\caption{Kilonova spectra at 1, 4, and 8 days for a single-component model "{\tt T}"
  (left) and two-component "{\tt TS}" morphologies (middle and right), for the full
  range of viewing angles. The middle and right components differ by the expansion
  velocities: in the middle column, both components have the same expansion velocity
  ${v_d=v_w=0.1\,c}$. In the right column, the velocity of the high-$Y_e$ "wind"
  ${v_w=0.05\,c}$ is much less than that for the low-$Y_e$ outflow ${v_d=0.2\,c}$.
  The differences in the velocities for the two outflows in the right column cause
  strong variation on the spectrum due to lanthanide curtaining or the high-$Y_e$
  outflows by the more extended lanthanide-rich toroidal component. This is the
  case that manifests the strongest angular dependence, while for the rest of the
  models, the angular dependence is very moderate.
  In the right column, the black solid line shows just the spherical high-$Y_e$ component
  "{\tt S5}," without the toroidal part. It illustrates the light focusing by the toroidal
  component, which causes the blue emission to preferentially diffuse toward the axis.
  The black vertical arrows in the bottom row point to the P~Cygni features in mid-IR
  generated by peaks of the wavelength-dependent opacity of Nd.}
\label{fig:T_spectra}
\end{figure*}

It is unlikely that measuring the light curves in a broad range of bands will be sufficient to lift these degeneracies.
Fig.~\ref{fig:twoc_broadband} shows the optical $g$- and nIR $K$-band light curves for all mixed morphologies and a low-$Y_e$ component mass of ${0.002\,M_\odot}$.
In the $g$-band, it is easy to distinguish models with more extended spherical lanthanide-rich ejecta ({\tt S2P1}, {\tt S2S1} and {\tt S2T1}, black lines): they are very strongly suppressed in this band for all orientations.
These models instead produce the brightest red kilonova in the $K$ band.
The rest of the models peak between 0.5 and 2 days in the $g$ band with magnitudes $-15$ to $-16$ mag, and rapidly decay by about 4~days.
Models {\tt T2P5} and {\tt T2S5} present a notable exception: they radiate longest in the $g$ band for the "top" orientation and are strongly suppressed for the "side" view.
This strong angular dependence is due to lanthanide curtaining.

\begin{figure}
  \begin{tabular}{c}
    \includegraphics[width=\columnwidth]{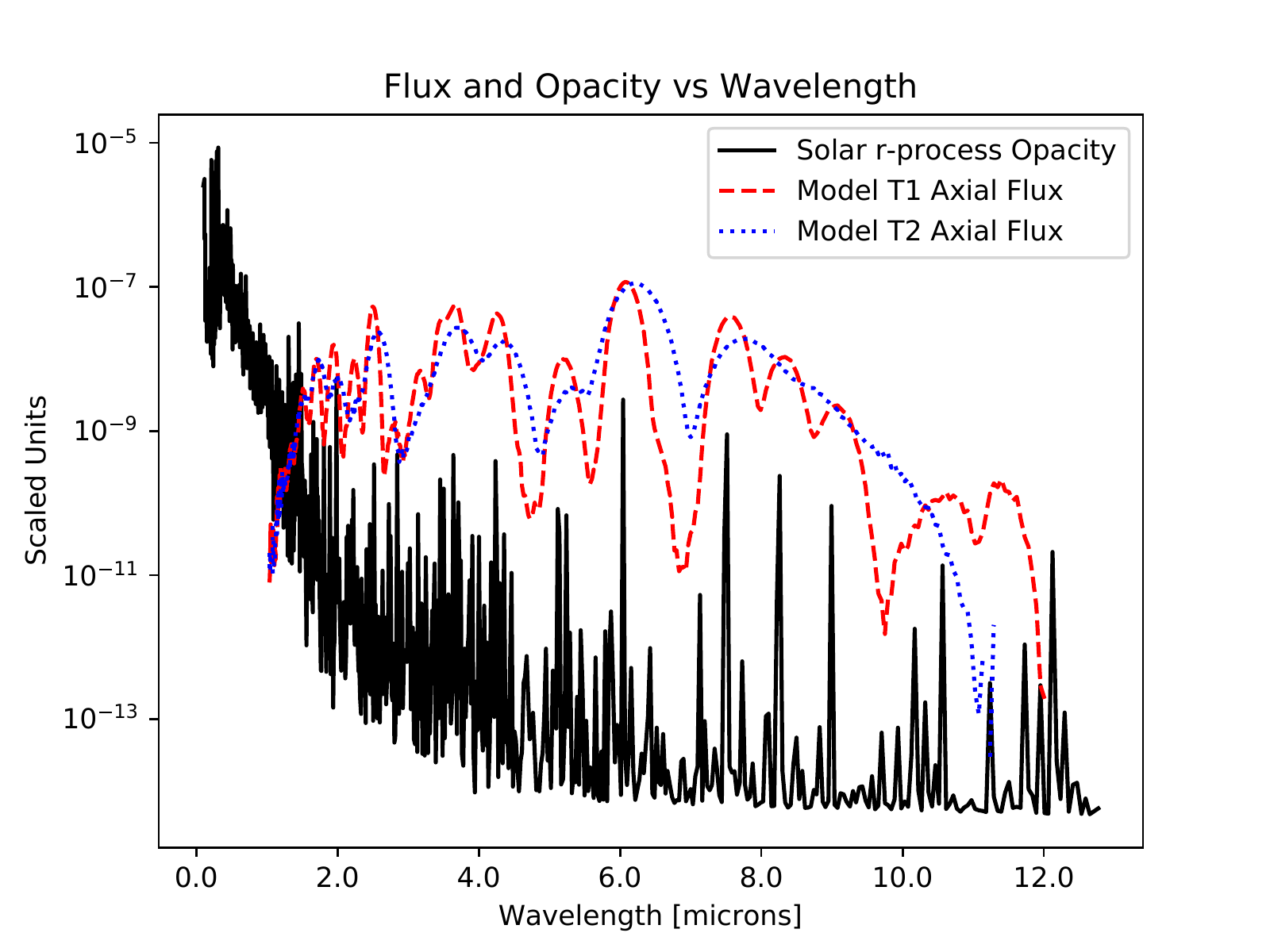} \\
    \includegraphics[width=\columnwidth]{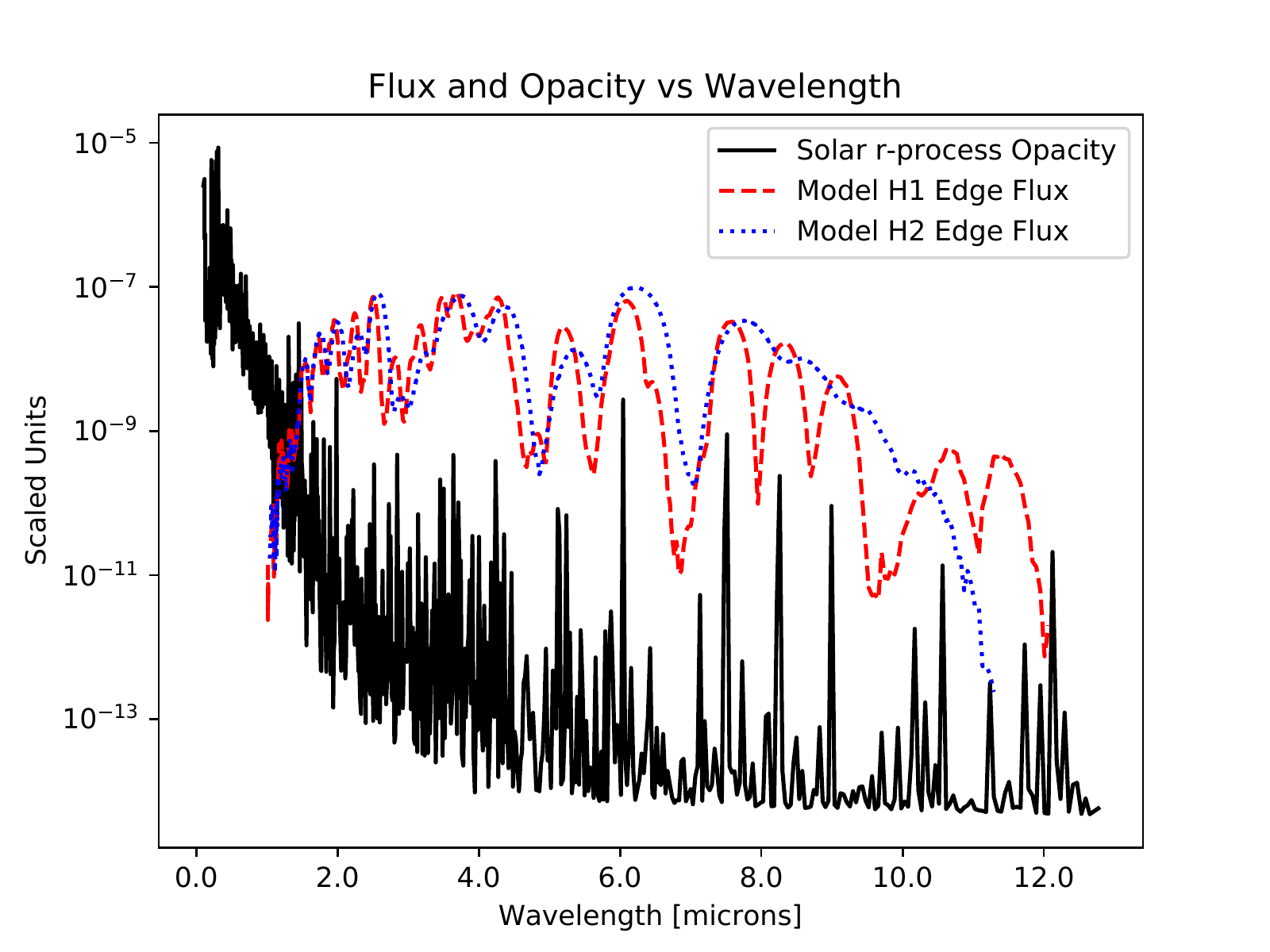}
  \end{tabular}
  \caption{
    Axial-view flux for model T and edge-view flux for model H, along with scaled
    opacity, versus wavelength.
    These models are moving at an average speed of $0.1c$ or $0.2c$.
    Relative to the $0.1c$ simulations, line emission blending is enhanced in the
    $0.2c$ simulations.
    Line features in the spectra are more visible in the axial (edge) view for
    model T (H), where the velocity toward the observer is slower.
    A majority of the tall lines above $\sim4\,\mu$m are from Nd, which is
    setting the IR emission at late time in our
    models~\citep[cf. Fig.~8 in][]{even20}.
  }
  \label{fig:kap_v_spectra}
\end{figure}

Most of our mixed models do not show pronounced dependence on the viewing angle beyond that of single-component models, where variability of about 1~mag can be attributed to the projected area of the photosphere~\citep{grossman14,darbha20}.
This is because lanthanide curtaining is only partial: {\tt T1S1} is a typical representative of this scenario.
In Figure~\ref{fig:T_spectra}, we compare its spectra with that of a single-component model {\tt T} which lacks the high-$Y_e$ contribution.
At early times, the spectra for model {\tt T1S1} are significantly brighter and bluer than for the model {\tt T}.
The impact of the high-$Y_e$ outflow also extends to late times, maintaining bluer emission relative to {\tt T}.
The latter behavior and the much weaker angular dependence of the blue wing of {\tt T1S1} at late time are the result of optical reprocessing by the lanthanide-free component.
At $\gtrsim$8~days, the viewing angle-dependent P~Cygni features of Nd lines start to form in the mid-IR.
However, these features only appear for the "top" orientation, as in this case the differential expansion velocity toward the observer is only about ${0.1\,c}$ (more about this at the end of this Section).

In some mixed-morphology scenarios, the viewing angle dramatically affects the spectrum and observed kilonova magnitudes.
This is illustrated with the model {\tt T2S5} (Fig.~\ref{fig:T_spectra}, right column).
Here the lanthanide-rich component is much more extended, curtaining the high-$Y_e$ outflow for some orientations.
This behavior is unlike that observed for {\tt T1S1}, where both components are always visible.
Moreover, the blue wing of {\tt T2S5} for the top orientations is much brighter than that of just the spherical low-opacity component {\tt S5} (its spectrum is shown with a solid black line in Fig.~\ref{fig:T_spectra}).
This appears to be because the lanthanide-rich toroidal component additionally redirects the flux toward low-opacity polar regions~\citep[also observed in the kilonova models of][]{kawaguchi18}.
No reddening of the redirected flux is observed, because the photons effectively return to the low-opacity region and get reemitted in blue wing before escaping down the gradient of the optical depth.
At $\gtrsim$8~days, pronounced P~Cygni features of Nd lines start to develop at the mid-IR wing of the spectrum, resembling those of model {\tt T}, but are twice as broad due to a higher line-of-sight expansion velocity.
This is in contrast with model {\tt T1S1} where these features are diluted by reprocessing in the enclosing high-$Y_e$ spherical component (bottom middle panel in Fig.~\ref{fig:T_spectra}).

The strong dependence on orientation for some mixed models directly projects onto broadband light curves.
As can be seen in Figure~\ref{fig:twoc_broadband}, models {\tt T2P5} and {\tt T2S5} (shown in red), are strongly suppressed in the optical bands for the "side" orientation.
On the other hand, for the "top" orientation, they show the same magnitude in the optical bands as the rest of the models.
Moreover, they produce the longest-lasting blue kilonovae---due to the flux redirection mentioned above.

Because of the very high opacity of the lanthanides, a strong angular dependence persists even if the mass of the lanthanide-rich component is very small.
In our models, it goes down to ${0.002\,M_\odot}$.
This is shown in Figure~\ref{fig:twoc_broadband}, where the difference for the $g$ band is 4~mag between the top and side orientations.
In nIR bands, such as the $K$ band shown in the right column of  Figure~\ref{fig:twoc_broadband}, the difference is only 1~mag.

At late times, we observe P~Cygni features \citep{castor79,robinson07} forming in the mid-IR wing of the spectra of one-component models and in the spectra of models where high-$Y_e$ and low-$Y_e$ components are well separated ({\tt T2P5} and {\tt T2S5}).
These features form around strong lines in the opacity and can be used to characterize both composition and morphology of the ejecta.
Figure~\ref{fig:kap_v_spectra} overplots the opacity ${\kappa_{\lambda}}$ and the late-time spectra for single-component models {\tt T} and {\tt H}, each with velocities of ${0.1\,c}$ (denoted {\tt H1}, {\tt T1}) or ${0.2\,c}$ (denoted {\tt H2}, {\tt T2}).
For the toroidal morphology {\tt T}, the spectrum is shown from a top view and for the hourglass {\tt H} from the side.
The opacity is evaluated at $\rho = 10^{-16}\,\gcc$, $T=1000$~K, and scaled by a constant factor to facilitate comparison with the spectra.
The low-velocity models {\tt H1} and {\tt T1} produce narrow features due to less Doppler broadening or blending between emission peaks.
The high-velocity models {\tt H2} and {\tt T2} appear to have a much less structured spectrum, in particular around $10\,\mu$m.
These features clearly reflect spikes in the opacity profile, which in turn are produced by Nd at this density and temperature~\citep[cf. Fig.~8 in][]{even20}.

\begin{figure}
  \begin{tabular}{c}
     \includegraphics[width=\columnwidth]{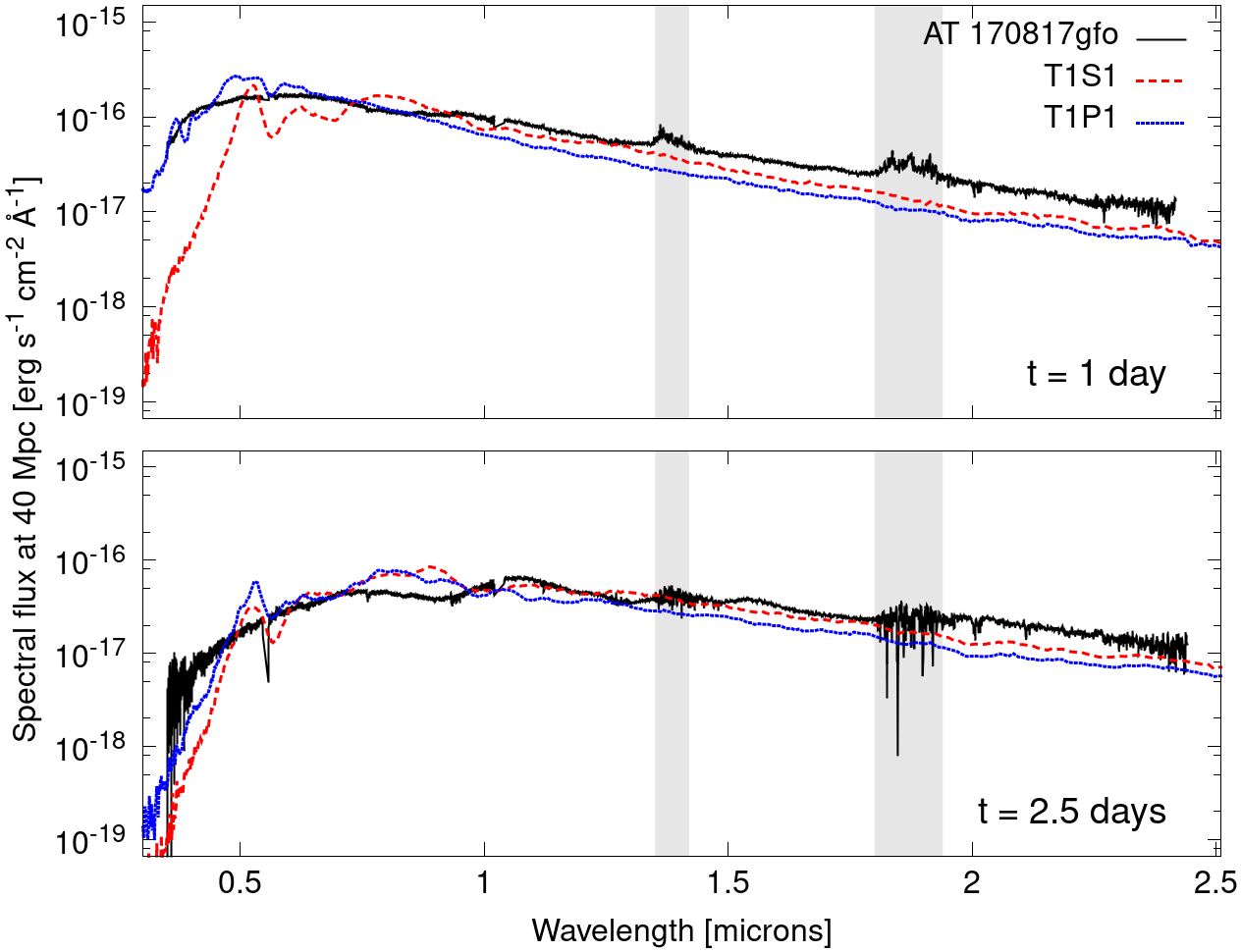} \\
  \end{tabular}
\caption{Model spectra for the two best-matching geometries
  at the epochs $t=1$~day and $2.5$~days in comparison with
  the spectrum measured for the kilonova {AT 2017gtfo} accompanying
  GW170817 \citep[instrument: VLT/X-shooter; ][]{pian17}.
  Both models have equal masses of the neutron-rich and
  neutron-poor components, $m_d = m_w = 0.01\ M_\odot$ and are shown with
  on-axis view. Wavelength ranges with poor data quality for the observed
  spectra are marked in gray~\citep[see][for details]{pian17}.
}
\label{fig:tantalizing}
\end{figure}

\section{Conclusions}
\label{sec:conclusions}

We study a set of analytically prescribed axisymmetric and spherically symmetric morphologies representative of neutron star merger ejecta using the multidimensional, multigroup Monte Carlo code \SuperNu\ \citep{wollaeger14} and detailed opacities from the LANL suite of atomic physics codes \citep{fontes15a,fontes19}.
Our five representative basic shapes are obtained from the family of Cassini ovals in axisymmetry (see Fig.~\ref{fig:basic_morphologies}), rescaled to a given total mass and average expansion velocity.
We model the merger ejecta with single- or two-component morphologies with two spatially uniform representative compositions producing ``red'' and ``blue'' contributions of the kilonova.
Specifically, we focus on two compositions: a low-electron-fraction ($Y_e$) lanthanide-rich solar r-process residuals \citep{even20}, and a high-$Y_e$ composition with ${Y_e=0.27}$ \citep[``wind 2'' in][]{wollaeger18}.
Compositions determine not only specific nuclear heating but also opacities of the material.
The former is computed with the \WinNET\ nuclear network \citep{winteler12} with energy partitioning between radiation species and spatially dependent thermalization \citep{barnes16,rosswog17,wollaeger18}.
For the latter, we employ a new suite of atomic opacities, which includes a complete set of lanthanides and uranium~\citep{fontes19}.

The results of our study can be summarized as follows:
\begin{enumerate}
\item When detailed, temperature-dependent opacities are used, the morphology affects the light curve more than the mass and velocity of the ejecta.
For the same mass and expansion velocity, switching between the different morphologies considered here leads to stronger variation in the peak time and luminosity than if we fixed the morphology and varied mass (by a factor of ten) or velocity (within $0.05 - 0.3\,c$ -- see Fig.~\ref{fig:onec_luminosities} and~\ref{fig:twoc_luminosities}).
\item Because of this result, the ejecta mass cannot be correctly inferred solely from the peak bolometric luminosity and time.
Our simulations show that the mass inferred from an effective gray-opacity formula can vary by an order of magnitude within one-component morphologies, and by three orders of magnitude for mixed models.
Without strong constraints on the geometry of the ejecta, artificially large or small masses can be inferred from nonspherical explosions (see Table~\ref{tab:mass}).
\item Morphological variability affects the peak luminosity in a way similar to varying the ejecta velocity or effective opacity. For a family of morphologies with the same mass and expansion velocity, the peak luminosity is inversely correlated with peak times: ${L_{\rm peak}\propto t_{\rm peak}^{-5/4}}$ (Fig.~\ref{fig:twoc_morphologies}).
The effect of mixing morphologies is degenerate with the effect of changing the average expansion velocity or effective gray opacity.
\item Unlike the gray-opacity case, in the models with detailed opacity, the temperature at the surface is more important in defining the peak properties than the projected area (see Appendix~\ref{sec:rad_structure} for detailed analysis).
\item Density-dependent thermalization of radioactive heating adds an extra boost to the bolometric luminosity, increasing it by almost a factor of two.
Therefore, a proper treatment of thermalization is as important as accurate composition-dependent nuclear heating rates~\citep[see also][]{barnes16,rosswog17,hotokezaka19}.
\item It is difficult to achieve the lanthanide curtaining effect if the two components have similar expansion velocities \citep{kasen15}. In this case, only partial lanthanide curtaining is observed. The majority of our models are quite isotropic, except when the high-$Y_e$ component is much slower than the lanthanide-rich component (models {\tt T2S5} and {\tt T2P5} -- see Fig.~\ref{fig:T_spectra}). Lanthanide curtaining is observed only in these models at low-latitude angles.
\item For the models with lanthanide curtaining, even a very small mass of $0.002\,M_\odot$ of lanthanide-rich material is sufficient to obscure the blue kilonova, because of the exceptionally high opacity of lanthanides~\citep{kasen13,tanaka13,fontes15a}. This is also consistent with the recent study of \cite{nativi20}.
\item At late epochs ($\gtrsim$8~days), the models in which lanthanide-rich ejecta is more extended exhibit pronounced P~Cygni features, allowing for the characterization of the composition and line-of-sight velocity of the ejecta (Fig.~\ref{fig:kap_v_spectra}).
Our results here confirm the findings of \cite{kawaguchi18}.
These P~Cygni features are unambiguously unidentified with peaks in opacity of Nd, which has been shown to dominate the opacity of lanthanide-rich material~\citep{fontes19,even20}.
\item Light reprocessing: in the case when the lanthanide-rich component is engulfed in a spherical high-$Y_e$ envelope, the blue kilonova becomes almost isotropic and the P Cygni features from Nd disappear (Fig.~\ref{fig:T_spectra}, middle column).
\item Light focusing by toroidal component: in the models with a more extended toroidal lanthanide-rich component, the light from the blue component is effectively funneled toward the axis such that the blue kilonova appears brighter (Fig.~\ref{fig:T_spectra}, right column).
\end{enumerate}

Morphological freedom creates sufficient variability to crudely fit the early spectrum of kilonova GW170817 with the limited set of models that we presented in this study.
In Figure~\ref{fig:tantalizing} we demonstrate a ``fit'' to GW170817 spectra at $t=1-2.5$~days.
The ``best-matching'' models appear to be {\tt T1S1} and {\tt T1P1}.
Clearly, tuning the velocities and masses of individual components, as well as the viewing angle, can significantly improve this result, which is beyond the scope of this paper~\citep[but see][]{heinzel21}.
Fitting the late-time spectra is a continuing goal.
This concept, yet again, demonstrates that a meaningful interpretation of kilonova light curves must properly account for morphological features, which should be informed by numerical simulations and observations.

We summarize the bolometric and broad band magnitudes at 1, 4, and 8 days for all models used in this study in Table~\ref{tb1:app} below. The complete suite of our models is available from the LANL CTA website.\footnote{\url{https://ccsweb.lanl.gov/astro/transient/transients_astro.html}}

\section*{Acknowledgements}

This work was supported by the US Department of Energy through the Los Alamos National Laboratory. Los Alamos National Laboratory is operated by Triad National Security, LLC, for the National Nuclear Security Administration of U.S.\ Department of Energy (Contract No.\ 89233218CNA000001).
Research presented in this article was supported by the Laboratory Directed Research and Development program of Los Alamos National Laboratory under project No. 20190021DR.
All LANL calculations were performed on LANL Institutional Computing resources.
Part of the work by C.L.F. was performed at the Aspen Center for Physics, which is supported by National Science Foundation grant PHY-1607611 and at the KITPl supported by NSF grant No. PHY-1748958, NIH grant No. R25GM067110, and the Gordon and Betty Moore Foundation grant No. 2919.01.
S.R. has been supported by the Swedish Research Council (VR) under grant No. 2016-03657-3, by the Swedish National Space Board under grant number Dnr. 107/16 and by the research environment grant "Gravitational Radiation and Electromagnetic Astrophysical Transients (GREAT)" funded by the Swedish Research council (VR) under Dnr 2016- 06012. We gratefully acknowledge support from COST Action CA16104 "Gravitational waves, black holes and fundamental physics" (GWverse) and from COST
Action CA16214 "The multi-messenger physics and astrophysics of neutron stars" (PHAROS).


\appendix
\section{Model tables: magnitudes at days 1, 4, and 8}
\label{sec:model_tables}

Table~\ref{tb1:app} lists the $r$- and $J$-band magnitudes and bolometric luminosity at days 1, 4, and 8 for top / side views.

\begin{table*}
  \centering
  \scriptsize
  \caption{Top / side viewing bin for absolute AB magnitudes in the $r$ and $J$ bands and angle-integrated bolometric luminosity ($L_{\rm bol}$ [$10^{40}$ erg s$^{-1}$]) for our models.}
  \begin{adjustbox}{width=0.95\textwidth}
  \begin{tabular}{rc|ccc|ccc|ccc}
    \hline\hline
    & $m_l + m_h$ & \multicolumn{3}{|c}{Day 1}
    & \multicolumn{3}{|c}{Day 4}
    & \multicolumn{3}{|c}{Day 8} \\
    Model & $[0.01\,M_\odot]$ & $r$ & $J$ & $L_{\rm bol}$ & $r$ & $J$ & $L_{\rm bol}$ & $r$ & $J$ & $L_{\rm bol}$ \\
    \hline
    low-$Y_e$ composition:
    H1  & 1 & -13.2 / -13.3 & -13.2 / -13.5 & 5.4 / 6.5  & -10.6 / -10.3 & -13.2 / -13.4 & 2.7 / 3.4 & -4.3 / -4.5 & -11.1 / -11.5 & 1.3 / 1.7  \\
    H2  & 1 & -13.6 / -13.0 & -13.9 / -13.8 & 7.2 / 6.2  & -8.0 / -6.9 & -13.0 / -12.7   & 3.0 / 3.1 & -0.0 / 0.2 & -10.0 / -8.6   & 0.8 / 1.0  \\
    P1  & 1 & -13.2 / -13.4 & -13.4 / -13.8 & 5.4 / 7.5  & -9.1 / -9.0 & -13.0 / -13.4   & 2.4 / 3.8 & -0.7 / -0.6 & -9.8 / -10.0  & 0.8 / 1.4  \\
    P2  & 1 & -13.0 / -12.3 & -14.1 / -14.1 & 6.2 / 6.3  & -4.9 / -3.8 & -12.1 / -11.5   & 2.2 / 2.8 & 0.1 / 0.1 & -9.4 / -7.8     & 0.6 / 0.7  \\
    B1  & 1 & -13.7 / -13.1 & -14.2 / -13.6 & 9.7 / 5.4  & -8.8 / -6.8 & -13.5 / -12.3   & 5.2 / 2.1 & 0.2 / 0.4 & -9.5 / -8.8     & 1.3 / 0.7  \\
    B2  & 1 & -12.2 / -12.3 & -14.4 / -14.1 & 8.7 / 6.0  & -2.7 / -2.6 & -11.1 / -11.1   & 3.0 / 1.7 & 0.0 / 0.0 & -7.7 / -8.6     & 0.5 / 0.5  \\
    T1  & 1 & -14.1 / -13.4 & -14.4 / -13.6 & 13.8 / 6.3 & -9.7 / -8.1 & -13.8 / -12.6   & 5.4 / 2.1 & 0.1 / 0.3 & -9.9 / -9.0     & 1.2 / 0.6  \\
    T2  & 1 & -13.0 / -12.9 & -14.6 / -14.1 & 10.4 / 6.1 & -3.9 / -3.3 & -11.7 / -11.3   & 3.0 / 1.6 & -0.0 / 0.0 & -8.2 / -9.0    & 0.6 / 0.5  \\
    high-$Y_e$ composition:
    H1  & 1 & -14.3 / -14.7 & -13.0 / -13.4 & 22.2 / 28.2 & -12.9 / -13.1 & -13.2 / -13.3 & 3.1 / 3.5 & -9.2 / -9.1 & -12.3 / -12.9 & 0.9 / 1.0  \\
    H2  & 1 & -14.4 / -14.9 & -13.7 / -13.7 & 16.9 / 17.7 & -11.8 / -11.1 & -12.8 / -13.2 & 2.0 / 1.9 & -8.8 / -6.8 & -11.9 / -12.6 & 0.7 / 0.7  \\
    P1  & 1 & -14.3 / -15.0 & -13.1 / -13.6 & 15.4 / 25.9 & -12.1 / -12.6 & -13.0 / -13.4 & 2.2 / 2.8 & -8.2 / -7.7 & -12.2 / -12.8 & 0.7 / 0.9  \\
    P2  & 1 & -14.4 / -14.9 & -14.1 / -14.4 & 13.6 / 14.7 & -10.3 / -8.9 & -12.8 / -13.5  & 1.8 / 1.7 & -8.2 / -6.4 & -11.8 / -12.7 & 0.6 / 0.6  \\
    B1  & 1 & -15.1 / -14.4 & -13.8 / -13.5 & 24.7 / 12.5 & -12.0 / -10.9 & -13.4 / -13.0 & 2.3 / 1.6 & -7.5 / -7.6 & -12.9 / -12.4 & 0.9 / 0.7  \\
    B2  & 1 & -14.9 / -14.5 & -14.8 / -14.7 & 15.1 / 12.3 & -8.4 / -9.2 & -13.7 / -13.1   & 1.9 / 1.7 & -6.2 / -7.1 & -12.9 / -12.2 & 0.5 / 0.5  \\
    T1  & 1 & -15.1 / -14.2 & -13.8 / -13.2 & 27.3 / 11.4 & -12.6 / -11.7 & -13.4 / -13.0 & 2.7 / 1.8 & -8.0 / -7.9 & -12.9 / -12.3 & 0.9 / 0.6  \\
    T2  & 1 & -15.0 / -14.4 & -14.4 / -14.3 & 15.6 / 11.5 & -8.8 / -9.6 & -13.7 / -12.9   & 2.0 / 1.6 & -6.7 / -7.6 & -12.9 / -12.1 & 0.6 / 0.5  \\
    \hline
    low-$Y_e$ + high-$Y_e$:
    P1S2 & $1 +1$   & -14.8 / -14.8 & -15.2 / -15.3 & 22.1 / 23.3      & -7.5 / -7.4  & -13.9 / -13.7  & 4.0 / 4.8    & -6.3 / -6.3   & -12.5 / -12.4 & 1.4 / 2.2  \\
    S2P1 & $1 +1$   & -7.3 / -6.9   & -12.8 / -12.7 & 3.1 / 2.9        & -0.5 / -0.1  & -11.4 / -10.0  & 5.1 / 3.9    & 0.0 / 0.0     & -7.4 / -6.5   & 2.7 / 2.9  \\
    S1S2 & $1 +1$   & -14.3 / -14.3 & -15.3 / -15.2 & 18.0 / 17.5      & -7.4 / -7.4  & -13.2 / -13.2  & 4.4 / 4.3    & -6.1 / -6.1   & -11.5 / -11.5 & 2.0 / 2.0  \\
    S2S1 & $1 +1$   & -7.4 / -7.0   & -12.8 / -12.7 & 3.1 / 2.9        & 0.4 / -0.2   & -10.8 / -10.6  & 4.7 / 4.6    & 0.0 / 0.0     & -6.5 / -6.5   & 2.6 / 2.6  \\
    S2T1 & $1 +1$   & -7.3 / -7.1   & -12.8 / -12.7 & 3.0 / 2.9        & 0.3 / 0.2     & -10.2 / -10.8 & 4.1 / 4.7    & 0.0 / 0.0     & -6.5 / -6.8   & 2.9 / 2.7  \\
    T1P1 & $1 +1$   & -15.4 / -15.4 & -14.7 / -14.2 & 41.3 / 37.4      & -11.1 / -10.9& -14.3 / -14.0  & 7.2 / 4.2    & -5.8 / -4.0   & -12.4 / -11.7 & 3.4 / 2.2  \\
    T2P1 & $1 +1$   & -15.7 / -14.0 & -15.1 / -13.7 & 55.2 / 12.3      & -10.6 / -8.8 & -14.1 / -12.8  & 7.5 / 2.7    & -5.5 / -2.2   & -12.3 / -10.7 & 2.3 / 1.6  \\
    T2P2 & $1 +1$   & -15.1 / -15.3 & -15.1 / -14.3 & 29.9 / 23.3      & -8.4 / -5.9  & -13.7 / -12.8  & 6.4 / 4.0    & -7.7 / -4.2   & -12.2 / -12.2 & 1.9 / 1.3  \\
    T1S1 & $1 +1$   & -15.7 / -14.6 & -15.3 / -14.2 & 39.4 / 15.6      & -8.0 / -7.4  & -14.1 / -13.9  & 7.0 / 3.8    & -5.3 / -5.0   & -12.7 / -12.2 & 3.8 / 1.7  \\
    T1S2 & $1 +1$   & -14.9 / -14.8 & -15.5 / -15.2 & 26.3 / 22.5      & -7.5 / -7.5  & -14.0 / -13.9  & 6.6 / 4.4    & -6.3 / -6.3   & -12.4 / -12.4 & 2.2 / 1.2  \\
    T2S5 & $1 +1$   & -15.3 / -11.2 & -14.8 / -13.3 & 43.4 / 3.1       & -12.7 / -2.0 & -14.5 / -10.9  & 11.0 / 1.9   & -8.4 / 0.0    & -13.0 / -7.7  & 3.0 / 0.9  \\
    T2S1 & $1 +1$   & -15.9 / -12.9 & -15.3 / -13.5 & 46.6 / 5.8       & -7.8 / -5.8  & -14.0 / -12.3  & 8.2 / 2.6    & -5.1 / -2.6   & -12.4 / -10.6 & 2.6 / 1.1  \\
    T2S2 & $1 +1$   & -14.8 / -14.1 & -15.5 / -15.0 & 28.0 / 14.3      & -7.5 / -7.0  & -13.5 / -13.1  & 6.9 / 3.5    & -6.3 / -5.4   & -11.7 / -11.6 & 1.6 / 1.0  \\
    P1S2 & $0.2+1$  & -14.8 / -14.8 & -15.2 / -15.3 & 20.7 / 22.3      & -7.4 / -7.4  & -13.4 / -13.4  & 2.2 / 2.6    & -6.2 / -6.3   & -12.3 / -12.3 & 0.6 / 0.7  \\
    S2P1 & $0.2+1$  & -6.0 / -3.5   & -12.2 / -11.0 & 2.0 / 1.3        & -3.7 / 0.1   & -12.5 / -11.1  & 5.2 / 4.8    & 0.0 / 0.0     & -9.8 / -6.6   & 1.3 / 1.5  \\
    S1S2 & $0.2+1$  & -14.4 / -14.3 & -15.3 / -15.2 & 18.4 / 18.1      & -7.7 / -7.6  & -13.6 / -13.5  & 3.9 / 3.9    & -7.7 / -7.6   & -13.6 / -13.5 & 3.9 / 3.9  \\
    S2S1 & $0.2+1$  & -6.1 / -5.1   & -12.2 / -11.8 & 2.3 / 2.0        & -0.2 / -0.4  & -11.2 / -11.2  & 4.5 / 4.5    & 0.0 / 0.0     & -8.0 / -8.0   & 1.3 / 1.3  \\
    T1P1 & $0.2+1$  & -15.2 / -15.3 & -14.4 / -13.9 & 36.8 / 35.3      & -10.7 / -10.4& -13.7 / -13.8  & 3.7 / 2.7    & -5.7 / -3.7   & -12.2 / -11.5 & 1.4 / 1.5  \\
    T2P2 & $0.2+1$  & -15.1 / -15.2 & -14.9 / -14.0 & 26.5 / 21.3      & -8.4 / -5.5  & -13.6 / -12.5  & 3.2 / 2.7    & -8.4 / -5.5   & -13.6 / -12.5 & 3.2 / 2.7  \\
    T1S1 & $0.2+1$  & -15.7 / -14.6 & -15.1 / -14.1 & 37.0 / 14.7      & -7.3 / -6.5  & -13.7 / -13.3  & 4.3 / 2.5    & -5.2 / -4.9   & -12.5 / -12.1 & 1.5 / 1.0  \\
    T1S2 & $0.2+1$  & -14.8 / -14.8 & -15.4 / -15.2 & 24.0 / 21.0      & -7.4 / -7.3  & -13.4 / -13.4  & 2.7 / 2.2    & -6.3 / -6.3   & -12.3 / -12.3 & 0.7 / 0.6  \\
    T2S2 & $0.2+1$  & -14.8 / -13.9 & -15.4 / -14.9 & 26.0 / 12.6      & -7.4 / -6.9  & -13.3 / -12.8  & 3.3 / 2.0    & -6.3 / -5.4   & -11.8 / -11.7 & 0.7 / 0.6  \\
    S2T1 & $0.2+1$  & -3.9 / -4.2   & -11.1 / -11.3 & 1.3 / 1.6        & 0.3 / -1.3   & -11.2 / -11.9  & 4.8 / 5.1    & 0.0 / 0.0     & -6.3 / -8.9   & 1.4 / 1.3  \\
    P1S2 & $0.5+1$  & -14.8 / -14.8 & -15.2 / -15.3 & 21.3 / 22.7      & -7.4 / -7.4  & -13.6 / -13.5  & 3.0 / 3.7    & -6.3 / -6.3   & -12.4 / -12.4 & 0.9 / 1.2  \\
    S2P1 & $0.5+1$  & -6.1 / -5.5   & -12.2 / -12.0 & 2.2 / 2.0        & -0.8 / -0.1  & -11.8 / -10.1  & 5.2 / 4.2    & 0.0 / 0.0     & -7.9 / -5.8   & 1.9 / 2.1  \\
    S1S2 & $0.5+1$  & -14.3 / -14.3 & -15.2 / -15.2 & 18.0 / 17.6      & -7.4 / -7.4  & -13.1 / -13.1  & 3.5 / 3.4    & -6.1 / -6.1   & -11.5 / -11.5 & 1.1 / 1.1  \\
    S2S1 & $0.5+1$  & -7.1 / -6.7   & -12.5 / -12.3 & 2.2 / 2.0        & -7.1 / -6.7  & -12.5 / -12.3  & 2.2 / 2.0    & -7.1 / -6.7   & -12.5 / -12.3 & 2.2 / 2.0  \\
    T1P1 & $0.5+1$  & -15.3 / -15.3 & -14.5 / -14.1 & 38.8 / 36.2      & -10.9 / -10.6& -14.0 / -13.9  & 5.3 / 3.3    & -5.7 / -3.8   & -12.3 / -11.5 & 2.0 / 1.7  \\
    T1S1 & $0.5+1$  & -15.7 / -14.6 & -15.2 / -14.1 & 38.1 / 15.0      & -7.6 / -6.8  & -13.9 / -13.5  & 5.6 / 3.1    & -5.2 / -4.9   & -12.5 / -12.1 & 2.3 / 1.2  \\
    T1S2 & $0.5+1$  & -14.8 / -14.8 & -15.4 / -15.2 & 25.0 / 21.7      & -7.4 / -7.4  & -13.6 / -13.6  & 4.5 / 3.1    & -6.3 / -6.3   & -12.3 / -12.3 & 1.1 / 0.8  \\
    T2S2 & $0.5+1$  & -14.8 / -14.0 & -15.4 / -15.0 & 26.9 / 13.2      & -7.4 / -6.9  & -13.4 / -12.9  & 4.7 / 2.5    & -6.2 / -5.4   & -11.7 / -11.6 & 1.0 / 0.7  \\
    S2T1 & $0.5+1$  & -5.9 / -5.7   & -12.1 / -12.0 & 2.1 / 2.0        & 0.5 / -0.1   & -10.3 / -11.1  & 4.5 / 5.0    & 0.0 / 0.0     & -5.6 / -6.7   & 2.0 / 1.9  \\
    T2P2 & $0.5+1$  & -15.0 / -15.1 & -15.0 / -14.3 & 30.0 / 20.0      & -11.1 / -9.4 & -13.1 / -13.3  & 4.3 / 2.4    & -9.0 / -6.8   & -12.2 / -12.8 & 1.3 / 1.1  \\
    T2P1 & $1 +2$   & -16.1 / -14.5 & -15.3 / -14.0 & 86.6 / 18.9      & -12.6 / -10.9& -15.0 / -13.5  & 13.6 / 4.1   & -6.4 / -3.6   & -13.1 / -11.3 & 3.8 / 2.6  \\
    T2P2 & $1 +2$   & -16.0 / -16.0 & -15.5 / -14.8 & 60.5 / 52.2      & -9.1 / -7.6  & -14.5 / -13.8  & 8.9 / 5.2    & -8.1 / -4.8   & -13.0 / -12.7 & 3.1 / 2.5  \\
    T2S1 & $1 +2$   & -16.4 / -13.5 & -15.7 / -13.6 & 81.8 / 8.8       & -15.5 / -13.7& -15.7 / -13.7  & 44.2 / 7.4   & -15.5 / -13.7 & -15.7 / -13.7 & 44.2 / 7.4 \\
    T2S2 & $1 +2$   & -15.6 / -15.0 & -16.0 / -15.3 & 49.4 / 26.0      & -8.2 / -7.8  & -14.3 / -13.8  & 10.4 / 6.0   & -7.1 / -6.4   & -12.9 / -12.6 & 2.8 / 1.8  \\
    T2P1 & $1 +3$   & -16.2 / -14.7 & -15.4 / -14.1 & 106.5 / 23.1     & -13.3 / -11.7& -15.4 / -13.9  & 19.7 / 5.5   & -7.5 / -5.2   & -13.7 / -12.1 & 5.2 / 3.2  \\
    T2P2 & $1 +3$   & -16.4 / -16.4 & -15.7 / -15.1 & 91.3 / 77.7      & -10.1 / -9.4 & -15.0 / -14.5  & 11.6 / 6.4   & -8.4 / -5.2   & -13.5 / -12.9 & 4.5 / 4.0  \\
    T2S1 & $1 +3$   & -16.7 / -13.9 & -15.9 / -13.7 & 108.5 / 12.0     & -11.9 / -10.0& -15.6 / -13.5  & 20.2 / 3.7   & -6.2 / -3.8   & -13.6 / -11.8 & 7.6 / 2.7  \\
    T2S2 & $1 +3$   & -16.1 / -15.5 & -16.3 / -15.5 & 70.6 / 35.7      & -8.7 / -8.3  & -14.9 / -14.4  & 13.5 / 8.5   & -7.6 / -6.9   & -13.5 / -13.1 & 4.3 / 2.6  \\
    T2P1 & $1+0.5$  & -15.1 / -13.5 & -14.9 / -13.6 & 31.3 / 7.5       & -8.1 / -6.0  & -13.2 / -11.9  & 5.3 / 2.3    & -5.0 / -1.5   & -11.6 / -10.3 & 1.5 / 0.9  \\
    T2S5 & $1+0.5$  & -15.0 / -11.2 & -14.7 / -13.3 & 35.1 / 3.1       & -11.4 / -1.4 & -13.8 / -10.9  & 6.8 / 1.9    & -5.0 / 0.0    & -11.4 / -7.4  & 1.8 / 0.7  \\
    T2S1 & $1+0.5$  & -15.4 / -11.2 & -14.8 / -13.3 & 39.2 / 3.1       & -8.9 / -1.8  & -13.5 / -10.9  & 5.9 / 1.9    & -3.8 / 0.0    & -11.5 / -7.4  & 1.8 / 0.7  \\
    T2P5 & $1+0.5$  & -15.0 / -11.2 & -14.7 / -13.3 & 46.2 / 3.1       & -11.9 / -1.7 & -13.3 / -10.9  & 6.1 / 1.9    & -8.5 / 0.0    & -12.1 / -7.4  & 1.7 / 0.7  \\
    \hline
  \end{tabular}
  \end{adjustbox}
  \label{tb1:app}
\end{table*}

\section{Radiative Structure in Single-component Morphologies}
\label{sec:rad_structure}

\begin{figure*}
\begin{center}
\begin{tabular}{ccc}
  \includegraphics[width=0.32\textwidth]{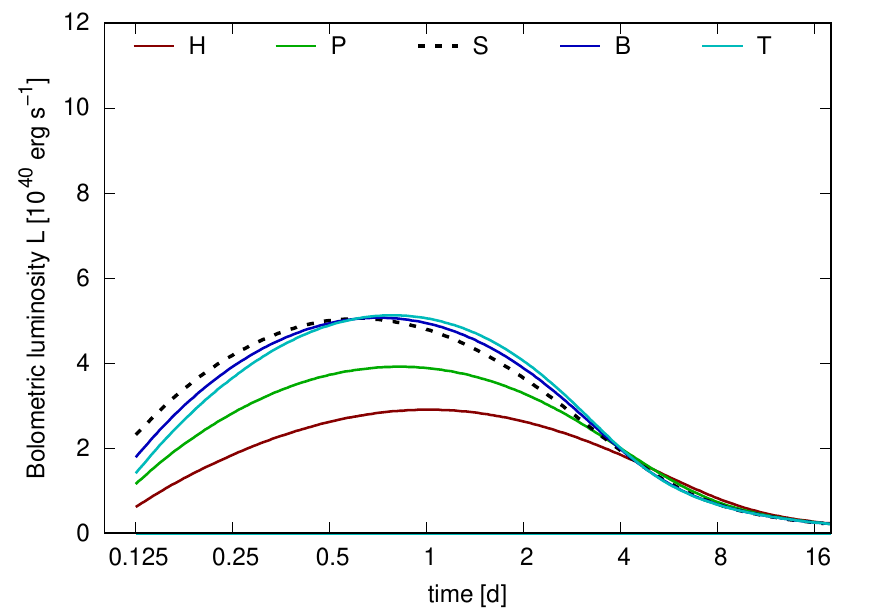} &
  \includegraphics[width=0.32\textwidth]{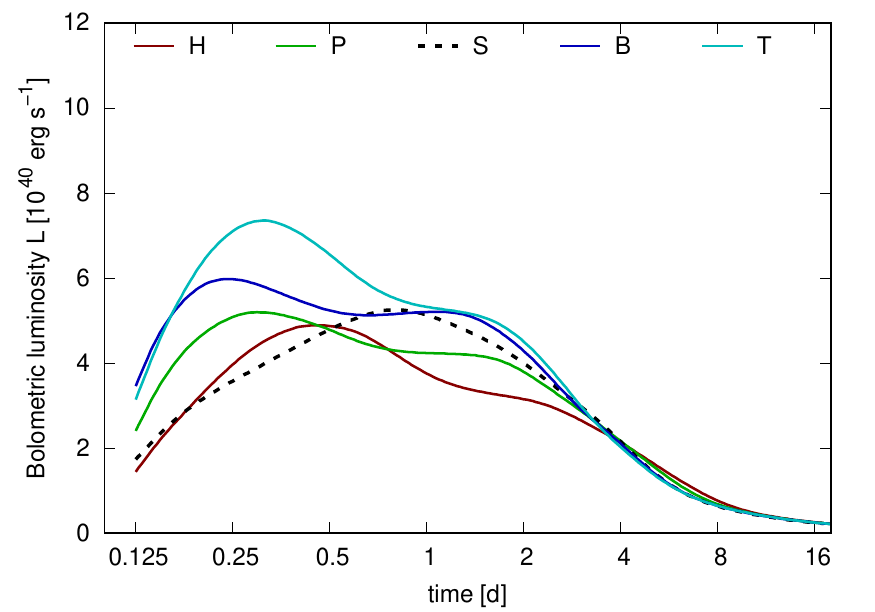} &
  \includegraphics[width=0.32\textwidth]{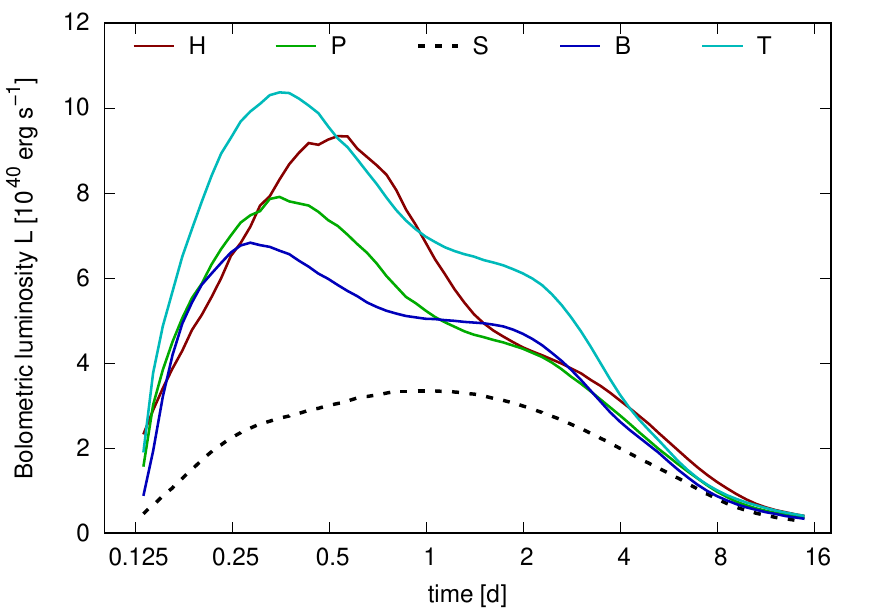}
  \\
  \includegraphics[width=0.32\textwidth]{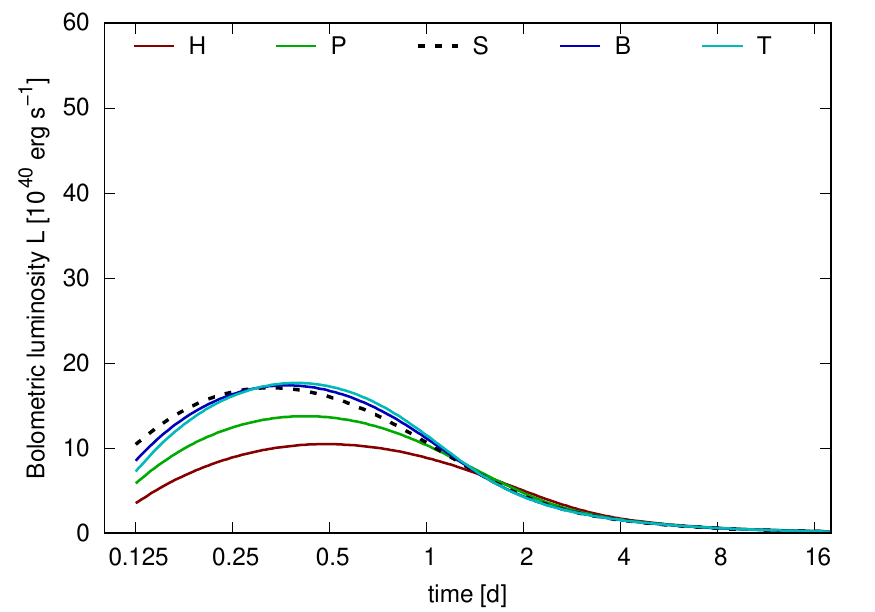} &
  \includegraphics[width=0.32\textwidth]{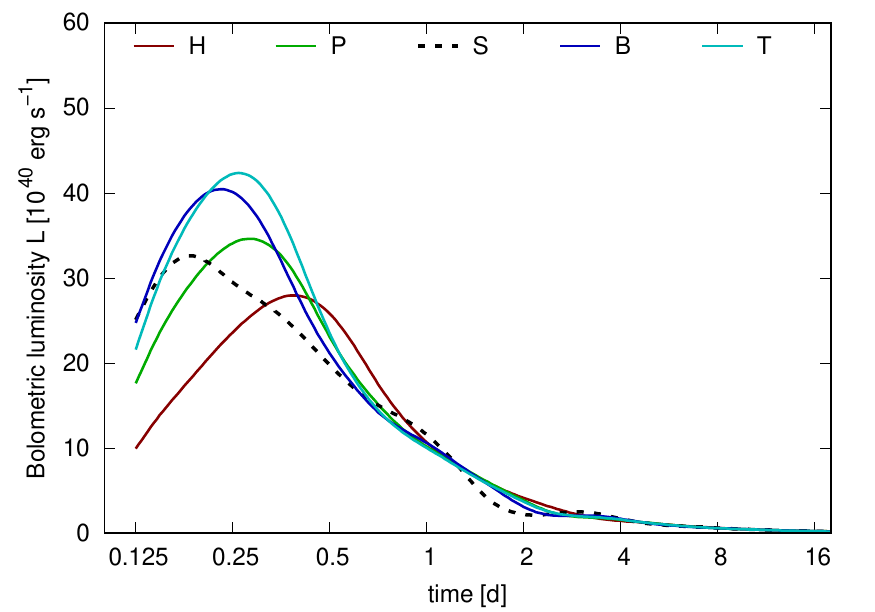} &
  \includegraphics[width=0.32\textwidth]{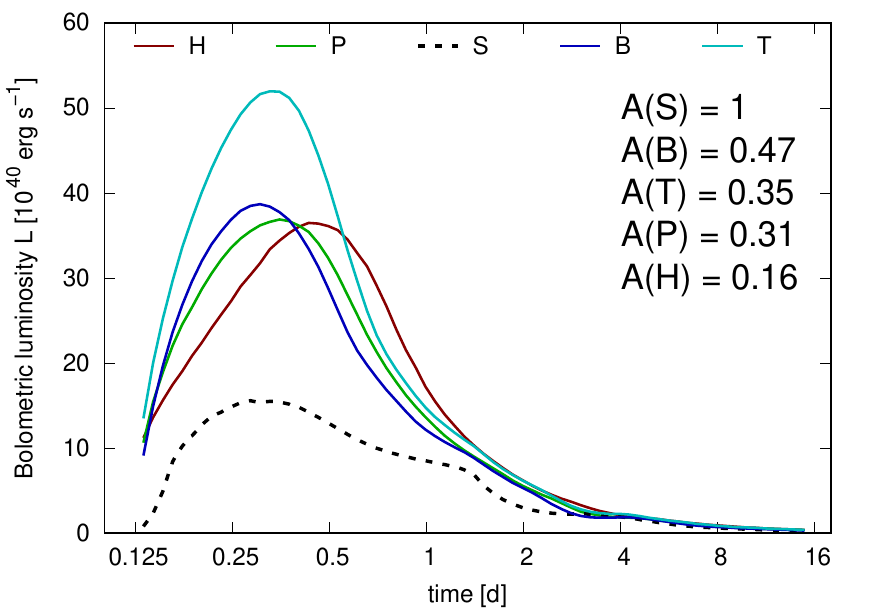}
\end{tabular}
\end{center}
\caption{Angle-integrated
  bolometric luminosity as a function of time for the single-component
  morphologies. The top and bottom rows represent different compositions:
  low-$Y_e$ lanthanide-rich solar r-process residuals (top) vs
  high-$Y_e$ composition (bottom).
  Left column: uniform gray opacities (10 and 1~{${\rm cm}^2\,{\rm g}^{-1}$} for
  the top and bottom panels, respectively)
  Middle column: simulations using detailed opacities.
  Right column: bolometric luminosity estimate based on the temperature and the
  area of the outer surface (see Eq.~\ref{eq:Lum_t}).
  All models have mass $m_{\rm ej} = 0.01\;M_{\odot}$, median expansion velocity
  $v_{\rm ej} = 0.2\;c$ and uniform analytic power-law heating (without
  thermalization). The labels on the right panel show the area of the outer surface
  for the morphologies relative to the spherical model.
} 
\label{fig:onec_lums_discussion}
\end{figure*}

One way to understand the features of the kilonova light curves is to break it up into components.
In this appendix, we study how the light curve of a single-component morphology is influenced by different factors, such as area of the photosphere or the temperature of the radiative layer.
We apply phenomenological analysis of the radiative structure of axisymmetric morphologies and disentangle effects of geometry, density and opacity.
For all models in this section, we assume uniform specific heating given by $\dot{\varepsilon}(t) = 2\times10^{10}\ t_d^{-1.3}\,\erggs$, where $t_d$ is the time since merger, in days \citep{metzger10,korobkin12}.

\subsection{Gray-opacity models}

In the gray-opacity approximation, models with more mass distributed at low optical depth are expected to be brighter.
This is illustrated in Figure~\ref{fig:onec_lums_discussion} (left column).
Total luminosity can be estimated then using a simple expression:
\begin{align}
    L_{\rm gray}(t) = \dot{\varepsilon}(t)\, m_{\rm unc}(t),
    \label{eq:Lum_t}
\end{align}
where $\dot{\varepsilon}(t)$ is the specific volumetric heating rate, and $m_{\rm unc}(t)$ is the ``uncovered'' mass, or the mass of the layer above the diffusion surface.
The latter is defined as a surface at which the diffusion velocity $v_{\rm diff} = c/\tau$ equals the velocity differential to the edge of ejecta.
Diffusion surface separates the bulk where photons are trapped from the outer envelope from where photons can escape, diffusing faster than the matter is expanding.
Here, $\tau$ is the optical depth.
In every (co-moving) point of homologously expanding ejecta with uniform gray opacity, $\tau(t)\propto t^{-2}$~\citep{grossman14}.
This model has limited applicability: it does not take into account thermalization in the thin optical region above the photosphere and thus overestimates luminosity after the peak is reached.

\begin{figure*}[!htb]
\centering
\begin{minipage}[t]{0.32\textwidth}
  \centering
  \includegraphics[width=\textwidth]{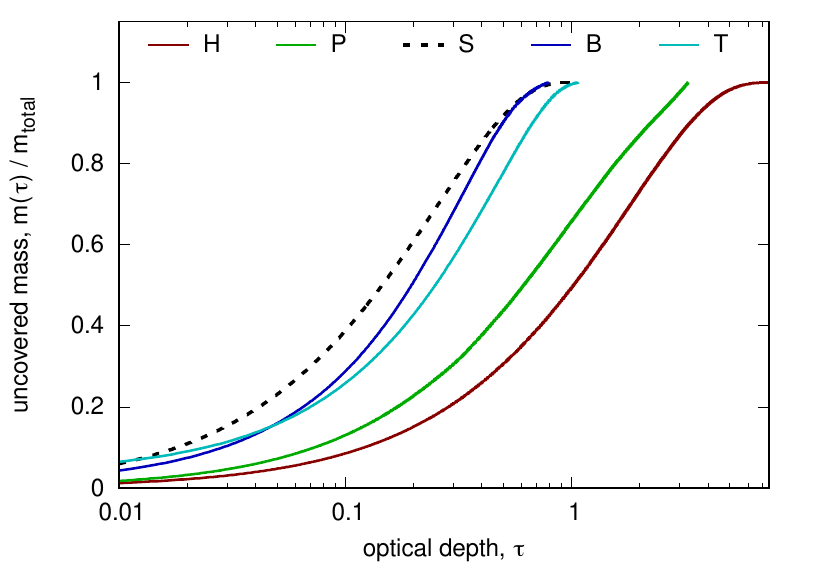}
   \caption{Fraction of mass uncovered above certain optical depth for models
   {\tt H}, {\tt P}, {\tt S}, {\tt B}, and {\tt T}.
   Gray opacity ${\kappa = 1\,{\rm cm}^2\,{\rm g}^{-1}}$ was used.
   The models are constrained to have the same total mass and average (RMS)
   expansion velocity.
   } 
   \label{fig:optical_depth_SHPRT}
\end{minipage}\;\;
\begin{minipage}[t]{0.65\textwidth}
  \centering
  \includegraphics[width=0.49\textwidth]{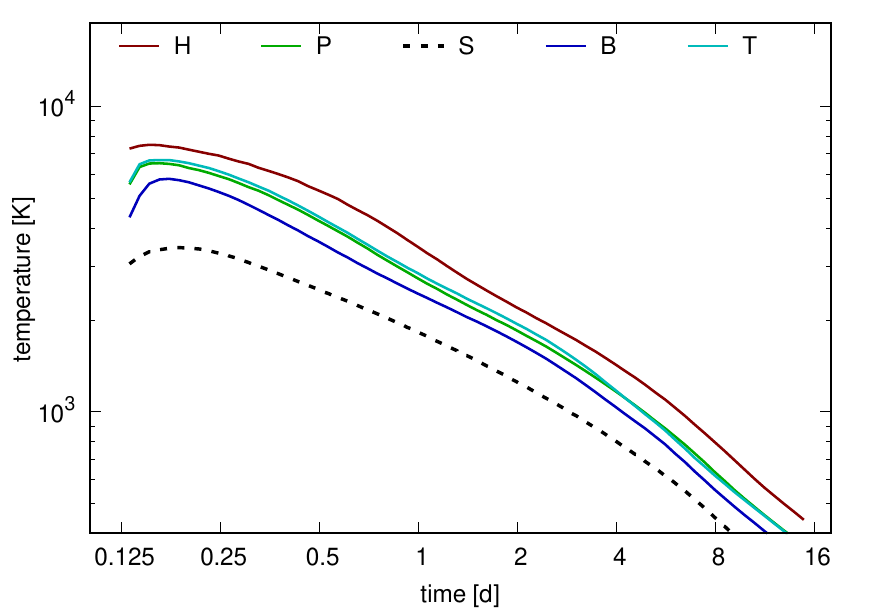}
  \includegraphics[width=0.49\textwidth]{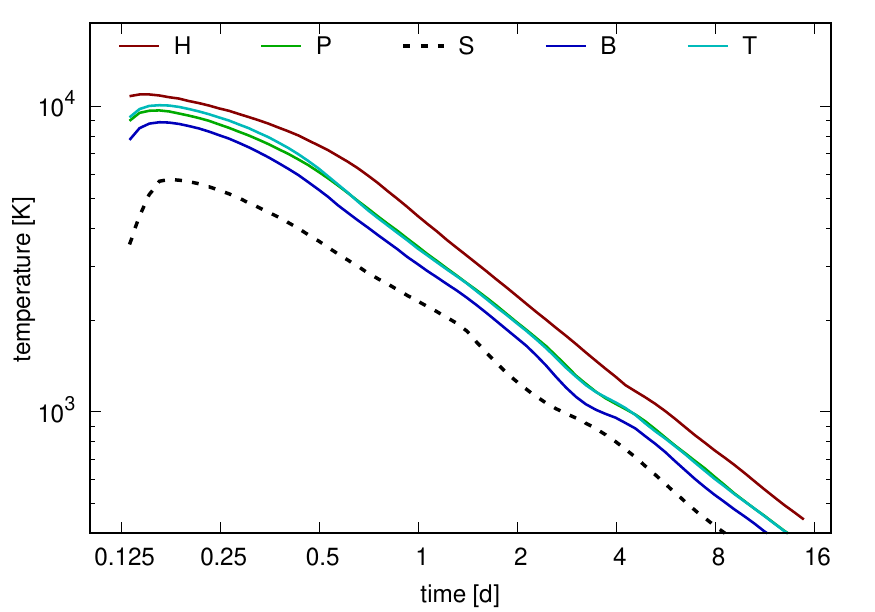}
  \caption{Radiation temperature at the matter boundary for the low-$Y_e$ (left)
  and high-$Y_e$ (right) compositions with analytic heating prescription.
  } 
  \label{fig:analysis_temp_lum}
\end{minipage}
\end{figure*}

Figure~\ref{fig:optical_depth_SHPRT} shows the fraction of uncovered mass as a function of optical depth for axisymmetric morphologies, normalized to the same mass and rms expansion velocity.
This profile $m_{\rm unc}(\tau)$ is only dependent on the density distribution inside each morphology, and is sufficient to estimate the bolometric light curve---using expression~(\ref{eq:Lum_t}) and the fact that $\tau(t)\propto t^{-2}$.
If more mass is ``buried'' at high optical depth, the kilonova will peak later and be dimmer.
Looking at Figure~\ref{fig:optical_depth_SHPRT}, we can anticipate that model {\tt S} will peak earliest, followed by models {\tt B} and {\tt T}, and the models {\tt P} and {\tt H} will peak last.
A similar trend is expected for the peak luminosity (in decreasing order).
This is indeed observed in simulations with gray opacity in Figure~\ref{fig:onec_lums_discussion} (left column), both for opacity $\kappa=10$ and $\kappa=1$.

\subsection{Thin-layer models}

When detailed opacities are introduced, their temperature dependence complicates the light curves, shifting peak time and luminosity by a factor of a few  (Fig.~\ref{fig:onec_lums_discussion}, middle column).
To understand these effects, we constructed a simple thin-layer approximation that uses snapshots of radiation temperature recorded by \SuperNu\ during our simulations.
Bolometric luminosity is computed as follows:
\begin{align}
    L_{\rm thin}(t) = A(t)\cdot c\ a\ T_{\rm out}^4(t),
\end{align}
where $T_{\rm out}$ is the recorded temperature at the surface of the morphology (plotted in Figure~\ref{fig:analysis_temp_lum}), and $A(t)$ is the uniformly expanding area of this surface: $A(t) = A_0 (t/t_0)^2$.
We use ${c\ a\ T_{\rm out}^4}$ for the surface flux instead of the usual $\sigma\ T_{\rm out}^4(\equiv{c\ a\ T_{\rm out}^4}/4)$ because in the thin photospheric layer, the radiation is free-streaming, such that intensity distribution is strongly peaked toward outward normal rather than being isotropic.
In this case, the entire radiative energy in the bulk of photospheric layer is escaping.
We also neglect photospheric recession and the corresponging decrease in the emitting area.
Figure~\ref{fig:onec_areas}, which shows the fractional area of the diffusion surface (discussed in detail below), proves this to be a reasonable assumption, as for nonspherical morphologies, the area of diffusion surface remains at least 0.8 of the area of the outer contour for up to about 2 days.
For the spherical morphology, this assumption is not very accurate and the ``thin-layer'' model is expected to overestimate the luminosity.

The resulting evolution of bolometric luminosity $L_{\rm thin}(t)$ is presented in the right column of Figure~\ref{fig:onec_lums_discussion}.
Comparing it to the middle column on the same plot, we see that our model correctly reproduces the features of the bolometric light curves, but overestimates luminosities by about $20-40\%$.
It also captures the order of peak times for different morphologies: {\tt B} peaking first and {\tt H} peaking the last.
The same is only partly true for the order in peak brightness, as this approximation overestimates it for morphology {\tt H}.
Overall, our numerical experiment demonstrates that the features in the bolometric light curve are primarily dictated by the behavior of the surface temperature, and less so by the emitting area. In particular, spherical morphology {\tt S} stands out by being the ``coldest'' and thus having the least luminous peak despite its having the largest surface area for the same mass and mean expansion velocity.

Next, we observe that the surface temperature ranking for different morphologies ($S < R < T\approx P < H$, Fig.~\ref{fig:analysis_temp_lum}) can be traced back to their ranking in density (Fig.~\ref{fig:density_SHPRT}).
Naturally, models with higher density produce more radioactive heat per volume and are expected to be hotter.
Moreover, if we compare bolometric luminosity of the low-$Y_e$ composition computed with thermalization taken into account (Fig.~\ref{fig:twoc_luminosities}), it boosts the higher-density models even further, as more radioactive energy is thermalized in denser regions.
Overall, density-dependent thermalization adds an extra boost by about factor of two in luminosity (cf. Fig.~\ref{fig:twoc_luminosities} and Fig.~\ref{fig:onec_lums_discussion})

\begin{figure}
\begin{tabular}{c}
  \includegraphics[width=0.95\textwidth]{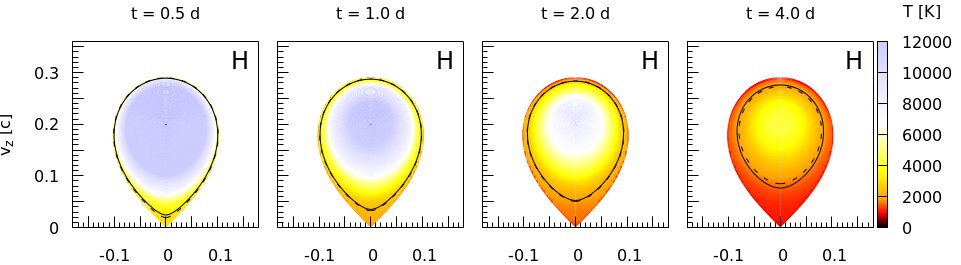} \\
  \includegraphics[width=0.95\textwidth]{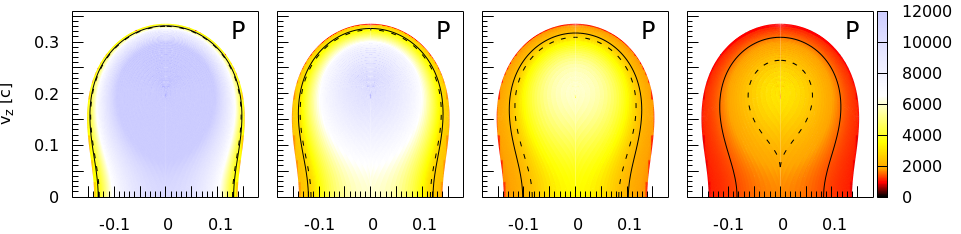} \\
  \includegraphics[width=0.95\textwidth]{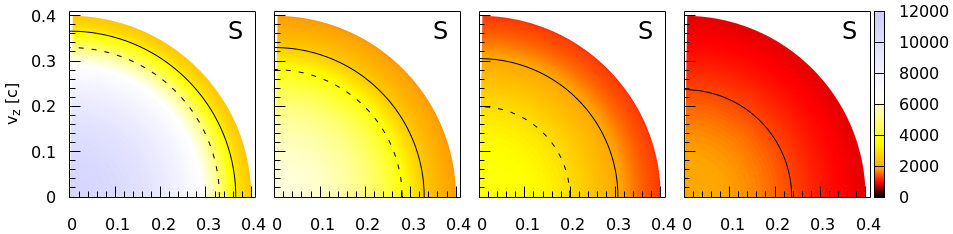} \\
  \includegraphics[width=0.95\textwidth]{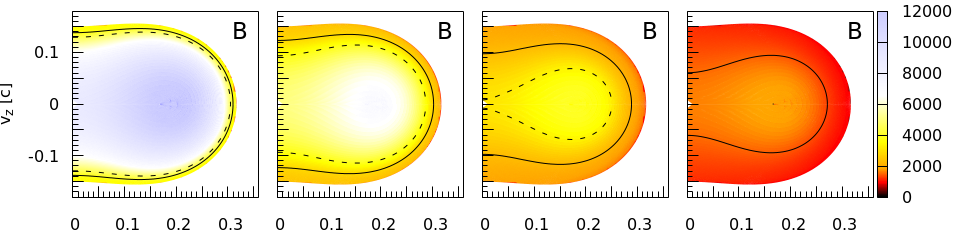} \\
  \includegraphics[width=0.95\textwidth]{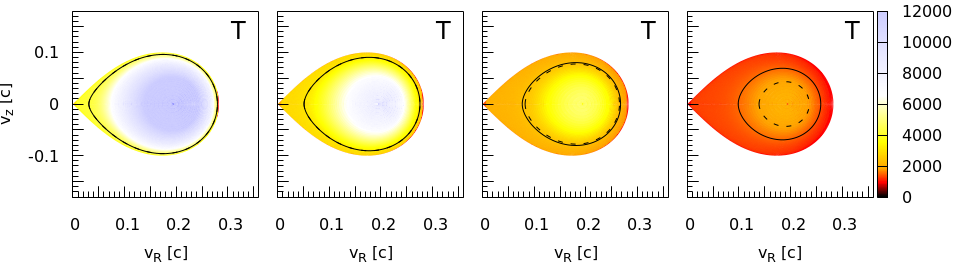}
\end{tabular}
\caption{Temperature maps for axisymmetric single-component morphologies with
  solar r-process residuals for four different epochs.
  The black contours on each model mark the estimated locations of the diffusion
  surface (solid line) and the photosphere (dashed line).
} 
\label{fig:temp_maps}
\end{figure}

The thin-layer model qualitatively reproduces the trends of the full radiative transfer model because it uses the surface temperatures from the simulations, which proves that the outbound Monte Carlo flux is consistent with the temperature evolution on the surface.
Below we attempt to further validate the ``thinness'' assumption by explicitly calculating an approximate position of the diffusion surface from simulation data.

\subsection{Estimating location of the diffusion surface and photosphere}

Figure~\ref{fig:temp_maps} shows the temperature color maps in our basic morphologies at different times, with overplotted estimated contours of the photosphere (dashed lines) and diffusion surface (solid lines).
To locate the photosphere, we used Rosseland mean opacities and integrated the optical depth $d\tau = \kappa_{\rm R}(\rho,T) \rho d\ell$ inwards from the outer edge of the expansion, adjusting the path of integration so that it always follows the local density gradient.
This results in a family of contours that determine the optical depth globally for every point as the optical depth minimized over all possible escape routes.

The photosphere is then given by the $\tau=2/3$ contour.
It turns out, however, that the Rosseland mean significantly underestimates true effective opacity and places the photosphere too deep.
Simple numerical integration of the radiative energy density in the layer above the photosphere computed in this manner gives numbers that greatly exceed the observed luminosity output of the kilonova.

A better way to pinpoint the location of the radiative layer is given by the diffusion surface.
We define the diffusion surface as enclosing the opaque ``core'' of the ejecta where photons are escaping slower than local expansion and are therefore trapped~\citep{grossman14}.
To simplify the analysis, we make further approximations and compute the diffusion surface as given by an optical depth $\tau_{\rm ds}(t)$ such that the integral of the radiative energy $E_{\rm rad}$ above this contour ($\tau < \tau_{\rm ds}$) is equal to the total bolometric luminosity:
\begin{align}
    \int_{\tau < \tau_{\rm ds}(t)} E_{\rm rad} dV = L_{\rm bol}(t).
    \label{eq:inttau}
\end{align}
This approximation ignores the anisotropy of radiation flux due to the asphericity of our models, but it is nevertheless sufficient to estimate the validity of the thin-layer approximation.

Indeed, Figure~\ref{fig:temp_maps} allows us to conclude that for all morphologies except {\tt S}, up to about 2~days the diffusion surface is not only very close to the outer edge of the expansion but also that the temperature in the radiative layer above it does not change significantly.
So taking the surface temperature as a proxy for the thin-layer model is justified.
On the other hand, contours of optical depth computed using the Rosseland mean fail to capture the photosphere: they instead place it inside the diffusion surface, which is unphysical for a typical kilonova scenario.

\begin{figure*}
\begin{tabular}{c}
  \includegraphics[width=0.95\textwidth]{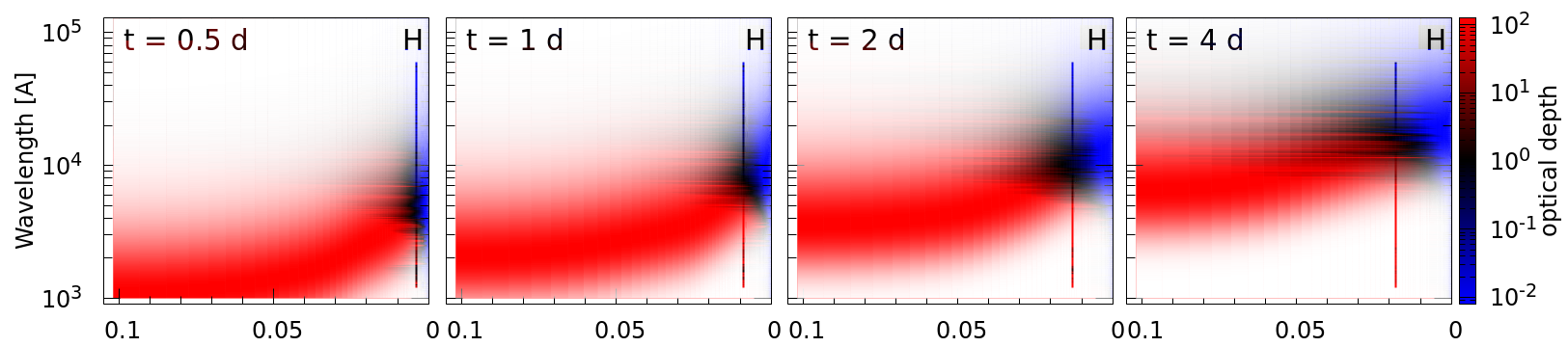} \\
  \includegraphics[width=0.95\textwidth]{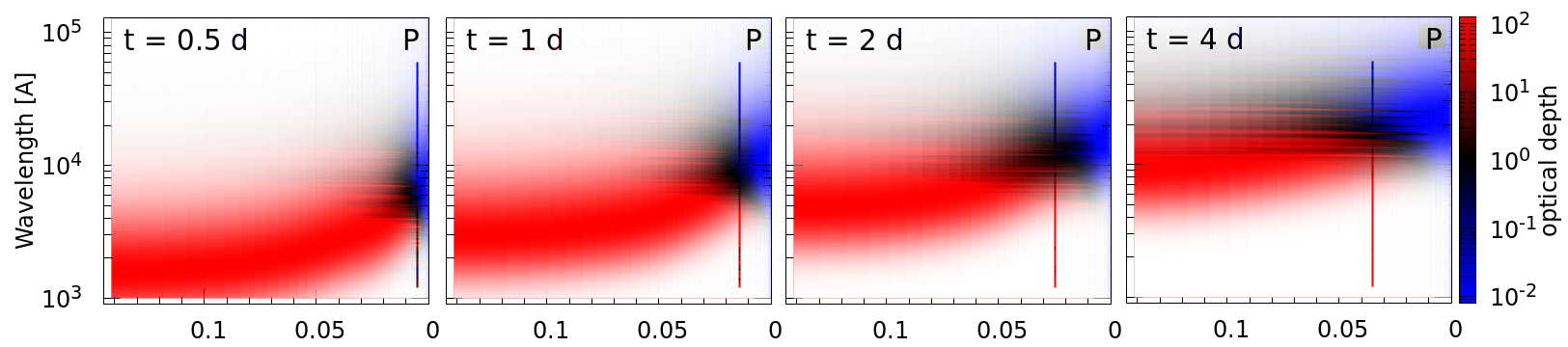} \\
  \includegraphics[width=0.95\textwidth]{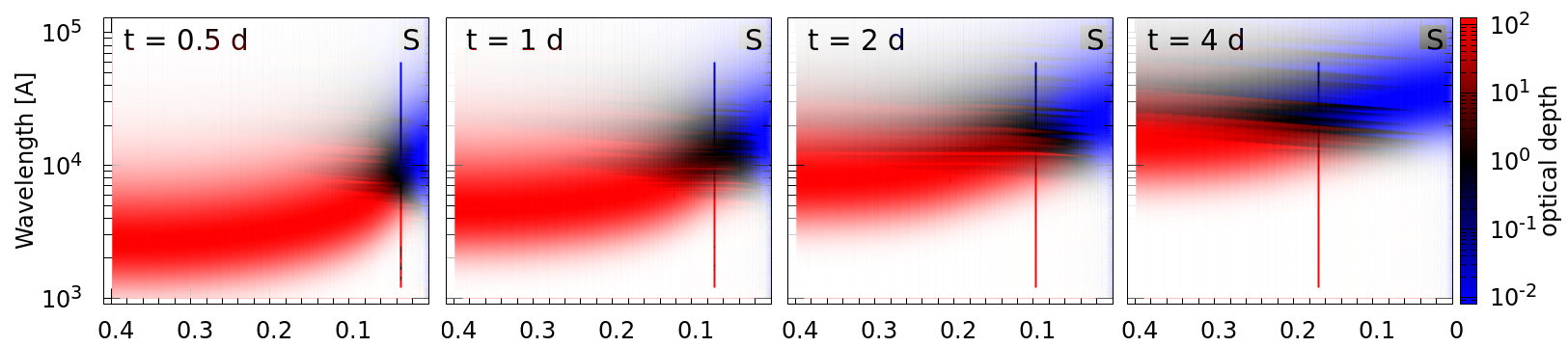} \\
  \includegraphics[width=0.95\textwidth]{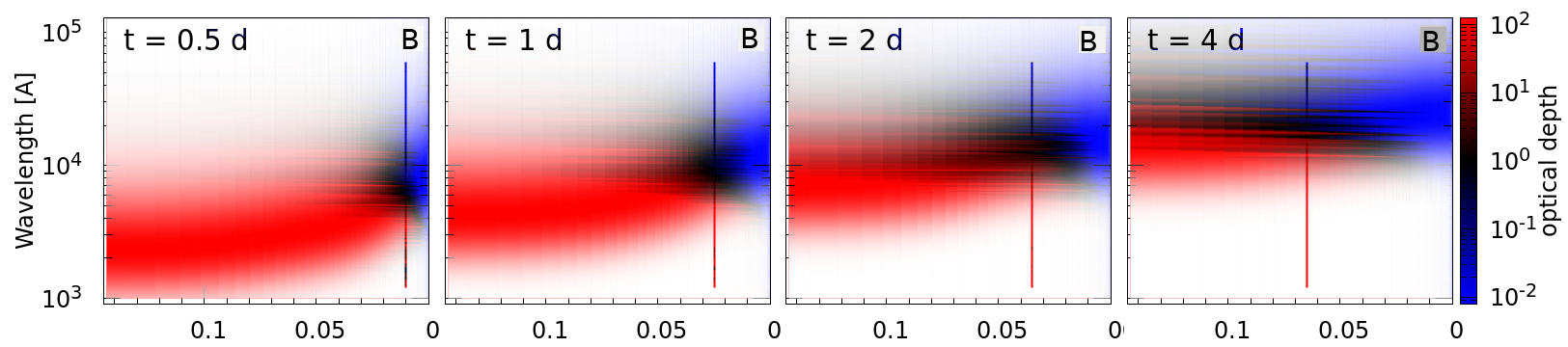} \\
  \includegraphics[width=0.95\textwidth]{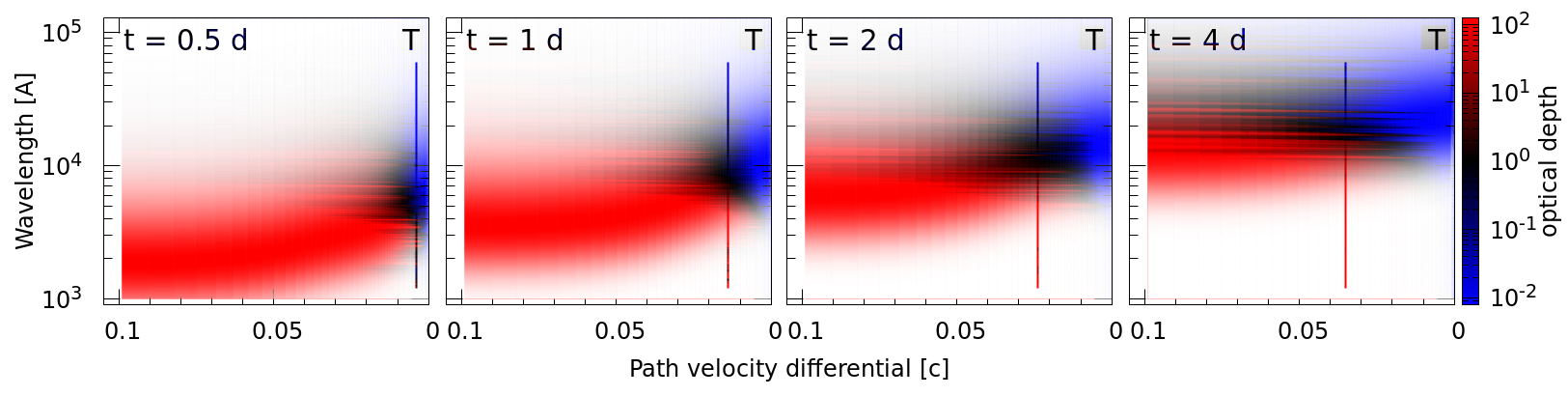}
\end{tabular}
\caption{Wavelength-dependent optical depth for single-component morphologies with
  solar r-process residuals for four different epochs. The solid lines
  show locations of the diffusion surface. See the main text for a
  detailed description of the plot.
} 
\label{fig:aimee_plots}
\end{figure*}

For any non-gray-opacity model, the location of the photosphere strongly depends on the wavelength.
This point is clearly illustrated on Figure~\ref{fig:aimee_plots} \citep[plots of this type were first introduced in][]{fontes19}.
Each plot represents a color map of the wavelength-dependent optical depth experienced by a photon exiting the ejecta and moving out from an initial point along a given path. For the path, we picked a straight line starting from the density maximum orthogonal to the longer axis of the morphology.
The horizontal axis starts at the density maximum and shows the overall velocity differential to the surface where the path terminates.
Peaks and troughs in the opacity landscape generate streaks of optical depth that appear tilted down to the right due to the expansion redshift.
As photons travel from left to right, they interact with progressively more redshifted opacity features, which creates the tilting effect.
The opacity landscape also changes with depth due to the changes in density and temperature.

For nonspherical morphologies, photons traverse a velocity differential of only ${0.1\,c}$, while for spherical morphology (middle row), it is ${0.4\,c}$ and the ``tilt'' is much more noticeable.

The three-color heat map was selected to show three different regions in optical depth: low (blue), medium and comparable to the diffusion speed (black), and high (red).
White matte regions on the plots are added to highlight the parts of the spectra where most of the photons would be emitted if spectra were blackbody.
Transparency of the matte layer is proportional to the Planck function with the local radiation temperature.
Solid vertical lines on each plot show the location of the diffusion surface.

The common feature on all plots is that the solid line crosses the optical depth map predominantly in the regions where the optical depth is ``black,'' i.e. $\sim1-10$.
In other words, the surface computed with Equation~(\ref{eq:inttau}) is located at an optical depth of about ${\tau_{\rm ds}\approx10}$, which is where the diffusion surface should be for an outflow with an expansion velocity of ${c/10}$.
This provides an additional argument in support of using line-binned opacities instead of expansion opacities for nonspherical models \citep[see][for detailed exposition of the argument]{fontes19}.


\begin{thebibliography}{}
\makeatletter
\relax
\def\mn@urlcharsother{\let\do\@makeother \do\$\do\&\do\#\do\^\do\_\do\%\do\~}
\def\mn@doi{\begingroup\mn@urlcharsother \@ifnextchar [ {\mn@doi@}
  {\mn@doi@[]}}
\def\mn@doi@[#1]#2{\def\@tempa{#1}\ifx\@tempa\@empty \href
  {http://dx.doi.org/#2} {doi:#2}\else \href {http://dx.doi.org/#2} {#1}\fi
  \endgroup}
\def\mn@eprint#1#2{\mn@eprint@#1:#2::\@nil}
\def\mn@eprint@arXiv#1{\href {http://arxiv.org/abs/#1} {{\tt arXiv:#1}}}
\def\mn@eprint@dblp#1{\href {http://dblp.uni-trier.de/rec/bibtex/#1.xml}
  {dblp:#1}}
\def\mn@eprint@#1:#2:#3:#4\@nil{\def\@tempa {#1}\def\@tempb {#2}\def\@tempc
  {#3}\ifx \@tempc \@empty \let \@tempc \@tempb \let \@tempb \@tempa \fi \ifx
  \@tempb \@empty \def\@tempb {arXiv}\fi \@ifundefined
  {mn@eprint@\@tempb}{\@tempb:\@tempc}{\expandafter \expandafter \csname
  mn@eprint@\@tempb\endcsname \expandafter{\@tempc}}}

\bibitem[\protect\citeauthoryear{{Abbott} et~al.,}{{Abbott}
  et~al.}{2017a}]{abbott17h}
{Abbott} B.~P.,  et~al., 2017a, \mn@doi [Physical Review Letters]
  {10.1103/PhysRevLett.119.161101}, \href
  {http://adsabs.harvard.edu/abs/2017PhRvL.119p1101A} {119, 161101}

\bibitem[\protect\citeauthoryear{{Abbott} et~al.,}{{Abbott}
  et~al.}{2017b}]{abbott17a}
{Abbott} B.~P.,  et~al., 2017b, \mn@doi [\apjl] {10.3847/2041-8213/aa91c9},
  \href {http://adsabs.harvard.edu/abs/2017ApJ...848L..12A} {848, L12}

\bibitem[\protect\citeauthoryear{{Abdikamalov}, {Burrows}, {Ott},
  {L{\"o}ffler}, {O'Connor}, {Dolence}  \& {Schnetter}}{{Abdikamalov}
  et~al.}{2012}]{abdikamalov12}
{Abdikamalov} E.,  {Burrows} A.,  {Ott} C.~D.,  {L{\"o}ffler} F.,  {O'Connor}
  E.,  {Dolence} J.~C.,   {Schnetter} E.,  2012, \mn@doi [\apj]
  {10.1088/0004-637X/755/2/111}, \href
  {https://ui.adsabs.harvard.edu/\#abs/2012ApJ...755..111A} {755, 111}

\bibitem[\protect\citeauthoryear{{Baiotti} \& {Rezzolla}}{{Baiotti} \&
  {Rezzolla}}{2017}]{baiotti17}
{Baiotti} L.,  {Rezzolla} L.,  2017, \mn@doi [Reports on Progress in Physics]
  {10.1088/1361-6633/aa67bb}, \href
  {https://ui.adsabs.harvard.edu/abs/2017RPPh...80i6901B} {80, 096901}

\bibitem[\protect\citeauthoryear{{Barbieri}, {Salafia}, {Perego}, {Colpi}  \&
  {Ghirlanda}}{{Barbieri} et~al.}{2019}]{barbieri19}
{Barbieri} C.,  {Salafia} O.~S.,  {Perego} A.,  {Colpi} M.,   {Ghirlanda} G.,
  2019, \mn@doi [\aap] {10.1051/0004-6361/201935443}, \href
  {https://ui.adsabs.harvard.edu/abs/2019A&A...625A.152B} {625, A152}

\bibitem[\protect\citeauthoryear{{Barnes} \& {Kasen}}{{Barnes} \&
  {Kasen}}{2013}]{barnes13}
{Barnes} J.,  {Kasen} D.,  2013, \mn@doi [\apj] {10.1088/0004-637X/775/1/18},
  \href {https://ui.adsabs.harvard.edu/abs/2013ApJ...775...18B} {775, 18}

\bibitem[\protect\citeauthoryear{{Barnes}, {Kasen}, {Wu}  \&
  {Mart{\'\i}nez-Pinedo}}{{Barnes} et~al.}{2016}]{barnes16}
{Barnes} J.,  {Kasen} D.,  {Wu} M.-R.,   {Mart{\'\i}nez-Pinedo} G.,  2016,
  \mn@doi [\apj] {10.3847/0004-637X/829/2/110}, \href
  {https://ui.adsabs.harvard.edu/abs/2016ApJ...829..110B} {829, 110}

\bibitem[\protect\citeauthoryear{{Barnes}, {Zhu}, {Lund}, {Sprouse}, {Vassh},
  {McLaughlin}, {Mumpower}  \& {Surman}}{{Barnes} et~al.}{2020}]{barnes20}
{Barnes} J.,  {Zhu} Y.~L.,  {Lund} K.~A.,  {Sprouse} T.~M.,  {Vassh} N.,
  {McLaughlin} G.~C.,  {Mumpower} M.~R.,   {Surman} R.,  2020, arXiv e-prints,
  \href {https://ui.adsabs.harvard.edu/abs/2020arXiv201011182B} {p.
  arXiv:2010.11182}

\bibitem[\protect\citeauthoryear{{Bauswein}, {Goriely}  \& {Janka}}{{Bauswein}
  et~al.}{2013}]{bauswein13}
{Bauswein} A.,  {Goriely} S.,   {Janka} H.~T.,  2013, \mn@doi [\apj]
  {10.1088/0004-637X/773/1/78}, \href
  {https://ui.adsabs.harvard.edu/abs/2013ApJ...773...78B} {773, 78}

\bibitem[\protect\citeauthoryear{{Bulla}}{{Bulla}}{2019}]{bulla19}
{Bulla} M.,  2019, \mn@doi [\mnras] {10.1093/mnras/stz2495}, \href
  {https://ui.adsabs.harvard.edu/abs/2019MNRAS.489.5037B} {489, 5037}

\bibitem[\protect\citeauthoryear{{Castor} \& {Lamers}}{{Castor} \&
  {Lamers}}{1979}]{castor79}
{Castor} J.~I.,  {Lamers} H.~J.~G.~L.~M.,  1979, \mn@doi [\apjs]
  {10.1086/190583}, \href
  {https://ui.adsabs.harvard.edu/abs/1979ApJS...39..481C} {39, 481}

\bibitem[\protect\citeauthoryear{{Chornock} et~al.,}{{Chornock}
  et~al.}{2017}]{chornock17}
{Chornock} R.,  et~al., 2017, \mn@doi [\apjl] {10.3847/2041-8213/aa905c}, \href
  {http://adsabs.harvard.edu/abs/2017ApJ...848L..19C} {848, L19}

\bibitem[\protect\citeauthoryear{Cleveland \& Gentile}{Cleveland \&
  Gentile}{2014}]{cleveland14}
Cleveland M.~A.,  Gentile N.,  2014, Journal of Computational and Theoretical
  Transport, 43, 6

\bibitem[\protect\citeauthoryear{{C{\^o}t{\'e}}, {Belczynski}, {Fryer},
  {Ritter}, {Paul}, {Wehmeyer}  \& {O'Shea}}{{C{\^o}t{\'e}}
  et~al.}{2017}]{cote17}
{C{\^o}t{\'e}} B.,  {Belczynski} K.,  {Fryer} C.~L.,  {Ritter} C.,  {Paul} A.,
  {Wehmeyer} B.,   {O'Shea} B.~W.,  2017, \mn@doi [\apj]
  {10.3847/1538-4357/aa5c8d}, \href
  {https://ui.adsabs.harvard.edu/abs/2017ApJ...836..230C} {836, 230}

\bibitem[\protect\citeauthoryear{{C{\^o}t{\'e}} et~al.,}{{C{\^o}t{\'e}}
  et~al.}{2018}]{cote18}
{C{\^o}t{\'e}} B.,  et~al., 2018, \mn@doi [\apj] {10.3847/1538-4357/aaad67},
  \href {https://ui.adsabs.harvard.edu/abs/2018ApJ...855...99C} {855, 99}

\bibitem[\protect\citeauthoryear{{C{\^o}t{\'e}} et~al.,}{{C{\^o}t{\'e}}
  et~al.}{2019}]{cote19}
{C{\^o}t{\'e}} B.,  et~al., 2019, \mn@doi [\apj] {10.3847/1538-4357/ab10db},
  \href {https://ui.adsabs.harvard.edu/abs/2019ApJ...875..106C} {875, 106}

\bibitem[\protect\citeauthoryear{{Cowan}, {Sneden}, {Lawler}, {Aprahamian},
  {Wiescher}, {Langanke}, {Mart{\'\i}nez-Pinedo}  \& {Thielemann}}{{Cowan}
  et~al.}{2019}]{cowan19}
{Cowan} J.~J.,  {Sneden} C.,  {Lawler} J.~E.,  {Aprahamian} A.,  {Wiescher} M.,
   {Langanke} K.,  {Mart{\'\i}nez-Pinedo} G.,   {Thielemann} F.-K.,  2019,
  arXiv e-prints, \href {https://ui.adsabs.harvard.edu/abs/2019arXiv190101410C}
  {p. arXiv:1901.01410}

\bibitem[\protect\citeauthoryear{{Cowperthwaite} et~al.,}{{Cowperthwaite}
  et~al.}{2017}]{cowperthwaite17}
{Cowperthwaite} P.~S.,  et~al., 2017, \mn@doi [\apj]
  {10.3847/2041-8213/aa8fc7}, \href
  {https://ui.adsabs.harvard.edu/abs/2017ApJ...848L..17C} {848, L17}

\bibitem[\protect\citeauthoryear{{Darbha} \& {Kasen}}{{Darbha} \&
  {Kasen}}{2020}]{darbha20}
{Darbha} S.,  {Kasen} D.,  2020, \mn@doi [\apj] {10.3847/1538-4357/ab9a34},
  \href {https://ui.adsabs.harvard.edu/abs/2020ApJ...897..150D} {897, 150}

\bibitem[\protect\citeauthoryear{{De La Rosa}, {Roming}  \& {Fryer}}{{De La
  Rosa} et~al.}{2017}]{delarosa17}
{De La Rosa} J.,  {Roming} P.,   {Fryer} C.,  2017, \mn@doi [\apj]
  {10.3847/1538-4357/aa93ee}, \href
  {https://ui.adsabs.harvard.edu/abs/2017ApJ...850..133D} {850, 133}

\bibitem[\protect\citeauthoryear{Densmore, Thompson  \& Urbatsch}{Densmore
  et~al.}{2012}]{densmore12}
Densmore J.~D.,  Thompson K.~G.,   Urbatsch T.~J.,  2012, Journal of
  Computational Physics, 231, 6924

\bibitem[\protect\citeauthoryear{{Dietrich}, {Bernuzzi}, {Ujevic}  \&
  {Tichy}}{{Dietrich} et~al.}{2017}]{dietrich17}
{Dietrich} T.,  {Bernuzzi} S.,  {Ujevic} M.,   {Tichy} W.,  2017, \mn@doi
  [\prd] {10.1103/PhysRevD.95.044045}, \href
  {https://ui.adsabs.harvard.edu/abs/2017PhRvD..95d4045D} {95, 044045}

\bibitem[\protect\citeauthoryear{{Eichler}, {Livio}, {Piran}  \&
  {Schramm}}{{Eichler} et~al.}{1989}]{eichler89}
{Eichler} D.,  {Livio} M.,  {Piran} T.,   {Schramm} D.~N.,  1989, \mn@doi
  [\nat] {10.1038/340126a0}, \href
  {https://ui.adsabs.harvard.edu/abs/1989Natur.340..126E} {340, 126}

\bibitem[\protect\citeauthoryear{{Evans} et~al.,}{{Evans}
  et~al.}{2017}]{evans17}
{Evans} P.~A.,  et~al., 2017, \mn@doi [Science] {10.1126/science.aap9580},
  \href {https://ui.adsabs.harvard.edu/abs/2017Sci...358.1565E} {358, 1565}

\bibitem[\protect\citeauthoryear{{Even} et~al.,}{{Even} et~al.}{2020}]{even20}
{Even} W.,  et~al., 2020, \mn@doi [\apj] {10.3847/1538-4357/ab70b9}, \href
  {https://ui.adsabs.harvard.edu/abs/2020ApJ...899...24E} {899, 24}

\bibitem[\protect\citeauthoryear{{Fahlman} \& {Fern{\'a}ndez}}{{Fahlman} \&
  {Fern{\'a}ndez}}{2018}]{fahlman18}
{Fahlman} S.,  {Fern{\'a}ndez} R.,  2018, \mn@doi [ApJ]
  {10.3847/2041-8213/aaf1ab}, \href
  {https://ui.adsabs.harvard.edu/abs/2018ApJ...869L...3F} {869, L3}

\bibitem[\protect\citeauthoryear{{Fern{\'a}ndez} \& {Metzger}}{{Fern{\'a}ndez}
  \& {Metzger}}{2013}]{fernandez13}
{Fern{\'a}ndez} R.,  {Metzger} B.~D.,  2013, \mn@doi [\mnras]
  {10.1093/mnras/stt1312}, \href
  {https://ui.adsabs.harvard.edu/abs/2013MNRAS.435..502F} {435, 502}

\bibitem[\protect\citeauthoryear{{Fern{\'a}ndez} \& {Metzger}}{{Fern{\'a}ndez}
  \& {Metzger}}{2016}]{fernandez16o}
{Fern{\'a}ndez} R.,  {Metzger} B.~D.,  2016, \mn@doi [Annual Review of Nuclear
  and Particle Science] {10.1146/annurev-nucl-102115-044819}, \href
  {http://adsabs.harvard.edu/abs/2016ARNPS..66...23F} {66, 23}

\bibitem[\protect\citeauthoryear{{Fern{\'a}ndez}, {Quataert}, {Schwab}, {Kasen}
   \& {Rosswog}}{{Fern{\'a}ndez} et~al.}{2015}]{fernandez15}
{Fern{\'a}ndez} R.,  {Quataert} E.,  {Schwab} J.,  {Kasen} D.,   {Rosswog} S.,
  2015, \mn@doi [\mnras] {10.1093/mnras/stv238}, \href
  {https://ui.adsabs.harvard.edu/abs/2015MNRAS.449..390F} {449, 390}

\bibitem[\protect\citeauthoryear{Fleck~Jr \& Cummings~Jr}{Fleck~Jr \&
  Cummings~Jr}{1971}]{fleck71}
Fleck~Jr J.,  Cummings~Jr J.,  1971, Journal of Computational Physics, 8, 313

\bibitem[\protect\citeauthoryear{{Fontes}, {Fryer}, {Hungerford}, {Hakel},
  {Colgan}, {Kilcrease}  \& {Sherrill}}{{Fontes} et~al.}{2015}]{fontes15a}
{Fontes} C.~J.,  {Fryer} C.~L.,  {Hungerford} A.~L.,  {Hakel} P.,  {Colgan} J.,
   {Kilcrease} D.~P.,   {Sherrill} M.~E.,  2015, \mn@doi [{High Energy Density
  Physics}] {http://dx.doi.org/10.1016/j.hedp.2015.06.002}, 16, 53

\bibitem[\protect\citeauthoryear{{Fontes}, {Fryer}, {Hungerford}, {Wollaeger},
  {Rosswog}  \& {Berger}}{{Fontes} et~al.}{2017}]{fontes17}
{Fontes} C.~J.,  {Fryer} C.~L.,  {Hungerford} A.~L.,  {Wollaeger} R.~T.,
  {Rosswog} S.,   {Berger} E.,  2017, arXiv e-prints, \href
  {https://ui.adsabs.harvard.edu/\#abs/2017arXiv170202990F} {p.
  arXiv:1702.02990}

\bibitem[\protect\citeauthoryear{{Fontes}, {Fryer}, {Hungerford}, {Wollaeger}
  \& {Korobkin}}{{Fontes} et~al.}{2020}]{fontes19}
{Fontes} C.~J.,  {Fryer} C.~L.,  {Hungerford} A.~L.,  {Wollaeger} R.~T.,
  {Korobkin} O.,  2020, \mn@doi [\mnras] {10.1093/mnras/staa485}, \href
  {https://ui.adsabs.harvard.edu/abs/2020MNRAS.493.4143F} {493, 4143}

\bibitem[\protect\citeauthoryear{{Freiburghaus}, {Rosswog}  \&
  {Thielemann}}{{Freiburghaus} et~al.}{1999}]{freiburghaus99}
{Freiburghaus} C.,  {Rosswog} S.,   {Thielemann} F.~K.,  1999, \mn@doi [\apj]
  {10.1086/312343}, \href
  {https://ui.adsabs.harvard.edu/abs/1999ApJ...525L.121F} {525, L121}

\bibitem[\protect\citeauthoryear{{Gaigalas}, {Kato}, {Rynkun},
  {Rad{\v{z}}i{\={u}}t{\.{e}}}  \& {Tanaka}}{{Gaigalas}
  et~al.}{2019}]{gaigalas19}
{Gaigalas} G.,  {Kato} D.,  {Rynkun} P.,  {Rad{\v{z}}i{\={u}}t{\.{e}}} L.,
  {Tanaka} M.,  2019, \mn@doi [\apjs] {10.3847/1538-4365/aaf9b8}, \href
  {https://ui.adsabs.harvard.edu/abs/2019ApJS..240...29G} {240, 29}

\bibitem[\protect\citeauthoryear{{Goriely}, {Bauswein}  \& {Janka}}{{Goriely}
  et~al.}{2011}]{goriely11}
{Goriely} S.,  {Bauswein} A.,   {Janka} H.-T.,  2011, \mn@doi [\apjl]
  {10.1088/2041-8205/738/2/L32}, \href
  {https://ui.adsabs.harvard.edu/abs/2011ApJ...738L..32G} {738, L32}

\bibitem[\protect\citeauthoryear{{Grossman}, {Korobkin}, {Rosswog}  \&
  {Piran}}{{Grossman} et~al.}{2014}]{grossman14}
{Grossman} D.,  {Korobkin} O.,  {Rosswog} S.,   {Piran} T.,  2014, \mn@doi
  [\mnras] {10.1093/mnras/stt2503}, \href
  {https://ui.adsabs.harvard.edu/abs/2014MNRAS.439..757G} {439, 757}

\bibitem[\protect\citeauthoryear{{Heinzel} et~al.,}{{Heinzel}
  et~al.}{2021}]{heinzel21}
{Heinzel} J.,  et~al., 2021, \mn@doi [\mnras] {10.1093/mnras/stab221}, \href
  {https://ui.adsabs.harvard.edu/abs/2021MNRAS.502.3057H} {502, 3057}

\bibitem[\protect\citeauthoryear{{Hotokezaka} \& {Nakar}}{{Hotokezaka} \&
  {Nakar}}{2019}]{hotokezaka19}
{Hotokezaka} K.,  {Nakar} E.,  2019, {HeatingRate: Radioactive heating rate and
  macronova (kilonova) light curve} (\mn@eprint {ascl} {1911.008})

\bibitem[\protect\citeauthoryear{{Hotokezaka}, {Piran}  \& {Paul}}{{Hotokezaka}
  et~al.}{2015}]{hotokezaka15}
{Hotokezaka} K.,  {Piran} T.,   {Paul} M.,  2015, \mn@doi [Nature Physics]
  {10.1038/nphys3574}, \href
  {https://ui.adsabs.harvard.edu/abs/2015NatPh..11.1042H} {11, 1042}

\bibitem[\protect\citeauthoryear{{Hotokezaka}, {Beniamini}  \&
  {Piran}}{{Hotokezaka} et~al.}{2018}]{hotokezaka18}
{Hotokezaka} K.,  {Beniamini} P.,   {Piran} T.,  2018, \mn@doi [International
  Journal of Modern Physics D] {10.1142/S0218271818420051}, \href
  {https://ui.adsabs.harvard.edu/abs/2018IJMPD..2742005H} {27, 1842005}

\bibitem[\protect\citeauthoryear{{Janiuk}}{{Janiuk}}{2014}]{janiuk14}
{Janiuk} A.,  2014, \mn@doi [\aap] {10.1051/0004-6361/201423822}, \href
  {https://ui.adsabs.harvard.edu/abs/2014A&A...568A.105J} {568, A105}

\bibitem[\protect\citeauthoryear{{Ji}, {Drout}  \& {Hansen}}{{Ji}
  et~al.}{2019}]{ji19}
{Ji} A.~P.,  {Drout} M.~R.,   {Hansen} T.~T.,  2019, \mn@doi [\apj]
  {10.3847/1538-4357/ab3291}, \href
  {https://ui.adsabs.harvard.edu/abs/2019ApJ...882...40J} {882, 40}

\bibitem[\protect\citeauthoryear{{Just}, {Bauswein}, {Ardevol Pulpillo},
  {Goriely}  \& {Janka}}{{Just} et~al.}{2015}]{just15}
{Just} O.,  {Bauswein} A.,  {Ardevol Pulpillo} R.,  {Goriely} S.,   {Janka}
  H.~T.,  2015, \mn@doi [\mnras] {10.1093/mnras/stv009}, \href
  {https://ui.adsabs.harvard.edu/abs/2015MNRAS.448..541J} {448, 541}

\bibitem[\protect\citeauthoryear{{Kasen} \& {Barnes}}{{Kasen} \&
  {Barnes}}{2019}]{kasen19}
{Kasen} D.,  {Barnes} J.,  2019, \mn@doi [\apj] {10.3847/1538-4357/ab06c2},
  \href {https://ui.adsabs.harvard.edu/abs/2019ApJ...876..128K} {876, 128}

\bibitem[\protect\citeauthoryear{{Kasen}, {Badnell}  \& {Barnes}}{{Kasen}
  et~al.}{2013}]{kasen13}
{Kasen} D.,  {Badnell} N.~R.,   {Barnes} J.,  2013, \mn@doi [\apj]
  {10.1088/0004-637X/774/1/25}, \href
  {https://ui.adsabs.harvard.edu/abs/2013ApJ...774...25K} {774, 25}

\bibitem[\protect\citeauthoryear{{Kasen}, {Fern{\'a}ndez}  \&
  {Metzger}}{{Kasen} et~al.}{2015}]{kasen15}
{Kasen} D.,  {Fern{\'a}ndez} R.,   {Metzger} B.~D.,  2015, \mn@doi [\mnras]
  {10.1093/mnras/stv721}, \href
  {https://ui.adsabs.harvard.edu/abs/2015MNRAS.450.1777K} {450, 1777}

\bibitem[\protect\citeauthoryear{{Kasen}, {Metzger}, {Barnes}, {Quataert}  \&
  {Ramirez-Ruiz}}{{Kasen} et~al.}{2017}]{kasen17}
{Kasen} D.,  {Metzger} B.,  {Barnes} J.,  {Quataert} E.,   {Ramirez-Ruiz} E.,
  2017, \mn@doi [\nat] {10.1038/nature24453}, \href
  {http://adsabs.harvard.edu/abs/2017Natur.551...80K} {551, 80}

\bibitem[\protect\citeauthoryear{{Kasliwal} et~al.,}{{Kasliwal}
  et~al.}{2017a}]{kasliwal17b}
{Kasliwal} M.~M.,  et~al., 2017a, \mn@doi [Science in press, available via
  doi:10.1126/science.aap9455] {10.1126/science.aap9455}, \href
  {http://adsabs.harvard.edu/abs/2017arXiv171005436K} {}

\bibitem[\protect\citeauthoryear{{Kasliwal}, {Korobkin}, {Lau}, {Wollaeger}  \&
  {Fryer}}{{Kasliwal} et~al.}{2017b}]{kasliwal17a}
{Kasliwal} M.~M.,  {Korobkin} O.,  {Lau} R.~M.,  {Wollaeger} R.,   {Fryer}
  C.~L.,  2017b, \mn@doi [\apj] {10.3847/2041-8213/aa799d}, \href
  {https://ui.adsabs.harvard.edu/\#abs/2017ApJ...843L..34K} {843, L34}

\bibitem[\protect\citeauthoryear{{Kawaguchi}, {Shibata}  \&
  {Tanaka}}{{Kawaguchi} et~al.}{2018}]{kawaguchi18}
{Kawaguchi} K.,  {Shibata} M.,   {Tanaka} M.,  2018, \mn@doi [\apj]
  {10.3847/2041-8213/aade02}, \href
  {https://ui.adsabs.harvard.edu/\#abs/2018ApJ...865L..21K} {865, L21}

\bibitem[\protect\citeauthoryear{{Kawaguchi}, {Shibata}  \&
  {Tanaka}}{{Kawaguchi} et~al.}{2020}]{kawaguchi20}
{Kawaguchi} K.,  {Shibata} M.,   {Tanaka} M.,  2020, \mn@doi [\apj]
  {10.3847/1538-4357/ab61f6}, \href
  {https://ui.adsabs.harvard.edu/abs/2020ApJ...889..171K} {889, 171}

\bibitem[\protect\citeauthoryear{{Kilpatrick} et~al.,}{{Kilpatrick}
  et~al.}{2017}]{kilpatrick17}
{Kilpatrick} C.~D.,  et~al., 2017, Science in press, available via
  doi:10.1126/science.aaq0073, \href
  {http://adsabs.harvard.edu/abs/2017arXiv171005434K} {}

\bibitem[\protect\citeauthoryear{{Kiuchi}, {Kawaguchi}, {Kyutoku}, {Sekiguchi},
  {Shibata}  \& {Taniguchi}}{{Kiuchi} et~al.}{2017}]{kiuchi17}
{Kiuchi} K.,  {Kawaguchi} K.,  {Kyutoku} K.,  {Sekiguchi} Y.,  {Shibata} M.,
  {Taniguchi} K.,  2017, \mn@doi [\prd] {10.1103/PhysRevD.96.084060}, \href
  {https://ui.adsabs.harvard.edu/abs/2017PhRvD..96h4060K} {96, 084060}

\bibitem[\protect\citeauthoryear{{Korobkin}, {Rosswog}, {Arcones}  \&
  {Winteler}}{{Korobkin} et~al.}{2012}]{korobkin12}
{Korobkin} O.,  {Rosswog} S.,  {Arcones} A.,   {Winteler} C.,  2012, \mn@doi
  [\mnras] {10.1111/j.1365-2966.2012.21859.x}, \href
  {https://ui.adsabs.harvard.edu/abs/2012MNRAS.426.1940K} {426, 1940}

\bibitem[\protect\citeauthoryear{{Kozyreva} et~al.,}{{Kozyreva}
  et~al.}{2017}]{kozyreva17}
{Kozyreva} A.,  et~al., 2017, \mn@doi [\mnras] {10.1093/mnras/stw2562}, \href
  {https://ui.adsabs.harvard.edu/\#abs/2017MNRAS.464.2854K} {464, 2854}

\bibitem[\protect\citeauthoryear{{Kr{\"u}ger} \& {Foucart}}{{Kr{\"u}ger} \&
  {Foucart}}{2020}]{krueger20}
{Kr{\"u}ger} C.~J.,  {Foucart} F.,  2020, \mn@doi [\prd]
  {10.1103/PhysRevD.101.103002}, \href
  {https://ui.adsabs.harvard.edu/abs/2020PhRvD.101j3002K} {101, 103002}

\bibitem[\protect\citeauthoryear{{Kyutoku}, {Ioka}  \& {Shibata}}{{Kyutoku}
  et~al.}{2013}]{kyutoku13}
{Kyutoku} K.,  {Ioka} K.,   {Shibata} M.,  2013, \mn@doi [\prd]
  {10.1103/PhysRevD.88.041503}, \href
  {https://ui.adsabs.harvard.edu/abs/2013PhRvD..88d1503K} {88, 041503}

\bibitem[\protect\citeauthoryear{{Lattimer} \& {Schramm}}{{Lattimer} \&
  {Schramm}}{1974}]{lattimer74}
{Lattimer} J.~M.,  {Schramm} D.~N.,  1974, \mn@doi [\apj] {10.1086/181612},
  \href {https://ui.adsabs.harvard.edu/abs/1974ApJ...192L.145L} {192, L145}

\bibitem[\protect\citeauthoryear{{Lippuner} \& {Roberts}}{{Lippuner} \&
  {Roberts}}{2015}]{lippuner15}
{Lippuner} J.,  {Roberts} L.~F.,  2015, \mn@doi [\apj]
  {10.1088/0004-637X/815/2/82}, \href
  {https://ui.adsabs.harvard.edu/abs/2015ApJ...815...82L} {815, 82}

\bibitem[\protect\citeauthoryear{{Martin}, {Perego}, {Arcones}, {Thielemann},
  {Korobkin}  \& {Rosswog}}{{Martin} et~al.}{2015}]{martin15}
{Martin} D.,  {Perego} A.,  {Arcones} A.,  {Thielemann} F.-K.,  {Korobkin} O.,
   {Rosswog} S.,  2015, \mn@doi [\apj] {10.1088/0004-637X/813/1/2}, \href
  {http://adsabs.harvard.edu/abs/2015ApJ...813....2M} {813, 2}

\bibitem[\protect\citeauthoryear{{McCully} et~al.,}{{McCully}
  et~al.}{2017}]{mccully17}
{McCully} C.,  et~al., 2017, \mn@doi [\apjl] {10.3847/2041-8213/aa9111}, \href
  {http://adsabs.harvard.edu/abs/2017ApJ...848L..32M} {848, L32}

\bibitem[\protect\citeauthoryear{{Metzger}}{{Metzger}}{2019}]{metzger19}
{Metzger} B.~D.,  2019, \mn@doi [Living Reviews in Relativity]
  {10.1007/s41114-019-0024-0}, \href
  {https://ui.adsabs.harvard.edu/abs/2019LRR....23....1M} {23, 1}

\bibitem[\protect\citeauthoryear{{Metzger} et~al.,}{{Metzger}
  et~al.}{2010}]{metzger10}
{Metzger} B.~D.,  et~al., 2010, \mn@doi [\mnras]
  {10.1111/j.1365-2966.2010.16864.x}, \href
  {https://ui.adsabs.harvard.edu/abs/2010MNRAS.406.2650M} {406, 2650}

\bibitem[\protect\citeauthoryear{{Metzger}, {Thompson}  \&
  {Quataert}}{{Metzger} et~al.}{2018}]{metzger18a}
{Metzger} B.~D.,  {Thompson} T.~A.,   {Quataert} E.,  2018, \mn@doi [\apj]
  {10.3847/1538-4357/aab095}, \href
  {https://ui.adsabs.harvard.edu/\#abs/2018ApJ...856..101M} {856, 101}

\bibitem[\protect\citeauthoryear{{Miller} et~al.,}{{Miller}
  et~al.}{2019}]{miller19b}
{Miller} J.~M.,  et~al., 2019, \mn@doi [\prd] {10.1103/PhysRevD.100.023008},
  \href {https://ui.adsabs.harvard.edu/abs/2019PhRvD.100b3008M} {100, 023008}

\bibitem[\protect\citeauthoryear{{Nativi}, {Bulla}, {Rosswog}, {Lundman},
  {Kowal}, {Gizzi}, {Lamb}  \& {Perego}}{{Nativi} et~al.}{2020}]{nativi20}
{Nativi} L.,  {Bulla} M.,  {Rosswog} S.,  {Lundman} C.,  {Kowal} G.,  {Gizzi}
  D.,  {Lamb} G.~P.,   {Perego} A.,  2020, \mn@doi [\mnras]
  {10.1093/mnras/staa3337}, \href
  {https://ui.adsabs.harvard.edu/abs/2020MNRAS.tmp.3146N} {}

\bibitem[\protect\citeauthoryear{{Oechslin}, {Janka}  \& {Marek}}{{Oechslin}
  et~al.}{2007}]{oechslin07}
{Oechslin} R.,  {Janka} H.-T.,   {Marek} A.,  2007, \mn@doi [\aap]
  {10.1051/0004-6361:20066682}, \href
  {http://adsabs.harvard.edu/abs/2007A%26A...467..395O} {467, 395}

\bibitem[\protect\citeauthoryear{{Papenfort}, {Gold}  \&
  {Rezzolla}}{{Papenfort} et~al.}{2018}]{papenfort18}
{Papenfort} L.~J.,  {Gold} R.,   {Rezzolla} L.,  2018, \mn@doi [\prd]
  {10.1103/PhysRevD.98.104028}, \href
  {https://ui.adsabs.harvard.edu/abs/2018PhRvD..98j4028P} {98, 104028}

\bibitem[\protect\citeauthoryear{{Perego}, {Rosswog}, {Cabez{\'o}n},
  {Korobkin}, {K{\"a}ppeli}, {Arcones}  \& {Liebend{\"o}rfer}}{{Perego}
  et~al.}{2014}]{perego14}
{Perego} A.,  {Rosswog} S.,  {Cabez{\'o}n} R.~M.,  {Korobkin} O.,
  {K{\"a}ppeli} R.,  {Arcones} A.,   {Liebend{\"o}rfer} M.,  2014, \mn@doi
  [\mnras] {10.1093/mnras/stu1352}, \href
  {https://ui.adsabs.harvard.edu/abs/2014MNRAS.443.3134P} {443, 3134}

\bibitem[\protect\citeauthoryear{{Perego}, {Radice}  \& {Bernuzzi}}{{Perego}
  et~al.}{2017}]{perego17}
{Perego} A.,  {Radice} D.,   {Bernuzzi} S.,  2017, \mn@doi [\apjl]
  {10.3847/2041-8213/aa9ab9}, \href
  {https://ui.adsabs.harvard.edu/abs/2017ApJ...850L..37P} {850, L37}

\bibitem[\protect\citeauthoryear{{Pian} et~al.,}{{Pian} et~al.}{2017}]{pian17}
{Pian} E.,  et~al., 2017, \mn@doi [\nat] {10.1038/nature24298}, \href
  {http://adsabs.harvard.edu/abs/2017Natur.551...67P} {551, 67}

\bibitem[\protect\citeauthoryear{{Radice}, {Galeazzi}, {Lippuner}, {Roberts},
  {Ott}  \& {Rezzolla}}{{Radice} et~al.}{2016}]{radice16}
{Radice} D.,  {Galeazzi} F.,  {Lippuner} J.,  {Roberts} L.~F.,  {Ott} C.~D.,
  {Rezzolla} L.,  2016, \mn@doi [\mnras] {10.1093/mnras/stw1227}, \href
  {https://ui.adsabs.harvard.edu/abs/2016MNRAS.460.3255R} {460, 3255}

\bibitem[\protect\citeauthoryear{{Radice}, {Perego}, {Hotokezaka}, {Fromm},
  {Bernuzzi}  \& {Roberts}}{{Radice} et~al.}{2018}]{radice18}
{Radice} D.,  {Perego} A.,  {Hotokezaka} K.,  {Fromm} S.~A.,  {Bernuzzi} S.,
  {Roberts} L.~F.,  2018, \mn@doi [ApJ] {10.3847/1538-4357/aaf054}, \href
  {https://ui.adsabs.harvard.edu/abs/2018ApJ...869..130R} {869, 130}

\bibitem[\protect\citeauthoryear{{Roberts}, {Kasen}, {Lee}  \&
  {Ramirez-Ruiz}}{{Roberts} et~al.}{2011}]{roberts11}
{Roberts} L.~F.,  {Kasen} D.,  {Lee} W.~H.,   {Ramirez-Ruiz} E.,  2011, \mn@doi
  [\apj] {10.1088/2041-8205/736/1/L21}, \href
  {https://ui.adsabs.harvard.edu/abs/2011ApJ...736L..21R} {736, L21}

\bibitem[\protect\citeauthoryear{Robinson}{Robinson}{2007}]{robinson07}
Robinson K.,  2007, The P Cygni Profile and Friends.
Springer New York, New York, NY, pp 119--125,
  \mn@doi{10.1007/978-0-387-68288-4_10}, \url
  {https://doi.org/10.1007/978-0-387-68288-4_10}

\bibitem[\protect\citeauthoryear{{Rosswog}, {Thielemann}, {Davies}, {Benz}  \&
  {Piran}}{{Rosswog} et~al.}{1998}]{rosswog98}
{Rosswog} S.,  {Thielemann} F.~K.,  {Davies} M.~B.,  {Benz} W.,   {Piran} T.,
  1998, in {Hillebrandt} W.,  {Muller} E.,  eds, Nuclear Astrophysics. Springer
  New York, p.~103 (\mn@eprint {arXiv} {astro-ph/9804332})

\bibitem[\protect\citeauthoryear{{Rosswog}, {Liebend{\"o}rfer}, {Thielemann},
  {Davies}, {Benz}  \& {Piran}}{{Rosswog} et~al.}{1999}]{rosswog99}
{Rosswog} S.,  {Liebend{\"o}rfer} M.,  {Thielemann} F.~K.,  {Davies} M.~B.,
  {Benz} W.,   {Piran} T.,  1999, \aap, \href
  {https://ui.adsabs.harvard.edu/abs/1999A&A...341..499R} {341, 499}

\bibitem[\protect\citeauthoryear{{Rosswog}, {Korobkin}, {Arcones}, {Thielemann}
   \& {Piran}}{{Rosswog} et~al.}{2014}]{rosswog14}
{Rosswog} S.,  {Korobkin} O.,  {Arcones} A.,  {Thielemann} F.~K.,   {Piran} T.,
   2014, \mn@doi [\mnras] {10.1093/mnras/stt2502}, \href
  {https://ui.adsabs.harvard.edu/\#abs/2014MNRAS.439..744R} {439, 744}

\bibitem[\protect\citeauthoryear{Rosswog, Feindt, Korobkin, Wu, Sollerman,
  Goobar  \& Martinez-Pinedo}{Rosswog et~al.}{2017}]{rosswog17}
Rosswog S.,  Feindt U.,  Korobkin O.,  Wu M.-R.,  Sollerman J.,  Goobar A.,
  Martinez-Pinedo G.,  2017, Classical and Quantum Gravity, 34, 104001

\bibitem[\protect\citeauthoryear{{Rosswog}, {Sollerman}, {Feindt}, {Goobar},
  {Korobkin}, {Wollaeger}, {Fremling}  \& {Kasliwal}}{{Rosswog}
  et~al.}{2018}]{rosswog18}
{Rosswog} S.,  {Sollerman} J.,  {Feindt} U.,  {Goobar} A.,  {Korobkin} O.,
  {Wollaeger} R.,  {Fremling} C.,   {Kasliwal} M.~M.,  2018, \mn@doi [\aap]
  {10.1051/0004-6361/201732117}, \href
  {https://ui.adsabs.harvard.edu/abs/2018A&A...615A.132R} {615, A132}

\bibitem[\protect\citeauthoryear{{Sekiguchi}, {Kiuchi}, {Kyutoku}, {Shibata}
  \& {Taniguchi}}{{Sekiguchi} et~al.}{2016}]{sekiguchi16}
{Sekiguchi} Y.,  {Kiuchi} K.,  {Kyutoku} K.,  {Shibata} M.,   {Taniguchi} K.,
  2016, \mn@doi [\prd] {10.1103/PhysRevD.93.124046}, \href
  {https://ui.adsabs.harvard.edu/abs/2016PhRvD..93l4046S} {93, 124046}

\bibitem[\protect\citeauthoryear{{Shibata} \& {Hotokezaka}}{{Shibata} \&
  {Hotokezaka}}{2019}]{shibata19}
{Shibata} M.,  {Hotokezaka} K.,  2019, \mn@doi [Annual Review of Nuclear and
  Particle Science] {10.1146/annurev-nucl-101918-023625}, \href
  {https://ui.adsabs.harvard.edu/abs/2019ARNPS..6901918S} {69, annurev}

\bibitem[\protect\citeauthoryear{{Shibata}, {Fujibayashi}, {Hotokezaka},
  {Kiuchi}, {Kyutoku}, {Sekiguchi}  \& {Tanaka}}{{Shibata}
  et~al.}{2017}]{shibata17}
{Shibata} M.,  {Fujibayashi} S.,  {Hotokezaka} K.,  {Kiuchi} K.,  {Kyutoku} K.,
   {Sekiguchi} Y.,   {Tanaka} M.,  2017, \mn@doi [\prd]
  {10.1103/PhysRevD.96.123012}, \href
  {https://ui.adsabs.harvard.edu/abs/2017PhRvD..96l3012S} {96, 123012}

\bibitem[\protect\citeauthoryear{{Siegel} \& {Metzger}}{{Siegel} \&
  {Metzger}}{2018}]{siegel18}
{Siegel} D.~M.,  {Metzger} B.~D.,  2018, \mn@doi [\apj]
  {10.3847/1538-4357/aabaec}, \href
  {https://ui.adsabs.harvard.edu/abs/2018ApJ...858...52S} {858, 52}

\bibitem[\protect\citeauthoryear{{Smartt} et~al.,}{{Smartt}
  et~al.}{2017}]{smartt17}
{Smartt} S.~J.,  et~al., 2017, \mn@doi [\nat] {10.1038/nature24303}, \href
  {http://adsabs.harvard.edu/abs/2017Natur.551...75S} {551, 75}

\bibitem[\protect\citeauthoryear{{Smith}, {Tsang}, {Bromm}  \&
  {Milosavljevi{\'c}}}{{Smith} et~al.}{2018}]{smith18}
{Smith} A.,  {Tsang} B. T.~H.,  {Bromm} V.,   {Milosavljevi{\'c}} M.,  2018,
  \mn@doi [\mnras] {10.1093/mnras/sty1509}, \href
  {https://ui.adsabs.harvard.edu/\#abs/2018MNRAS.479.2065S} {479, 2065}

\bibitem[\protect\citeauthoryear{{Tanaka} \& {Hotokezaka}}{{Tanaka} \&
  {Hotokezaka}}{2013}]{tanaka13}
{Tanaka} M.,  {Hotokezaka} K.,  2013, \mn@doi [\apj]
  {10.1088/0004-637X/775/2/113}, \href
  {https://ui.adsabs.harvard.edu/abs/2013ApJ...775..113T} {775, 113}

\bibitem[\protect\citeauthoryear{{Tanaka} et~al.,}{{Tanaka}
  et~al.}{2017}]{tanaka17}
{Tanaka} M.,  et~al., 2017, \mn@doi [PASJ] {doi:10.1093/pasj/psx121}, \href
  {http://adsabs.harvard.edu/abs/2017arXiv171005850T} {}

\bibitem[\protect\citeauthoryear{{Tanaka}, {Kato}, {Gaigalas}  \&
  {Kawaguchi}}{{Tanaka} et~al.}{2020}]{tanaka20}
{Tanaka} M.,  {Kato} D.,  {Gaigalas} G.,   {Kawaguchi} K.,  2020, \mn@doi
  [\mnras] {10.1093/mnras/staa1576}, \href
  {https://ui.adsabs.harvard.edu/abs/2020MNRAS.496.1369T} {496, 1369}

\bibitem[\protect\citeauthoryear{{Tanvir} et~al.,}{{Tanvir}
  et~al.}{2017}]{tanvir17}
{Tanvir} N.~R.,  et~al., 2017, \mn@doi [\apjl] {10.3847/2041-8213/aa90b6},
  \href {http://adsabs.harvard.edu/abs/2017ApJ...848L..27T} {848, L27}

\bibitem[\protect\citeauthoryear{{Troja} et~al.,}{{Troja}
  et~al.}{2017}]{troja17}
{Troja} E.,  et~al., 2017, \mn@doi [\nat] {10.1038/nature24290}, \href
  {http://adsabs.harvard.edu/abs/2017Natur.551...71T} {551, 71}

\bibitem[\protect\citeauthoryear{{Villar} et~al.,}{{Villar}
  et~al.}{2017}]{villar17}
{Villar} V.~A.,  et~al., 2017, \mn@doi [\apj] {10.3847/2041-8213/aa9c84}, \href
  {https://ui.adsabs.harvard.edu/abs/2017ApJ...851L..21V} {851, L21}

\bibitem[\protect\citeauthoryear{{Wallner} et~al.,}{{Wallner}
  et~al.}{2015}]{wallner15}
{Wallner} A.,  et~al., 2015, \mn@doi [Nature Communications]
  {10.1038/ncomms6956}, \href
  {https://ui.adsabs.harvard.edu/abs/2015NatCo...6.5956W} {6, 5956}

\bibitem[\protect\citeauthoryear{{Wanajo}, {Sekiguchi}, {Nishimura}, {Kiuchi},
  {Kyutoku}  \& {Shibata}}{{Wanajo} et~al.}{2014}]{wanajo14}
{Wanajo} S.,  {Sekiguchi} Y.,  {Nishimura} N.,  {Kiuchi} K.,  {Kyutoku} K.,
  {Shibata} M.,  2014, \mn@doi [\apj] {10.1088/2041-8205/789/2/L39}, \href
  {https://ui.adsabs.harvard.edu/abs/2014ApJ...789L..39W} {789, L39}

\bibitem[\protect\citeauthoryear{{Wehmeyer}, {Fr{\"o}hlich}, {C{\^o}t{\'e}},
  {Pignatari}  \& {Thielemann}}{{Wehmeyer} et~al.}{2019}]{wehmeyer19}
{Wehmeyer} B.,  {Fr{\"o}hlich} C.,  {C{\^o}t{\'e}} B.,  {Pignatari} M.,
  {Thielemann} F.~K.,  2019, \mn@doi [\mnras] {10.1093/mnras/stz1310}, \href
  {https://ui.adsabs.harvard.edu/abs/2019MNRAS.487.1745W} {487, 1745}

\bibitem[\protect\citeauthoryear{{Winteler}, {K{\"a}ppeli}, {Perego},
  {Arcones}, {Vasset}, {Nishimura}, {Liebend{\"o}rfer}  \&
  {Thielemann}}{{Winteler} et~al.}{2012}]{winteler12}
{Winteler} C.,  {K{\"a}ppeli} R.,  {Perego} A.,  {Arcones} A.,  {Vasset} N.,
  {Nishimura} N.,  {Liebend{\"o}rfer} M.,   {Thielemann} F.~K.,  2012, \mn@doi
  [\apjl] {10.1088/2041-8205/750/1/L22}, \href
  {https://ui.adsabs.harvard.edu/abs/2012ApJ...750L..22W} {750, L22}

\bibitem[\protect\citeauthoryear{{Wollaeger} \& {van Rossum}}{{Wollaeger} \&
  {van Rossum}}{2014}]{wollaeger14}
{Wollaeger} R.~T.,  {van Rossum} D.~R.,  2014, \mn@doi [The Astrophysical
  Journal Supplement Series] {10.1088/0067-0049/214/2/28}, \href
  {https://ui.adsabs.harvard.edu/\#abs/2014ApJS..214...28W} {214, 28}

\bibitem[\protect\citeauthoryear{{Wollaeger}, {Hungerford}, {Fryer},
  {Wollaber}, {van Rossum}  \& {Even}}{{Wollaeger} et~al.}{2017}]{wollaeger17}
{Wollaeger} R.~T.,  {Hungerford} A.~L.,  {Fryer} C.~L.,  {Wollaber} A.~B.,
  {van Rossum} D.~R.,   {Even} W.,  2017, \mn@doi [\apj]
  {10.3847/1538-4357/aa82bd}, \href
  {https://ui.adsabs.harvard.edu/abs/2017ApJ...845..168W} {845, 168}

\bibitem[\protect\citeauthoryear{{Wollaeger} et~al.,}{{Wollaeger}
  et~al.}{2018}]{wollaeger18}
{Wollaeger} R.~T.,  et~al., 2018, \mn@doi [\mnras] {10.1093/mnras/sty1018},
  \href {http://adsabs.harvard.edu/abs/2018MNRAS.478.3298W} {478, 3298}

\bibitem[\protect\citeauthoryear{{Wollaeger} et~al.,}{{Wollaeger}
  et~al.}{2019}]{wollaeger19}
{Wollaeger} R.~T.,  et~al., 2019, \mn@doi [\apj] {10.3847/1538-4357/ab25f5},
  \href {https://ui.adsabs.harvard.edu/abs/2019ApJ...880...22W} {880, 22}

\bibitem[\protect\citeauthoryear{{Zhu} et~al.,}{{Zhu} et~al.}{2018}]{zhuyl18}
{Zhu} Y.,  et~al., 2018, \mn@doi [\apjl] {10.3847/2041-8213/aad5de}, \href
  {https://ui.adsabs.harvard.edu/abs/2018ApJ...863L..23Z} {863, L23}

\bibitem[\protect\citeauthoryear{{Zhu}, {Barnes}, {Lund}, {Sprouse}, {Vassh},
  {Mumpower}, {McLaughlin}  \& {Surman}}{{Zhu} et~al.}{2019}]{zhuyl20}
{Zhu} Y.,  {Barnes} J.,  {Lund} K.,  {Sprouse} T.,  {Vassh} N.,  {Mumpower} M.,
   {McLaughlin} G.,   {Surman} R.,  2019, in APS Division of Nuclear Physics
  Meeting Abstracts. p. SM.003

\bibitem[\protect\citeauthoryear{{Zhu}, {Yang}, {Liu}, {Huang}, {Zhang}, {Li},
  {Yu}  \& {Gao}}{{Zhu} et~al.}{2020}]{zhujp20}
{Zhu} J.-P.,  {Yang} Y.-P.,  {Liu} L.-D.,  {Huang} Y.,  {Zhang} B.,  {Li} Z.,
  {Yu} Y.-W.,   {Gao} H.,  2020, \mn@doi [\apj] {10.3847/1538-4357/ab93bf},
  \href {https://ui.adsabs.harvard.edu/abs/2020ApJ...897...20Z} {897, 20}

\bibitem[\protect\citeauthoryear{{van Rossum}, {Kashyap}, {Fisher},
  {Wollaeger}, {Garc{\'\i}a-Berro}, {Aznar-Sigu{\'a}n}, {Ji}  \&
  {Lor{\'e}n-Aguilar}}{{van Rossum} et~al.}{2016}]{vanrossum16}
{van Rossum} D.~R.,  {Kashyap} R.,  {Fisher} R.,  {Wollaeger} R.~T.,
  {Garc{\'\i}a-Berro} E.,  {Aznar-Sigu{\'a}n} G.,  {Ji} S.,
  {Lor{\'e}n-Aguilar} P.,  2016, \mn@doi [\apj] {10.3847/0004-637X/827/2/128},
  \href {https://ui.adsabs.harvard.edu/\#abs/2016ApJ...827..128V} {827, 128}

\makeatother
\end{thebibliography}
\end{document}